\documentclass[9pt,superscriptaddress,groupedaddress,amsmath,amssymb,aps,twocolumn]{revtex4}
\renewcommand{\epsilon}{\varepsilon}
\usepackage{graphicx}
\usepackage[colorlinks=true,linkcolor=blue,urlcolor=blue,citecolor=blue,anchorcolor=blue]{hyperref}

\begin{document}

\title{Stochastic sampling effects favor manual over digital contact tracing}

\author{Marco Mancastroppa}
\affiliation{Dipartimento di Scienze Matematiche, Fisiche e Informatiche,
  Universit\`a degli Studi di Parma, Parco Area delle Scienze, 7/A 43124 Parma, Italy}
\affiliation{INFN, Sezione di Milano Bicocca, Gruppo Collegato di Parma, Parco Area delle Scienze, 7/A 43124 Parma, Italy}
\author{Claudio Castellano} 
\affiliation{Istituto dei Sistemi Complessi (ISC-CNR), Via dei Taurini 19, I-00185 Roma, Italy}
\author{Alessandro Vezzani}
\affiliation{Istituto dei Materiali per l'Elettronica ed il Magnetismo (IMEM-CNR), Parco Area delle Scienze, 37/A-43124 Parma, Italy}
\affiliation{Dipartimento di Scienze Matematiche, Fisiche e Informatiche,
  Universit\`a degli Studi di Parma, Parco Area delle Scienze, 7/A 43124 Parma, Italy}
\author{Raffaella Burioni}
\affiliation{Dipartimento di Scienze Matematiche, Fisiche e Informatiche,
  Universit\`a degli Studi di Parma, Parco Area delle Scienze, 7/A 43124 Parma, Italy}
\affiliation{INFN, Sezione di Milano Bicocca, Gruppo Collegato di Parma, Parco Area delle Scienze, 7/A 43124 Parma, Italy}

\begin{abstract}
Isolation of symptomatic individuals, tracing and testing of their
nonsymptomatic contacts are fundamental strategies for mitigating the current COVID-19 pandemic. The breaking of contagion chains relies on two complementary strategies: manual reconstruction of contacts based on interviews and a digital (app-based) privacy-preserving contact tracing. We compare their effectiveness using model parameters tailored to describe SARS-CoV-2 diffusion within the activity-driven model, a general empirically validated framework for network dynamics.
We show that, even for equal probability of tracing a contact, manual
tracing robustly performs better than the digital protocol, also taking into account the intrinsic delay and limited scalability of the manual procedure.
This result is explained in terms of the stochastic sampling occurring during the case-by-case manual reconstruction of contacts, contrasted with the intrinsically
prearranged nature of digital tracing, determined by the decision to adopt the app or not by each individual. The better performance of manual tracing is enhanced by heterogeneity in agent behavior: superspreaders not adopting the app are completely invisible to digital contact tracing, while they can be easily traced manually, due to their multiple contacts. We show that this intrinsic difference makes the manual procedure dominant in realistic hybrid protocols.
\end{abstract}

\maketitle

\section{Introduction}
The current COVID-19 pandemic is impacting daily life worldwide at an
unprecedented scale. Among the features that have contributed to
transform the emerging diffusion of SARS-CoV-2 coronavirus into such a
global scale crisis, a prominent role is played by the high rates of
virus transmission mediated by presymptomatic and asymptomatic
individuals~\cite{Lavezzo2020, Ferretti2020,Rothe2020,Pinotti2020,Li2020,wei2020presymptomatic}.
Given the absence of effective
pharmaceutical interventions, this feature makes the mitigation of the pandemic a highly nontrivial
task, that has been tackled with various strategies, none of them
devoid of drawbacks.\\
\indent Initially, governments resorted to very restrictive limitations of non
strictly necessary activities (lock-downs) to curb the diffusion of
the infection. Such measures turned out to be effective from an
epidemiological point of view, but exceedingly costly in other
respects, for their economic and social
consequences~\cite{Bonaccorsi2020,Nicola2020}. Recently, such
restrictive measures have been progressively lifted, and
we now rely on other tools to contain the pandemic: social distancing,
reinforced hygiene and the use of individual protection devices. Along
with these provisions, aimed at preventing single virus transmission
events, another set of measures points at breaking contagion chains:
the isolation of infected individuals (symptomatic or found via some
testing), followed by the tracing of their contacts (contact tracing,
CT), the testing of the latter and the possible isolation of the
infected~\cite{Eames2003,Fraser2004,Klinkenberg2006,kojaku2020}.

This CT procedure has proven effective in the past in various
contexts~\cite{Mclean2010,Crook2017,Bell2004,WilderSmith2020}
but it comes, in its standard manual implementation, with important
limitations~\cite{Hellewell2020}.  It requires the set up of a
physical infrastructure, needed to find infected individuals,
interview them and reconstruct their contacts in a temporal window,
call these contacts, convince them to get tested and eventually
isolated. Apart from the evident problems of practical feasibility and
economic cost, the manual CT procedure intrinsically implies a
delay between the moment an individual is found infected (and
isolated) and the time her contacts are tested and possibly isolated.
For an epidemic such as COVID-19, characterized by a rather long
presymptomatic infectious stage and a high relevance of transmission
by asymptomatics, the delay implied by manual CT risks to undermine
the effectiveness of the whole procedure.\\
\indent For this reason the additional strategy of a digital CT procedure,
based in particular on the installation of apps on smartphones
(app-based), has been proposed alongside manual CT~\cite{Ferretti2020}. The rationale is
that proximity sensors installed on these ubiquitous devices allow the
detection of contacts of epidemiological significance among
individuals. When an individual is found infected, the app permits to
instantaneously trace all contacts in the recent past, thus allowing
for much quicker testing and isolation. A quantitative comparison
between manual and digital CT applied to an epidemiological model
describing COVID-19 diffusion suggested that already a delay of the
order of 3 days completely spoils the
ability of manual CT to prevent the initial exponential growth of the
epidemic~\cite{Ferretti2020}. The conclusion was that only a digital
CT avoiding this delay could be a viable strategy to control the
current epidemic. The proposal for digital CT rapidly gained momentum,
leading to the development of technical
solutions~\cite{PPCT2020,Troncoso2020, Oliver2020} and to the
deployment of app-based CT infrastructures in many
countries~\cite{Ahmed2020}.

Many works have scrutinized the actual validity of this solution and
investigated the possible shortcomings of app-based CT, casting doubts
over many of the assumptions underlying such
strategy~\cite{Sapiezynski2020,Braithwaite2020,Zastrow2020}:
there are too few modern enough smartphones; Bluetooth based proximity
measurements are unreliable; co-location is not always a good proxy
for epidemiological contact. The potential risks for privacy breaches
have also been exposed.

Other papers have tried to evaluate the impact of digital CT on the
current COVID-19 epidemic, attempting to precisely determine, by means
of detailed data-driven epidemiological models, to what extent such a
strategy is able to suppress virus
diffusion~\cite{Cencetti2020,Almagor2020,Nielsen2020,Kretzschmar2020,Barrat2020}.
A critical role is played by the fraction $f$ of individuals in a
population that actually use the app. Fairly high values
of $f$ (of the order of $60\%$) are required for the digital CT
protocol to lead to global
protection~\cite{Eames2003,Ferretti2020,Braithwaite2020}. These values
are in striking contrast with the low app adoption rates observed so
far in most countries~\cite{Longo2020,MIT2020,MIT2020_2}.

In this paper we take a different approach. 
We compare the effectiveness of the two CT protocols in
exactly the same conditions, i.e., in the very same realistic
epidemiological scenario, without making claims on their absolute
performance, and we estimate their relative contribution in realistic hybrid protocols, where the two strategies are complementary. We consider a sensible epidemiological model
incorporating all main ingredients of the current epidemic, with
parameters tuned to values derived from empirical observations about
COVID-19 spreading.  Within this single framework we consider the
impact of both manual and digital CT strategies, working in similar
manner but with their own specific features: delayed isolation of
contacts, limited scalability and imperfect recall for the manual procedure; dependence on the predetermined app adoption decision for the digital CT.

The comparison reveals that even when the number of reconstructed
contacts is the same, manual CT performs better than digital CT in
practically all realistic cases. The manual protocol is more efficient
in increasing the epidemic threshold (i.e., the value of the effective
infection rate above which the infection spreads diffusely), in
limiting the height of the epidemic peaks and in reducing the number
of isolated individuals.  This surprising result is due to the
stochastic {\it annealed} nature of the manual CT procedure, in which
each symptomatic node randomly recalls a fraction of her contacts, in
contrast with the digital CT where the traced nodes belong
deterministically to the prearranged {\it quenched} fraction of the
population adopting the app.  In the latter case, the individuals not
adopting the app can never be reached by the CT protocol, while the
entire population is potentially detectable through the stochastic
sampling of the manual procedure. The better performance of manual CT
is already evident in homogeneous populations and it is strongly
enhanced in the presence of a heterogeneous distribution of
contacts. Superspreaders not adopting the app are invisible to the
digital tracing while they are very likely to be detected by a manual
tracing originating from one of their many contacts.
  In a realistic setup where both protocols are adopted
  simultaneously, manual CT leads to a considerable reduction of the
  transmission even considering delays and scalability, while the
  digital protocol produces a relevant contribution only for large
  adoption rates, as suggested by previous literature. 

\section{Results}

\subsection{Epidemic spreading on heterogeneous dynamical networks}
We consider an activity-driven network model with attractiveness, taking
into account both the temporal dynamics of social contacts and the
heterogeneity in the propensity to establish social
ties~\cite{Perra2012,Ubaldi2016,Pozzana2017,Tizzani2018}.  Each susceptible node
$S$ is assigned with an activity $a_S$ and an attractiveness parameter
$b_S$, drawn from the joint distribution $\rho(a_S,b_S)$: the
activation rate $a_S$ describes the Poissonian activation dynamics of
the node; the attractiveness $b_S$ sets the probability $p_{b_S}
\varpropto b_S$ for a node to be contacted by an active agent. At the
beginning all nodes are disconnected and when a node activates it
creates $m$ links with $m$ randomly selected nodes (hereafter we set
$m=1$); then all links are destroyed and the procedure is iterated.
The functional form of $\rho(a_S,b_S)$ encodes the correlations between
activity and attractiveness in a population with a given distribution
of activity. It has been observed that several social systems feature
positive correlations between activity and attractiveness and a broad
power-law distribution of
activity~\cite{Tizzani2018,Perra2012,Pozzana2017,Ubaldi2016,Ghoshal2006,Ubaldi2017,Starnini2013}:
\begin{equation}
\rho(a_S,b_S) \sim a_S^{-(\nu+1)} \delta(b_S- a_S)
\label{eq:act}
\end{equation}
with $\nu$ typically ranging between $0.5$ and $2$.

\begin{figure}
\centering
\includegraphics[width=80mm]{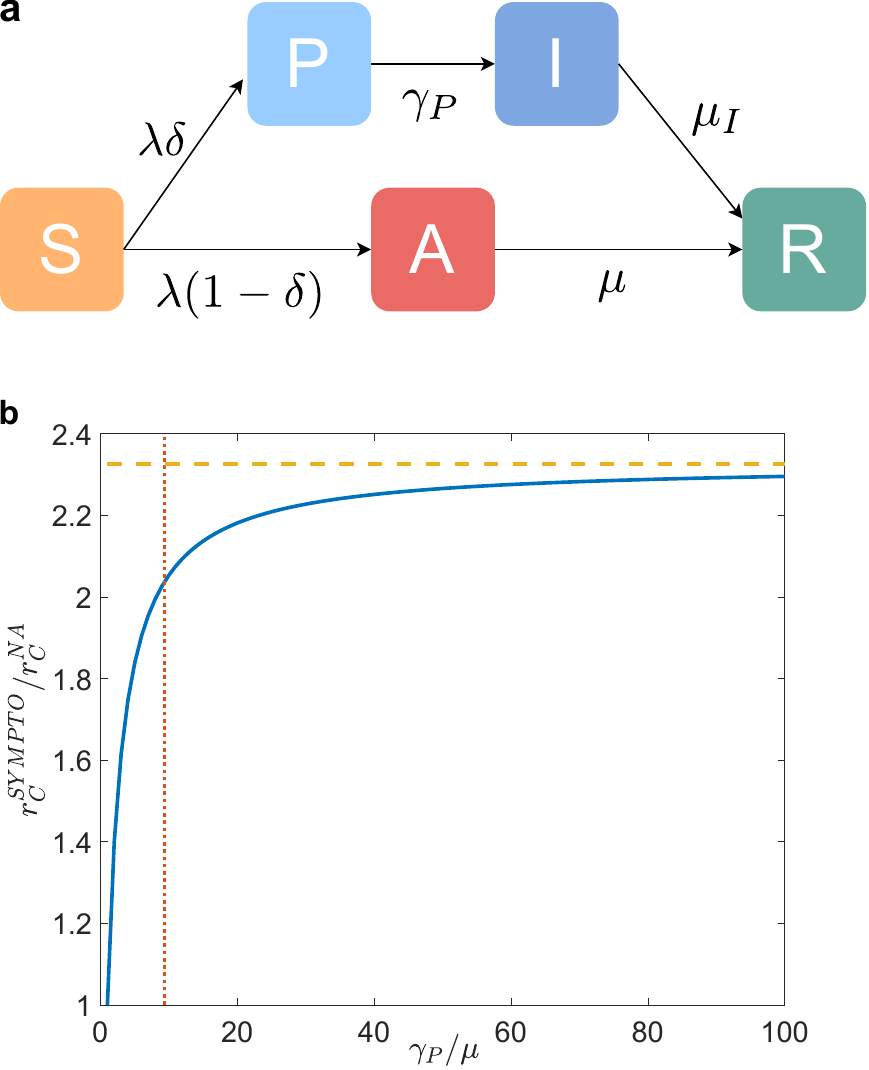}
  \caption{{\bf Epidemic model without contact tracing.} {\bf a} Diagram of  the compartmental epidemic model without
    CT. {\bf b} We  plot, as a function of $\gamma_P/\mu$,
    the ratio between the epidemic
    threshold $r_C^{SYMPTO}$ when symptomatic nodes are isolated and
    the epidemic threshold of the non-adaptive case $r_C^{NA}$.
    The horizontal dashed orange line indicates the
    value of $r_C^{SYMPTO}/r_C^{NA}$ for $\gamma_P=\infty$
    (instantaneous onset of symptoms).
    The vertical red dash-dotted line indicates the value of $\gamma_P/\mu$
    we consider in the rest of the paper.
    We set $\rho(a_S,b_S) = \rho_S(a_S) \delta(b_S-a_S)$. The curve does
    not depend on the specific form of $\rho_S(a_S)$.}
 \label{fig:scheme_sympto_homo}
\end{figure}

On top of the activity-driven dynamics, we consider a compartmental
epidemic model which includes the main phases of clinical progression
of the SARS-CoV-2 infection~\cite{wei2020presymptomatic,Guan2020,DiDomenico2020,who_china_report}, also
applicable to other infectious diseases with asymptomatic and
presymptomatic transmission. The model is composed by five
compartments: $S$ susceptible, $P$ presymptomatic, $A$ infected
asymptomatic, $I$ infected symptomatic, $R$ recovered. A contagion
process (see Fig.~\ref{fig:scheme_sympto_homo}a)
occurs with probability $\lambda$ when a link is established
between an infected (either $P$, $A$ and $I$) and a susceptible
node $S$ (contact-driven transition): a node has probability $\delta$
to become presymptomatic after infection and probability $(1-\delta)$ to
become asymptomatic, thus $S \xrightarrow[]{\lambda \delta} P$ and $S
\xrightarrow[]{\lambda (1-\delta)} A$.  A presymptomatic node
spontaneously develops symptoms with rate $\gamma_P=1/\tau_P$, thus
with a Poissonian process $P \xrightarrow[]{\gamma_P} I$; both
asymptomatic and symptomatic nodes spontaneously recover respectively
with rate $\mu=1/\tau$ and with rate $\mu_I=\mu \gamma_P/(\gamma_P -
\mu)$, so that the average infectious period for both symptomatic and
asymptomatic is $\tau$. We neglect states of hospitalization and consider recovery without death: this choice does not
affect the infection dynamics.

Adaptive behavior of populations exposed to epidemics can be
modelled within the activity-driven network framework: infected nodes
experience a reduction in activity, due to isolation or the
appearance of symptoms; similarly, other individuals undertake
self-protective behavior to reduce the probability of contact with an
infected node, and this is modelled as a reduction in the
attractiveness of infected nodes~\cite{Vankerkove2013,Mancastroppa2020,Fenichel2010}.
We assume that symptomatic infected nodes $I$ are immediately isolated
$(a_I,b_I)=(0,0)$, therefore not being able to infect anymore.
On the contrary, we assume that recovered $R$, asymptomatic $A$
and presymptomatic $P$ individuals behave as when they were susceptible
$(a_A,b_A)=(a_P,b_P)=(a_R,b_R)=(a_S,b_S)$.
The adaptive behavior is implemented without affecting the activity of
nodes which are not isolated~\cite{Mancastroppa2020}.

The control parameter $r=\lambda/\mu$ is the effective infection rate,
whose critical value $r_C$ -- the epidemic threshold -- sets the transition
point between the absorbing and the active phase of the epidemic.
The increase in the value of $r_C$ is an indicator of the effectiveness
of mitigation strategies.
Within the adaptive activity-driven framework, the epidemic threshold can
be calculated analytically via a mean-field approach
(see Methods).

The effect of isolating symptomatic nodes as the only containment
measure is shown in Fig.~\ref{fig:scheme_sympto_homo}b. We compare
the epidemic threshold $r_C$, obtained with the isolation of
symptomatic nodes only, with the epidemic threshold $r_C^{NA}$ of the
non-adaptive (NA) case, in which no containment measures are taken on
infected individuals, as a function of $\gamma_P/\mu$ (see Methods for the explicit expression).
In the case of instantaneous symptoms development
($\gamma_P/\mu \to \infty$), the threshold is increased by a factor of
$1/(1-\delta)$, while for smaller $\gamma_P/\mu$ the gain is
reduced. For example, for $(1-\delta)=0.43$, that is $43\%$ of
asymptomatic individuals, $\tau_P = 1.5$ days and $\tau=14$ days
as observed for
SARS-CoV-2 (see Methods for details on the parameters
used in all figures), the
threshold is doubled by the isolation of symptomatic nodes.
This is the baseline reference for the evaluation of the performance
of CT strategies. 

\begin{figure}
\centering
\includegraphics[width=80mm]{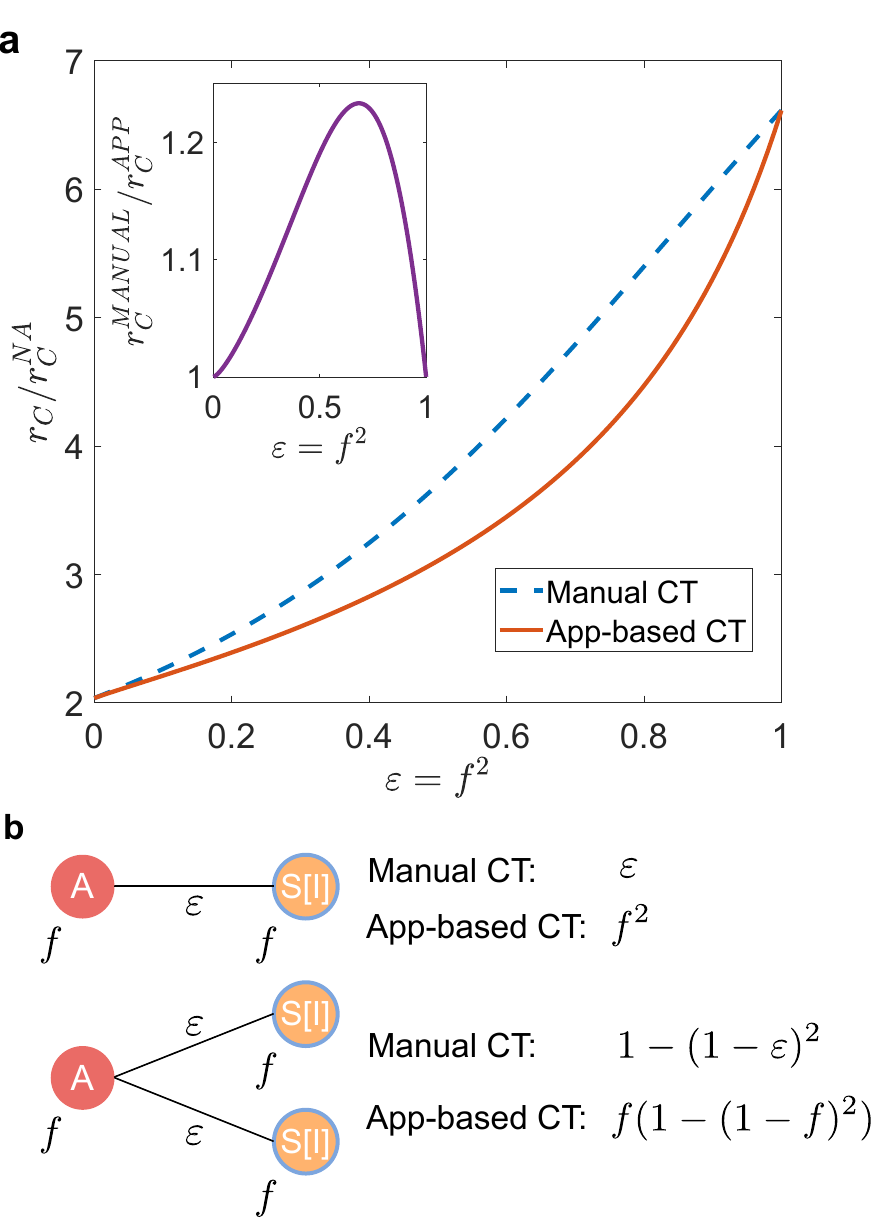}
  \caption{{\bf Stochastic vs. prearranged sampling in contact tracing.} 
    {\bf a} Plot, as a function of $\epsilon=f^2$,
     of the ratio between the epidemic threshold $r_C$ in
    the presence of CT protocols and the epidemic threshold of the
    non-adaptive case $r_C^{NA}$. In the inset we
    plot the ratio between the epidemic threshold of the manual CT
    $r_C^{MANUAL}$ and that of the app-based CT $r_C^{APP}$, as a
    function of $\epsilon=f^2$.
    We consider homogeneous activity and attractiveness, setting
    $\rho(a_S,b_S)=\delta(a_S-a) \delta(b_S-b)$.
    {\bf b} If
    an asymptomatic node $A$ infects a single
    susceptible node $S$ (which subsequently becomes symptomatic, $I$),
    the infector is traced with probability $\epsilon$ in
    the manual CT and with probability $f^2$ in the digital CT.
    Thus the probability is the same if we impose $f^2=\epsilon$.
    However, if an asymptomatic node infects two susceptible nodes (which
    subsequently become symptomatic), the probability of tracing
    the infector with the manual protocol is $1-(1-\epsilon)^2=2f^2-f^4$
    (still considering $\epsilon=f^2$). This value is always larger than
    that of the digital protocol $f(1-(1-f)^2)=2f^2-f^3$.
}
 \label{fig:stoch}
\end{figure}

\subsection{Manual and digital contact tracing protocols}
The CT protocols differ in their practical implementation as well in
their exploration properties.

Manual tracing is performed by personnel who, through
interviews, collects information, contacts individuals who may have been
infected and arranges for testing.  In manual CT, as soon as an
individual develops symptoms (i.e. $P \to I$), her contacts
in the previous $T_{CT}$ days
are traced
with recall probability $\epsilon(a_S)$, where $a_S$ is the activity of the
symptomatic individual. A traced contact is tested and, if found in
state $A$ (infected asymptomatic), isolated ($a=b=0$): the
average time between the isolation of the symptomatic individual and
the isolation of her asymptomatic infected contacts is $\tau_C$. Such
delay can be quite large, due to the time required for the collection
of the diary, the execution of the diagnostic test and the subsequent
isolation~\cite{Ferretti2020,Hellewell2020}.  Moreover, the manual
protocol depends on $\epsilon(a_S)$, which takes into account the
limited resources allocated for tracing and the limited memory/knowledge
of symptomatic individuals in reconstructing their contacts.
Low activity nodes make few contacts over time and a fraction of
their contacts will be traced; on the other hand
high activity nodes will only remember a finite number of
their contacts so that, also because of limitations of the tracing
capacity, we expect that at most a number $k_c$ of contacts can be
traced~\cite{ECDC_ct,CDC_ct}.
This translates into the limited scalability property:

\begin{equation}
\epsilon(a_S)=
\begin{cases}
\epsilon^*, & \mbox{if} \, a_S \leq a^*\\
\epsilon^* \frac{a^*}{a_S}=\frac{k_c}{2T_{CT}a_S}, & \mbox{if} \, a_S>a^*
\end{cases}
\label{eq:scal}
\end{equation}

where $a^*=k_c/2T_{CT}\epsilon^*$. Manual CT also
  suffers from a global scalability limitation. Indeed in the active
  phase with a large number of infected individuals, the tracing
  system could be ineffective due to the excessive number of contacts
  to be followed. However, here we focus on the epidemic threshold and
  we evaluate the features of the CT procedure that keep the
  epidemic spreading under control before widespread diffusion occurs.\\

Digital CT is based on the download of an app which allows the tracing
of close contacts equipped with the same app.  We assume that each of
the individuals has a probability $f$ to download the app before the
epidemic starts. As soon as an individual develops symptoms (i.e. $P
\to I$), if she downloaded the app, her contacts are traced only if
they downloaded the app as well. A traced contact is tested and, if
found in state $A$ (infected asymptomatic), is isolated ($a=b=0$).
The time passing between the isolation of a symptomatic individual and
the isolation of her asymptomatic infected contacts is taken to be
$0$, thus assuming an idealized scenario of instantaneous notification
and isolation. We finally consider a more realistic
  scenario where the two procedures are combined into a hybrid
  protocol in which digital CT supports manual CT, potentially
  reaching individuals not traced
  manually~\cite{Troncoso2020,Cencetti2020,Barrat2020}.
See Methods for details on the implementation of the 
CT protocols.

\subsection{Stochastic vs. prearranged sampling}

\begin{figure*}
\includegraphics[width=180mm]{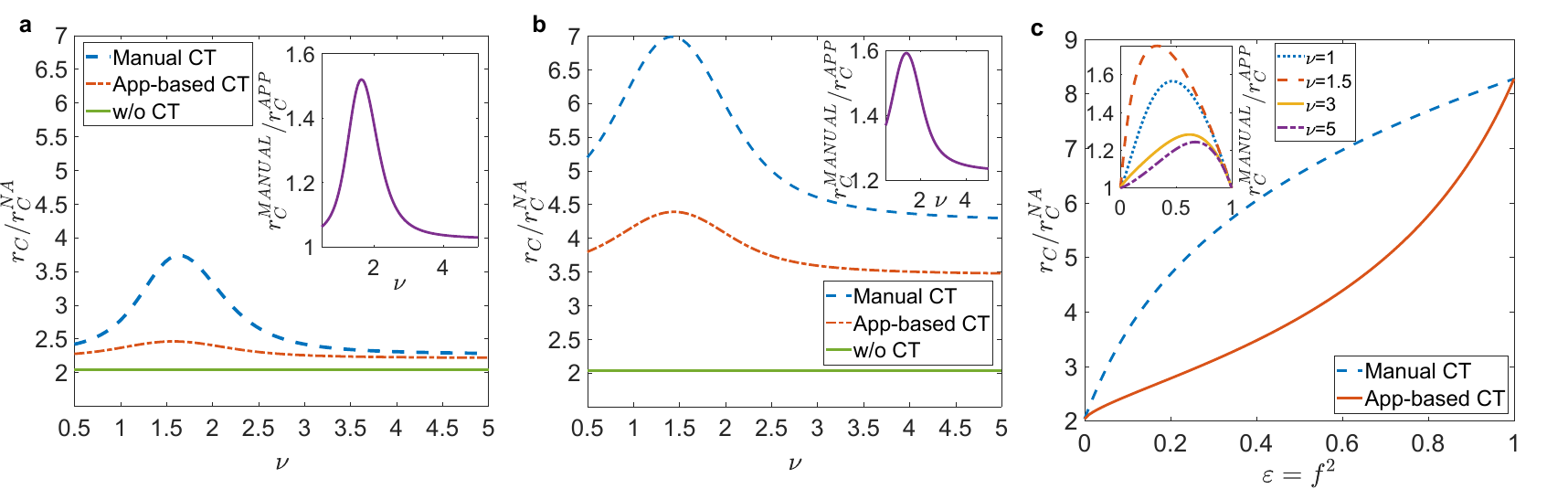}
\caption{{\bf Effects of heterogeneity.} 
    {\bf a} Plot, as a function of 
    the exponent $\nu$, of 
  the ratio between the epidemic threshold $r_C$ in
  the presence of CT protocols and the epidemic threshold of the
  non-adaptive case $r_C^{NA}$. We set $\epsilon=f^2=0.1$ ($f \approx 0.316$).
  In the inset we plot the ratio between the epidemic threshold of the manual CT
  $r_C^{MANUAL}$ and that of the app-based CT $r_C^{APP}$, as a
   function of $\nu$. {\bf b} Same plot as {\bf a} with $\epsilon=f^2=0.6$ ($f \approx 0.775$). {\bf c}  The
  ratio $r_C/r_C^{NA}$ is plotted as a function of $\epsilon=f^2$ for both 
  CT protocols, with $\nu=1.5$. In the inset we plot the ratio
  $r_C^{MANUAL}/r_C^{APP}$ as a function of $\epsilon=f^2$ for
  several $\nu$ values.
  In all panels the distribution $\rho(a_S,b_S)$ is given by Eq.~\eqref{eq:act}.}
 \label{fig:heterog}
\end{figure*}

We first compare the manual and digital CT protocols in the case of a population with
homogeneous activity and attractiveness, $\rho(a_S,b_S)=\delta(a_S-a)
\delta(b_S-b)$, without delay even in manual CT, i.e., $\tau_C=0$.  We
set $\epsilon=f^2$ so that the probability that a single contact is
traced in the two protocols is the same. However, it should be
emphasized that typical values of $f^2$ range between $0.01$ and $0.1$
in many countries~\cite{MIT2020_2,MIT2020,Longo2020}, while $\epsilon$
is usually larger ($\approx 0.3-0.5$), since typically $30\%-50\%$ of contacts are close and occur at home, at work or at school and thus are easily
traceable~\cite{Mossong2008,Keeling2020}.  An exact analytical estimate of the
epidemic threshold is obtained through a linear stability analysis
around the absorbing state (see Methods for explicit
expressions and Supplementary Method 1 for detailed derivation): in Fig.~\ref{fig:stoch}a we compare the threshold for
manual and app-based CT, compared to the non-adaptive case (NA), for
realistic COVID-19 parameters.  Both protocols feature the same
epidemic threshold when $\epsilon=f=0$ and $\epsilon=f=1$: indeed the
former corresponds to the isolation of symptomatic individuals only,
without CT, while the latter limit corresponds to the case in which
all contacts are traced. For intermediates values of $\epsilon = f^2$,
manual tracing is strongly and surprisingly more effective than
digital tracing, as it increases significantly the epidemic threshold,
compared to the app-based protocol. Since we are considering no
heterogeneity or delays, the difference is due only to sampling
effects in the CT dynamics.  In practice, in app-based CT the
population to be tested is prearranged, based on whether or not the
app was downloaded before the outbreak started. On the contrary,
manual CT performs a stochastic sampling of the population: the random
exploration can potentially reach the entire population, since anyone
who has come in contact with a symptomatic node can be traced.
The simplest example of the difference in tracing multiple
infections processes in the two protocols is illustrated in
Fig.~\ref{fig:stoch}b.

\subsection{Effects of heterogeneous activity}
We now consider a heterogeneous activity distribution, as observed in
several human systems, and we consider a positive
activity-attractiveness correlation, as defined in
Eq.~\eqref{eq:act}~\cite{Perra2012,Pozzana2017,Ubaldi2016,Ghoshal2006,Ubaldi2017,Starnini2013}.
The power-law distribution for the activity implies the presence of
hubs with high activity and high attractiveness.  We
investigate the pure effect of heterogeneities in contact tracing~\cite{Nielsen2020,kojaku2020} setting now $\epsilon(a_S)=\epsilon, \, \forall a_S$ and not considering delays in
manual CT, $\tau_C=0$.  We perform again a mean-field approach,
obtaining an analytical closed form for the epidemic threshold (see
Methods and Supplementary Method 1): in
Fig.~\ref{fig:heterog} we compare the epidemic threshold with the two
protocols as a function of the exponent $\nu$ of the activity
distribution, for realistic parameters and setting an average activity
$\overline{a_S}=6.7$ days$^{-1}$~\cite{Mossong2008,Vankerkove2013}.
Both protocols are more effective in heterogeneous populations, that is at $\nu
\sim 1-1.5$. Note that due to the cut-offs and the constraint on the average, the fluctuations of the activity distribution are maximum for $\nu=1$, and the epidemic thresholds depend both on activity fluctuations and higher order moments of $\rho_S(a_S)$ (see Methods and Supplementary Method 1 for details). However heterogeneity greatly amplifies stochastic
effects, further increasing the advantage of the manual tracing over
the app-based prearranged protocol.  Indeed, in heterogeneous
populations nodes with high activity and attractiveness
(super-spreaders) drive and sustain the spread of the epidemic. Manual
CT is far more effective in identifying and isolating them than
app-based CT: in digital CT, hubs which have not downloaded the app
will never be traced, despite the high number of their contacts. On
the contrary, manual CT is very effective in tracing super-spreaders,
because they are engaged in many contacts and are traced very
effectively by stochastic exploration.

\subsection{Limited scalability and delay in manual contact tracing}
\begin{figure*}
\includegraphics[width=180mm]{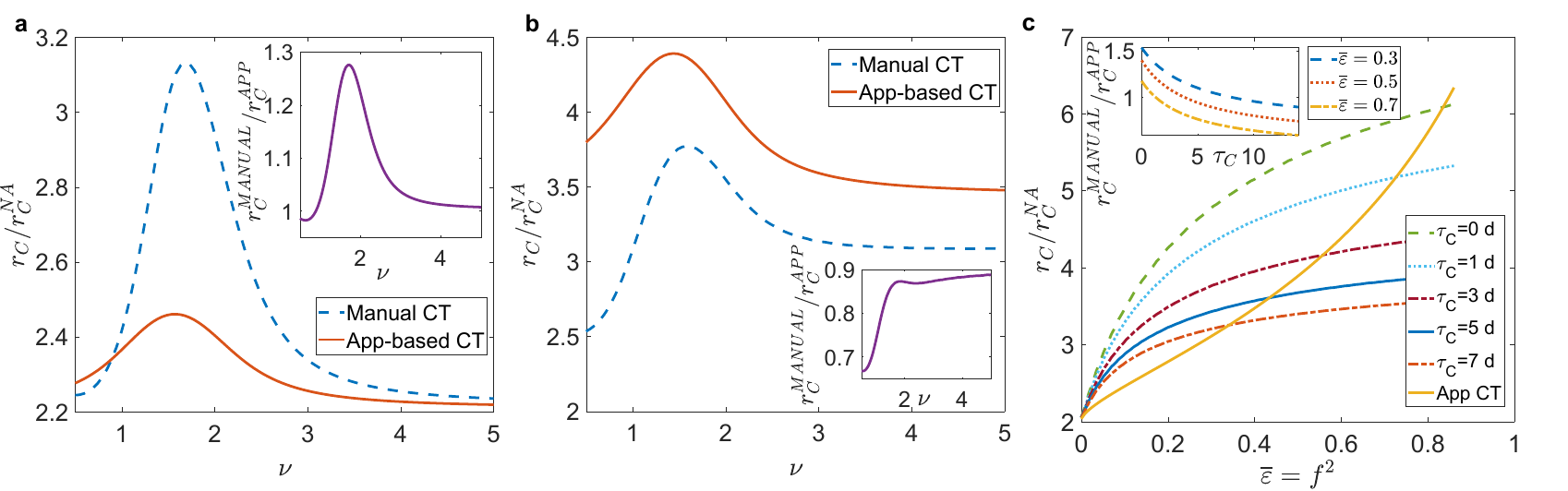}
\caption{{\bf Effects of limited scalability and isolation delay in manual
    contact tracing.} 
{\bf a} Plot, as a function of 
    the exponent $\nu$, of 
  the ratio between the epidemic threshold $r_C$ in
  the presence of CT protocols and the epidemic threshold of the
  non-adaptive case $r_C^{NA}$ , with
  limited scalability and delay in manual CT. We set $\overline{\epsilon}=f^2=0.1$ ($f \approx 0.316$), $\tau_C=3$ days. 
  In the insets we plot
  the ratio between the epidemic threshold of the manual CT
  $r_C^{MANUAL}$ and that of the app-based CT $r_C^{APP}$, as a
    function of $\nu$. 
    {\bf b} Same plot as {\bf a} with
  $\overline{\epsilon}=f^2=0.6$ ($f \approx 0.775$), $\tau_C=5$ days. 
{\bf c} Plot of
  the ratio $r_C/r_C^{NA}$ as a function of
  $\overline{\epsilon}=f^2$ for both CT protocols, setting
  $\nu=1.5$ and for several values of $\tau_C$. In the inset we plot the ratio
  $r_C^{MANUAL}/r_C^{APP}$ as a function of $\tau_C$ for several
  $\overline{\epsilon}$ values.
  In all panels the distribution $\rho(a_S,b_S)$ is given by Eq.~\eqref{eq:act}.}
 \label{fig:scal-delay}
\end{figure*}

We now consider some features of manual CT that can reduce
its effectiveness: the limited scalability of the tracing
capacity~\cite{ECDC_ct,CDC_ct} and the delays in CT and
isolation~\cite{Ferretti2020,Braithwaite2020,Kretzschmar2020}. We set
$\epsilon(a_S)$ as in Eq.~\eqref{eq:scal} and consider a large delay
$\tau_C=3$ days~\cite{Ferretti2020} in manual CT.
In Fig.~\ref{fig:scal-delay} we compare the epidemic threshold for
the two CT protocols, setting equal probabilities of tracing a contact
$\overline{\epsilon}= f^2$, where
$\overline{\epsilon} = \int d a_S  \epsilon(a_S) \rho_S(a_S)$.
For small values of $\overline{\epsilon}$ (note
that this however corresponds to a quite large adoption rate
$f = \sqrt{\overline{\epsilon}}\approx 0.316$)
manual CT is still more effective (see Fig.~\ref{fig:scal-delay}a):
the delay in isolation and the limited scalability are not able
to significantly reduce the advantage provided by the stochastic exploration
of contacts.
Fig.~\ref{fig:scal-delay}b shows that digital CT can become more effective
than manual CT, but this occurs only for very large values of $f^2$ and
$\tau_C$.
This indicates that for realistic settings, the advantage of the manual
protocol over the app-based protocol is robust even including delays and
limited scalability.
Fig.~\ref{fig:scal-delay}c further illustrates for which
(unrealistically large) values of $f^2$ and $\tau_C$, digital CT
outperforms manual CT.
Note that realistic values of $f$ correspond to $f^2$ at most of
the order of 0.1.
Hence, an extremely high adoption of the app is necessary in
order to obtain an effective advantage of the digital CT.

\subsection{Manual and app-based CT in the epidemic phase}
\begin{figure}
\centering
\includegraphics[width=80mm]{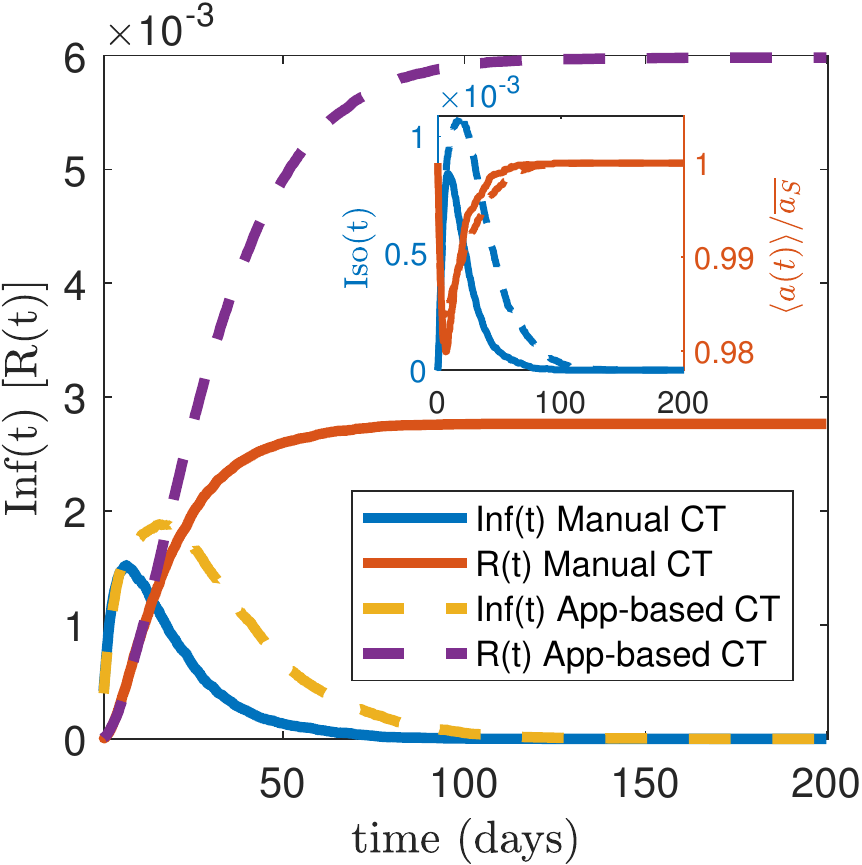}
\caption{{\bf Effects of manual and digital contact tracing on the epidemic active
  phase.} We plot the temporal evolution of the
  fraction of infected nodes $\textrm{Inf}(t)$, i.e. infected asymptomatic and
  infected symptomatic, and of the fraction of removed nodes $R(t)$,
  for both manual and digital CT. In the inset
  we plot the temporal evolution of the fraction of isolated nodes
  $\textrm{Iso}(t)$ (right y-axis) and of the average activity of the
  population $\langle a(t) \rangle$ (left y-axis), normalized with $\overline{a_S}$.  All curves are
  averaged on several realizations of the disorder and of the
  temporal evolution.
  We set $\overline{\epsilon}=f^2=0.1$, $\tau_C=3$ days, $r/r_C^{NA}=3.1$ and $N=5 \, 10^3$.
  The distribution $\rho(a_S,b_S)$ is given by Eq.~\eqref{eq:act}, with $\nu=1.5$.
  The curves for digital and manual CT are averaged respectively over $554$ and $681$ realizations and the errors, evaluated through the standard deviation, are smaller or comparable with the curves thickness.}
\label{fig:active_phase}
\end{figure}

We now explore with numerical simulations (see Methods,
Supplementary Method 2) the effects of manual and digital CT protocols in the active phase of the
epidemic.  We consider an optimistic value of $f=0.316$, setting
$\overline{\epsilon}=f^2=0.1$ that is a very low value for the recall
probability, and we consider the system above the epidemic threshold
$r>r_c$,  in the conditions of Fig.~\ref{fig:scal-delay}a.
Fig.~\ref{fig:active_phase} shows that the infection peak with manual
tracing is lower than the app-based one. Moreover, in the manual CT
the duration of the epidemic is reduced: this strongly impacts on the
final epidemic size, which is about half of the one observed in the
app-based CT.  We also plot the temporal evolution of the average
activity of the system $\langle a(t) \rangle$ and of the fraction of
isolated nodes $\textrm{Iso(t)}$ (see inset).  In general, the average activity
$\langle a(t) \rangle$ features a minimum, however its value remains
very large (about $98\%$ of the case without any tracing
measure). This implies that both protocols do not disrupt the
functionality of the system.  Interestingly, the fraction of isolated
nodes is coherent with the infection peak, and in particular it is
lower when the infection peak is lowered. This means that the most
effective procedure, i.e. manual CT for realistic values of the
parameters, not only lowers the infection peak but it is also able to
isolate a smaller number of nodes, a key feature of any effective CT
strategy.

\subsection{Robustness}
In the Supplementary Notes we show that the advantage of the manual CT is robust with
respect to the relaxation of several assumptions and to changes of
parameters. In particular, we consider the case where all nodes have
equal attractiveness $\rho(a_S,b_S)=\rho_S(a_S) \delta(b_S-b)$, we
take into account very long delays $\tau_C$ and we change the maximum
number of traceable contacts $k_c$.  We also show that
  the advantage of manual contact tracing in the presence of
  heterogeneous activity is robust, if one takes into account that
  contacts belonging to the close social circle of the index case
  (same household) are always traced and isolated manually, also in
  the digital protocol. This can be verified by using a hybrid
  procedure with a fixed small number of contacts always
  traced (see Supplementary Notes).

\subsection{Hybrid contact tracing protocols: manual plus digital CT}
We now consider the implementation of a realistic hybrid CT protocol
with manual and digital contact tracing working at the same time.  We
assume that each individual has downloaded the app with probability
$f$ before the epidemic starts. As soon as an individual with activity
$a_S$ develops symptoms (i.e. $P \to I$), contact tracing is activated
with reference to the previous $T_{CT}$ days: if the infected individual and
also her contact downloaded the app,
the contacted agent is immediately traced and isolated with
no delay.  Otherwise, the contact can only be traced manually, with
probability $\epsilon(a_S)$ (as in Eq.~\eqref{eq:scal}) and, if found
in state $A$, the contact is isolated with a delay $\tau_C$. We set a
realistically large delay $\tau_C=3$ days.

\begin{figure*}
\includegraphics[width=180mm]{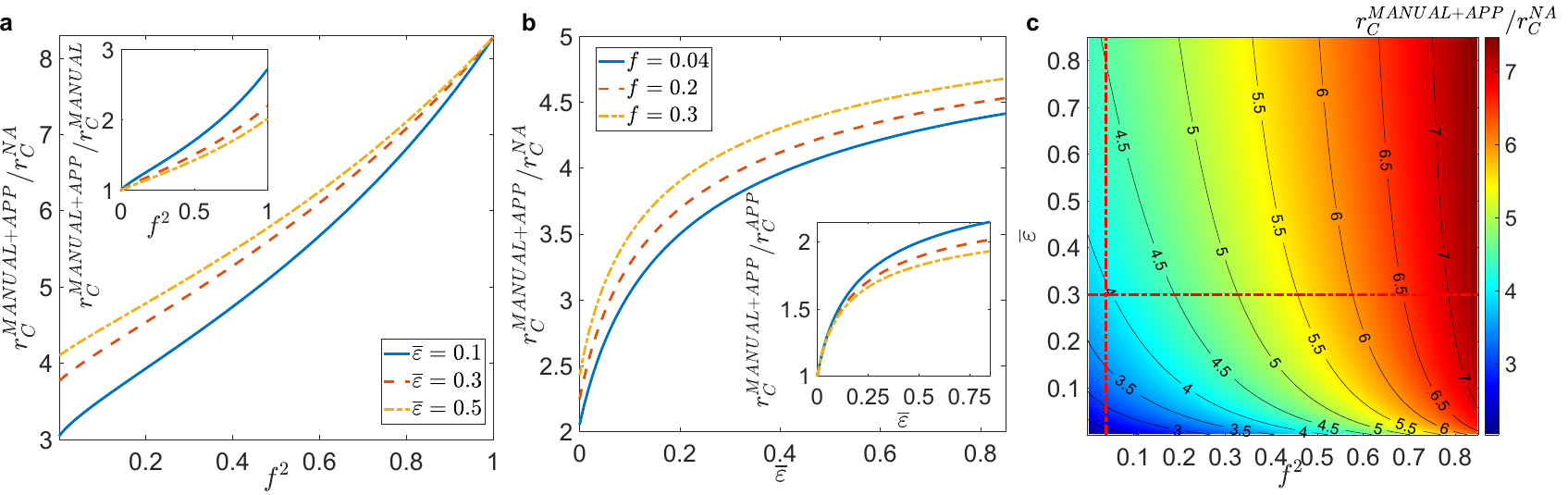}
\caption{{\bf Effects of hybrid contact tracing protocol.} 
	{\bf a} Plot, as a function of $f^2$, of the ratio
 	between the
    epidemic threshold in the presence of a hybrid CT protocol,
    $r_C^{MANUAL+APP}$, and the epidemic threshold of the non-adaptive
    case, $r_C^{NA}$, for realistic values of $\overline{\epsilon}$
    (see legend). In the inset we plot, as a function of $f^2$, the
    ratio between $r_C^{MANUAL+APP}$ and the epidemic threshold when
    only manual CT is implemented, $r_C^{MANUAL}$.  
	{\bf b} Plot of the ratio    
     $r_C^{MANUAL+APP}/r_C^{NA}$ as a function of
    $\overline{\epsilon}$, for realistic app adoption levels $f$
    (see legend). In the inset we plot, as a function of
    $\overline{\epsilon}$, the ratio between $r_C^{MANUAL+APP}$ and
    the epidemic threshold when only app-based CT is implemented,
    $r_C^{APP}$.  
    {\bf c} The ratio $r_C^{MANUAL+APP}/r_C^{NA}$
    is plotted as a function of $f^2$ and $\overline{\epsilon}$,
    through a heat map: the red dash-dotted lines are plotted for
    $\overline{\epsilon}=0.3$ and $f=0.2$, i.e. they correspond to the
    red dashed curves of panel {\bf a} and {\bf b}. In all panels the
    distribution $\rho(a_S,b_S)$ is given by Eq.~\eqref{eq:act}, the
    manual CT features a delay of $\tau_C=3$ days and limited
    scalability is considered.}
 \label{fig:hybrid}
\end{figure*}

The threshold in the hybrid case can be determined
  through the solution of a complex set of equations, that we solve
  numerically in Supplementary Method 1.
  In the presence of heterogeneous activity (i.e. superspreaders),
  starting form a purely digital protocol (Fig.~\ref{fig:hybrid}b,c)
  the addition of
  manual contact tracing rapidly increases the threshold, leading to
  an improvement by a large factor (about 80\%), for realistic values
  $\overline{\epsilon} \gtrsim 0.3$.
  Instead, starting from a realistic manual contact tracing
  setup $\overline{\epsilon} \gtrsim 0.3$,  a significant
  improvement (i.e. a 50\% of threshold increase) is obtained
  by implementing a digital contact tracing only if $f \gtrsim 0.60-0.75$
  (Fig.~\ref{fig:hybrid}a,c).
  These values are consistent with other results in the
  literature~\cite{Eames2003,Ferretti2020,Braithwaite2020}; however,
  in most countries, the app adoption rates do not reach these values~\cite{Longo2020,MIT2020,MIT2020_2}.

\section{Discussion}
Our results indicate that manual CT, despite its
drawbacks, can be an efficient protocol in heterogeneous populations,
more efficient than its digital counterpart, due to its specific
sampling properties. This conclusion is robust with respect
to variations in several model assumptions, including correlations
between activity and attractiveness or the limited scalability of the
manual CT protocol.
However, 
epidemic propagation and strategies to mitigate it are very complex
processes and several of their features have been left out from our
modelling scheme.
Some of these features
(the possibility that isolation is non complete, that some individuals
do not report symptoms or the existence of testing campaigns detecting
infected nonsymptomatic individuals)
act similarly on both types of CT and hence do not modify the
relative performance.
Other more realistic features (the presence of delays even in digital CT
and the existence of additional sources of
heterogeneity in viral shedding~\cite{Liu2020}, recovery
rates~\cite{Arruda2020} and activity temporal
patterns~\cite{Mancastroppa2019}) would even reduce the relative
performance of digital CT.

Our results put forward several directions to increase the effectiveness of tracing. An important aspect is the correlation between app adoption rates and activity of nodes: our analysis in the Supplementary Notes shows that, as expected,
in heterogeneous populations a positive correlation strongly increases the success of digital CT. This is a direction that could be pursued in campaigns aimed at driving app adoption
among potential superspreaders. However, this represents a challenge for policy makers. Evidence is currently emerging that those who download the app are individuals who adopt very cautious behavior, i.e. $f$ and $a_S$ are typically anticorrelated~\cite{Wyl2020,Saw2020}.
Another road that could enhance the efficacy of CT protocols is to follow chains of transmissions along multiple steps, so that when a traced contact is found infected also her contacts are reconstructed and tested. This additional step improves the overall effectiveness of CT protocols, but also increases the delay associated to the manual procedure with respect to the digital one.
Digital CT allows in principle to extend the tracing procedure to an arbitrary number of steps, however, strong concerns related to privacy issues~\cite{Firth2020}
make this path difficult to follow.

In summary, even if additional features of CT can be
  considered, the weakness of digital CT, originated by the nature of
  the sampling of contacts and worsened by heterogeneities, seems to
  be an intrinsic unavoidable property of the procedure. The manual CT
  protocol, with its higher intrinsic stochasticity, does not suffer
  from this problem and samples contacts effectively, especially in
  realistic heterogeneous populations: thus, digital CT cannot be
  considered simply as a cheaper and more rapid way of implementing
  standard contact tracing and should only be considered in
  combination with manual protocols. Manual CT must necessarily play
  an important role in any strategy to mitigate the current pandemic.
  Considerations about costs and practical feasibility of the two
  approaches (which have not been taken into account here) suggest
  that a careful integration of the two protocols may be the key for
  more effective mitigation strategies. As suggested in the last part
  of our results, an optimal set up should include both procedures,
  adopted simultaneously.  In this respect, the availability of
  detailed data about the rates of app adoption in various population
  groups (and correlations with age or activity levels), as well as
  more precise estimates of other parameters, such as the recall
  probability, would be highly beneficial.  The design of optimal
  hybrid CT protocols, including decisions on how
  to allocate resources and how to target recommendations for app
  adoption, is a very promising direction for future work.

\section{Methods}
\subsection{Mean-field equations and implementation of CT protocols}
We consider an activity-attractiveness based mean-field approach,
dividing the population into classes of nodes with the same activity
$a_S$ and attractiveness $b_S$ and treating them as if they were
statistically equivalent. For each class $(a_S,b_S)$ we consider the
probability that at time $t$ a node is in one of the epidemic
compartments.
For arbitrary
$\rho(a_S,b_S)$ distribution and arbitrary functional form of $f(a_S)$
and $\epsilon(a_S)$ we build the mean-field equations which describe
the temporal evolution of the network, the epidemic spreading and the
adaptive behavior due to isolation and CT.
In particular, in order to model the manual and
app-based CT we introduce two further compartments: $T$ traced
asymptomatic and $Q$ isolated asymptomatic. An asymptomatic individual
became traced $T$ when infected by a presymptomatic node or when
she infects a susceptible node that eventually develops symptoms.  In
the manual case the tracing is effective with probability
$\epsilon(a)$. In the app-based case, tracing occurs only if both
nodes involved in the contact downloaded the app. A traced node is
still infective $(a_T,b_T)=(a_S,b_S)$ and with rate
$\gamma_A=1/\tau_A\gg \mu$ it is quarantined, $T
\xrightarrow[]{\gamma_A} Q$; while a quarantined node is no more
infective since $(a_Q,b_Q)=(0,0)$.  In order to take into account the
delay in the manual CT we set: $\tau_A=\tau_C+\tau_P$ for the manual
case and $\tau_A=\tau_P$ for the app-based process. See Supplementary Method 1 for the detailed equations of manual, digital and hybrid CT protocols.

\subsection{Epidemic thresholds}
We perform a linear stability analysis of the mean field equations
around the absorbing state obtaining the conditions for its stability
and then the epidemic threshold $r_C$ (see Supplementary Method 1 for details).
For the non-adaptive case
(NA), when infected individuals do not modify their behavior, the
threshold is equal to the one obtained in
Refs.~\cite{Pozzana2017,Mancastroppa2020}. Indeed, setting
$\rho(a_S,b_S)=\rho_S(a_S) \delta(b_S- a_S)$ we obtain:
\begin{equation}
r_C^{NA}=\frac{\overline{a_S}}{2 \overline{a_S^2}}
\end{equation}
If  only symptomatic nodes are isolated as soon as they develop symptoms,
we obtain, setting $\rho(a_S,b_S)=\rho_S(a_S) \delta(b_S- a_S)$:
\begin{equation}
r_C^{SYMPTO}=r_C^{NA} \frac{\frac{\gamma_P}{\mu}}{\delta+(1-\delta)\frac{\gamma_P}{\mu}}
\end{equation}
For the case with CT on homogeneous population, we
set $\rho(a_S,b_S)=\delta(a_S-a) \delta(b_S-b)$,
$\epsilon(a_S)=\epsilon$ and no delay in manual CT $\tau_C=0$. We
obtain the epidemic threshold for both the manual and digital CT:
\begin{widetext}
\begin{equation}
r_C^{MANUAL}=\frac{2 r_C^{NA} \frac{\gamma_P}{\mu}}{\delta+(1-\delta-\epsilon \delta)\frac{\gamma_P}{\mu}+\sqrt{(\delta+(1-\delta-\epsilon \delta)\frac{\gamma_P}{\mu})^2+4 \delta^2 \epsilon \frac{\gamma_P}{\mu}}}
\end{equation}
\begin{equation}
r_C^{APP}=\frac{2 r_C^{NA} \frac{\gamma_P}{\mu}}{\delta+(1-\delta-f \delta)\frac{\gamma_P}{\mu}+\sqrt{(\delta+(1-\delta-f \delta)\frac{\gamma_P}{\mu})^2+4 \delta f \frac{\gamma_P}{\mu} (\delta + \frac{\gamma_P}{\mu} (1-f)(1-\delta))}}
\end{equation}
\end{widetext}
The more general case of populations with arbitrary distribution
$\rho(a_S,b_S)$, arbitrary delays $\tau_C$ and general form of
$\epsilon(a_S)$ and $f(a_S)$, is reported in Supplementary Method 1.
In Supplementary Method 1 we also report the derivation of the
  epidemic threshold for the hybrid protocol. Its values are derived from
  the stability conditions of a complex set of $22$ differential
  equations, that we can solve numerically.

\subsection{Model parameters}

Figures present results where one of the model
parameters is varied and all the others are fixed. Here we report the
parameter values used throughout the paper, unless specified
otherwise.
They are tailored to describe the current COVID-19 pandemic.

The fraction of infected individuals who develop symptoms is
$\delta=0.57$~\cite{Lavezzo2020}.  The time after which a
presymptomatic individual spontaneously develops symptoms is $\tau_P =
1.5$ days~\cite{DiDomenico2020,wei2020presymptomatic,Keeling2020},
while infected individuals recover on average after $\tau=14$
days~\cite{who_china_report,Liu2020}.  The time window over which
contacts are reconstructed is $T_{CT} = 14$ days~\cite{Keeling2020}:
it is fixed equal to $\tau$ to track both nodes infected by the index
case in the pre-symptomatic phase (forward CT) and the primary case
who infected the index case (backward CT)~\cite{kojaku2020}.  The maximum number of
contacts engaged in $T_{CT}$ by a single individual that can be
reconstructed with the manual CT procedure is $k_C = 130$, according to reasonable estimates of the number of contacts manually traced for very active individuals~\cite{Keeling2020}. Moreover, fixing $k_C=130$ and for realistic $\overline{\epsilon} \sim 0.1-0.5$~\cite{Mossong2008}, the average number of traced contacts for each index case is approximatively $10-60$, consistently with reported data on manual CT and with estimates for resources allocation~\cite{ECDC_ct,ECDC_ct2,Keeling2020,Guardian2020} (see Supplementary Notes for the distribution $P(k_T)$ of contacts $k_T$ traced manually by each index case). The activity-driven network
parameters are fixed so that the average value of the activity is
always the same, i.e.  $\overline{a_S} = 6.7$ days$^{-1}$
~\cite{Mossong2008, Vankerkove2013}.  In particular for a power-law
distribution $\rho_S(a_S) \sim a_S^{-(\nu+1)}$, the values of $a_S$
are constrained between a minimum and a maximum value
($a_m<a_S<a_M$). We keep $a_M=10^3 a_m$ 
and then we tune $a_m$ to set
$\overline{a_S}$.

\subsection{Numerical simulations}
We perform numerical simulations of the epidemic model on the
adaptive activity-driven network: the network dynamics and epidemic
spreading are implemented by a continuous time Gillespie-like
algorithm. We consider an activity-driven
network of $N$ nodes. The results are averaged over several
realizations of the disorder and of the dynamical evolution, so that the error on the infection peak height is lower than $6\%$. The
initial conditions are imposed by infecting the node with the highest
activity $a_S$ and the CT protocols are immediately adopted.  A
detailed description of the simulations is reported in Supplementary Method 2.

\end{document}


\title{Supplementary Information for "Stochastic sampling effects favor manual over digital contact tracing"}

\author{Marco Mancastroppa}
\affiliation{Dipartimento di Scienze Matematiche, Fisiche e Informatiche,
  Universit\`a degli Studi di Parma, Parco Area delle Scienze, 7/A 43124 Parma, Italy}
\affiliation{INFN, Sezione di Milano Bicocca, Gruppo Collegato di Parma, Parco Area delle Scienze, 7/A 43124 Parma, Italy}
\author{Claudio Castellano} 
\affiliation{Istituto dei Sistemi Complessi (ISC-CNR), Via dei Taurini 19, I-00185 Roma, Italy}
\author{Alessandro Vezzani}
\affiliation{Istituto dei Materiali per l'Elettronica ed il Magnetismo (IMEM-CNR), Parco Area delle Scienze, 37/A-43124 Parma, Italy}
\affiliation{Dipartimento di Scienze Matematiche, Fisiche e Informatiche,
  Universit\`a degli Studi di Parma, Parco Area delle Scienze, 7/A 43124 Parma, Italy}
\author{Raffaella Burioni}
\affiliation{Dipartimento di Scienze Matematiche, Fisiche e Informatiche,
  Universit\`a degli Studi di Parma, Parco Area delle Scienze, 7/A 43124 Parma, Italy}
\affiliation{INFN, Sezione di Milano Bicocca, Gruppo Collegato di Parma, Parco Area delle Scienze, 7/A 43124 Parma, Italy}

\maketitle

\tableofcontents

\bigskip

In this supplementary information we derive the mean-field equations for the temporal evolution of the epidemic model on adaptive activity-driven networks.  We also derive analytically the epidemic threshold and we describe the scheme of the continuous-time Gillespie-like algorithm used for the numerical approach to the temporal evolution of the network and of the epidemic. Finally, we discuss in detail the robustness of the results, relaxing assumptions in activity distribution, changing most CT parameters in the modelling scheme, introducing further realistic features such as deterministic household CT and correlation between app adoption and individual activity.

\section{Supplementary Method 1: Mean-field equations and analytical derivation of the epidemic thresholds} 
We consider the epidemic model proposed in the main text evolving on an adaptive activity-driven network in the presence of contact tracing of asymptomatic nodes. 

The epidemic model proposed is a Susceptible-Infected-Recovered (SIR) model, with a further distinction for the states of infection $I$. The distinction is  based on the presence of symptoms, on tracing and isolation and it models changes in social behaviour depending on nodes' health status. We do not consider here burstiness effects~\cite{Mancastroppa2019} nor memory~\cite{Tizzani2018}.

\subsection{Manual CT}
We first focus on the manual CT. In the original epidemic model, individuals can be in states $S$ (susceptible), $P$ (infected presymptomatic), $I$ (infected symptomatic), $A$ (infected asymptomatic) and $R$ (recovered). In the presence of CT,  two additional compartments appear: $T$ (asymptomatic traced) and $Q$ (asymptomatic quarantined). The possible transitions among these states are depicted in Supplementary Fig.~\ref{fig:bn_small}.

\begin{figure}
  \includegraphics[width=0.6\textwidth]{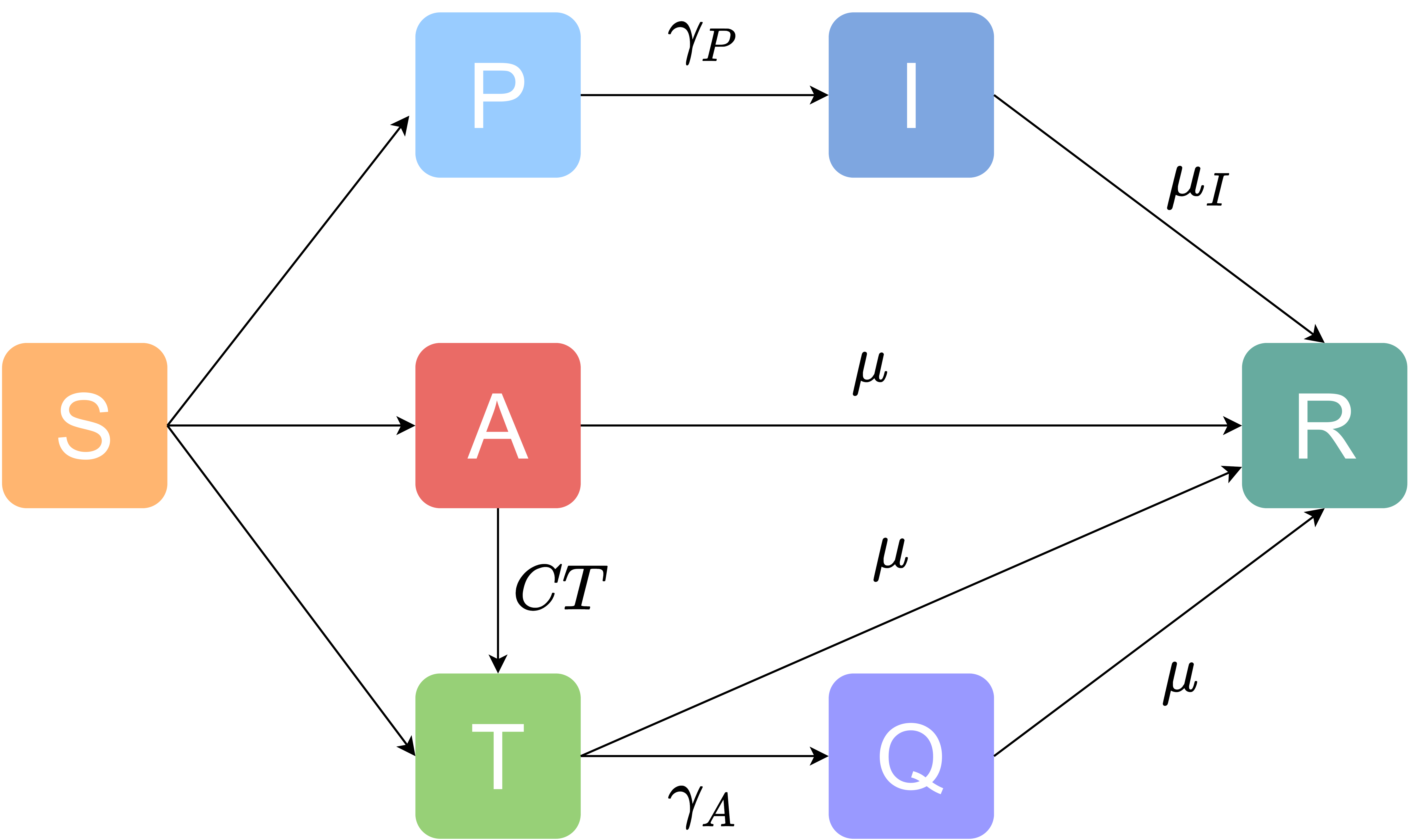}
  \caption{\textbf{Epidemic model with contact tracing.} We plot the scheme of the epidemic model with the transitions for CT. The rates for the infection events (emanating from $S$) and for the CT events are not indicated, as the relevant events, indicated in the text, are complicated.}
  \label{fig:bn_small}
\end{figure}

There are two types of events related to CT: an asymptomatic individual $A$
becomes traced $T$ when infected by a presymptomatic node $P$ (forward
CT) or when she infects a susceptible node that eventually develops
symptoms (backward CT). A traced individual remains infectious $(a_T,b_T)=(a_S,b_S)$ until the presymptomatic starts developing symptoms. Only after that
moment she enters quarantine $(a_Q,b_Q)=(0,0)$. For this reason, the rate for the transition $T \to Q$ is $\gamma_A=1/\tau_A$, with $\tau_A=\tau_P+\tau_C$, where $\tau_P=1/\gamma_P$ and $\tau_C$ is the delay in manual CT.\\

In the manual case the tracing is effective
with probability $\epsilon(a_S)$, with $a_S$ activity of the index
case. The infection and CT transitions are as follows:
\begin{align}
P+S& \xrightarrow[]{\lambda \delta} P+P& \qquad \qquad \qquad &A+S \xrightarrow[]{\lambda \delta \epsilon} T+P&  \qquad \qquad \qquad &T+S \xrightarrow[]{\lambda \delta \epsilon} T+P& \\
P+S& \xrightarrow[]{\lambda (1-\delta) \epsilon} P+T& \qquad \qquad \qquad  &A+S \xrightarrow[]{\lambda (1-\delta)} A+A& \qquad \qquad \qquad &T+S \xrightarrow[]{\lambda (1-\delta)} T+A&\\
P+S& \xrightarrow[]{\lambda (1-\delta) (1-\epsilon)} P+A& \qquad \qquad \qquad  &A+S \xrightarrow[]{\lambda \delta (1-\epsilon)} A+P&\qquad \qquad \qquad &T+S \xrightarrow[]{\lambda \delta (1-\epsilon)} T+P&
\end{align}
while the spontaneous transitions are:
\begin{align}
&P \xrightarrow[]{\gamma_P} I& \qquad \qquad \qquad  &A \xrightarrow[]{\mu} R& \qquad \qquad \qquad &I \xrightarrow[]{\mu_I} R&\\
&T \xrightarrow[]{\gamma_A} Q& \qquad \qquad \qquad  &T \xrightarrow[]{\mu} R& \qquad \qquad \qquad &Q \xrightarrow[]{\mu} R&
\end{align}
Notice that in the event $A+S \to T+P$ both individuals change state; in the event $T+S \xrightarrow[]{\lambda \delta \epsilon} T+P$ the individual $T$ is traced two times (the second time from the individual she infects while already traced), but this has actually no consequences.

We apply an \textit{activity-attractiveness based mean-field} approach, dividing the population in classes of nodes with same $(a_S,b_S)$ and considering them statistically equivalent. The model is an exact mean-field since local correlations are continuously destroyed due to link reshuffling: thus the epidemic threshold of the $SIR$ and $SIS$ epidemic models are the same~\cite{Tizzani2018}. Therefore, to obtain the epidemic threshold we consider the mean-field equations for the $SIS$ version of the model, in which the recovered nodes become susceptible again without gaining immunity.
We assign initially to each node the status of symptomatic (with probability $\delta$) or asymptomatic (with probability $1-\delta$), instead of assigning it at the time of infection. This choice is completely equivalent to the epidemic model described and allows us to write the mean-field equations in a simpler way. Thus, at the mean-field level, the epidemic dynamics is described by the probabilities:
\begin{itemize}
\item $P_{a_S,b_S}(t)$ for a symptomatic node to be infected pre-symptomatic at time $t$;
\item $I_{a_S,b_S}(t)$ for a symptomatic node to be infected symptomatic at time $t$;
\item $1-I_{a_S,b_S}(t)-P_{a_S,b_S}(t)$ to be susceptible at time $t$, for a node which will develop symptoms;
\item $A_{a_S,b_S}(t)$ for an asymptomatic node to be infected asymptomatic at time $t$;
\item $T_{a_S,b_S}(t)$ for an asymptomatic node to be infected traced at time $t$;
\item $Q_{a_S,b_S}(t)$ for an asymptomatic node to be infected isolated at time $t$;
\item $1-A_{a_S,b_S}(t)-Q_{a_S,b_S}(t)-T_{a_S,b_S}(t)$ to be susceptible at time $t$, for a node which will not develop symptoms.
\end{itemize}

In this case the average attractiveness at time $t$ is $\langle b(t) \rangle = \overline{b_S} - (1-\delta)\overline{b_S Q}(t)- \delta \overline{b_SI}(t) $
where we define in general $\overline{g}=\int da_S db_S \rho(a_S,b_S) g_{a_S,b_S}$.
We consider the system in the thermodynamic limit. The probabilities previously introduced evolve accordingly to the following equations:

\begin{equation}
\begin{aligned}
\partial_t P_{a_S,b_S}(t)= -\gamma_P P_{a_S,b_S}(t) &+ \lambda a_S (1-I_{a_S,b_S}(t)-P_{a_S,b_S}(t)) \frac{\delta \overline{b_S P}(t)+(1-\delta)[\overline{b_S T}(t)+\overline{b_S A}(t)]}{\overline{b_S}- (1-\delta) \overline{b_S Q}(t)- \delta \overline{b_S I}(t)}\\
&+ \lambda b_S (1-I_{a_S,b_S}(t)-P_{a_S,b_S}(t)) \frac{\delta \overline{a_S P}(t)+(1-\delta)[\overline{a_S T}(t)+\overline{a_S A}(t)]}{\overline{b_S}- (1-\delta) \overline{b_S Q}(t)- \delta \overline{b_S I}(t)}
\end{aligned}
\label{eq:CT_m_P}
\end{equation}
where the first term on right hand side accounts for symptoms onset; the second and third terms account for contagion processes of a susceptible node who engages a contact with a pre-symptomatic or a non-isolated asymptomatic infected node, respectively for the activation of the susceptible and of the infected node. Both terms are averaged over all the activity-attractiveness classes of the infected node.

\begin{equation}
\partial_t I_{a_S,b_S}(t)=-\mu_I I_{a_S,b_S}(t) + \gamma_P P_{a_S,b_S}(t)
\label{eq:CT_m_Is}
\end{equation}
where the first term on the right hand side accounts for spontaneous recovery and the second term for spontaneous symptoms development.

\begin{small}
\begin{equation}
\begin{aligned} 
\partial_t A_{a_S,b_S}(t)=-\mu A_{a_S,b_S}(t) &+\lambda a_S (1-A_{a_S,b_S}(t)-T_{a_S,b_S}(t)-Q_{a_S,b_S}(t)) \frac{\delta [\overline{b_S P}(t)-\overline{\epsilon b_S P}(t)]+(1-\delta)[\overline{b_S T}(t)+\overline{b_S A}(t)]}{\overline{b_S}- (1-\delta) \overline{b_S Q}(t)- \delta \overline{b_S I}(t)}\\
&+ \lambda b_S (1-A_{a_S,b_S}(t)-T_{a_S,b_S}(t)-Q_{a_S,b_S}(t)) \frac{\delta [\overline{a_S P}(t)-\overline{\epsilon a_S P}(t)]+(1-\delta)[\overline{a_S T}(t)+\overline{a_S A}(t)]}{\overline{b_S}- (1-\delta) \overline{b_S Q}(t)- \delta \overline{b_S I}(t)}\\
&- \lambda a_S \delta A_{a_S,b_S}(t) \frac{\overline{\epsilon b_S}-\overline{\epsilon b_S I}(t)-\overline{\epsilon b_S P}(t)}{\overline{b_S}- (1-\delta) \overline{b_S Q}(t)- \delta \overline{b_S I}(t)} - \lambda b_S \delta A_{a_S,b_S}(t) \frac{\overline{\epsilon a_S}-\overline{\epsilon a_S I}(t)-\overline{\epsilon a_S P}(t)}{\overline{b_S}- (1-\delta) \overline{b_S Q}(t)- \delta \overline{b_S I}(t)}
\end{aligned}
\label{eq:CT_m_Ia}
\end{equation}
\end{small}
where the first term on right hand side accounts for spontaneous recovery; the second and third terms account for contagion processes of a susceptible node who engages a contact with a non-isolated asymptomatic infected node or with a pre-symptomatic node and their contact is not traced ($(1-\epsilon(a_S'))$, with $a_S'$ activity of the pre-symptomatic node). Both terms are averaged over all the activity-attractiveness classes of the infected node. The fourth and fifth terms correspond to contact tracing of infected asymptomatic due to infection of a susceptible symptomatic and effective CT of the link ($\epsilon(a_S')$, with $a_S'$ activity of the pre-symptomatic node). Both terms are averaged over all the activity-attractiveness classes of the susceptible node.

\begin{equation}
\begin{aligned} 
\partial_t T_{a_S,b_S}(t)=-(\mu + \gamma_A) T_{a_S,b_S}(t) &+ \lambda b_S (1-A_{a_S,b_S}(t)-T_{a_S,b_S}(t)-Q_{a_S,b_S}(t)) \frac{\delta \overline{\epsilon a_S P}(t)}{\overline{b_S}- (1-\delta) \overline{b_S Q}(t)- \delta \overline{b_S I}(t)}\\
&+ \lambda a_S (1-A_{a_S,b_S}(t)-T_{a_S,b_S}(t)-Q_{a_S,b_S}(t)) \frac{\delta \overline{\epsilon b_S P}(t)}{\overline{b_S}- (1-\delta) \overline{b_S Q}(t)- \delta \overline{b_S I}(t)}\\
&+ \lambda b_S \delta A_{a_S,b_S}(t) \frac{\overline{\epsilon a_S}-\overline{\epsilon a_S I}(t)-\overline{\epsilon a_S P}(t)}{\overline{b_S}- (1-\delta) \overline{b_S Q}(t)- \delta \overline{b_S I}(t)}\\
&+ \lambda a_S \delta A_{a_S,b_S}(t) \frac{\overline{\epsilon b_S}-\overline{\epsilon b_S I}(t)-\overline{\epsilon b_S P}(t)}{\overline{b_S}- (1-\delta) \overline{b_S Q}(t)- \delta \overline{b_S I}(t)} 
\end{aligned}
\label{eq:CT_m_T}
\end{equation}
where the first term on right hand side accounts for isolation and recovery. The second and third terms account for contagion processes of a susceptible node who engage a contact with a pre-symptomatic node and their contact is traced ($\epsilon(a_S')$, with $a_S'$ activity of the pre-symptomatic node). Both terms are averaged over all the activity-attractiveness classes of the infected node. The fourth and fifth terms correspond to contact tracing of infected asymptomatic due to infection of a susceptible symptomatic and effective CT of the link ($\epsilon(a_S')$, with $a_S'$ activity of the pre-symptomatic node). Both terms are averaged over all the activity-attractiveness classes of the susceptible node.

\begin{equation}
\partial_t Q_{a_S,b_S}(t)=-\mu Q_{a_S,b_S}(t) + \gamma_A T_{a_S,b_S}(t)
\label{eq:CT_m_Q}
\end{equation}
where the first term on right hand side accounts for spontaneous recovery and the second term for isolation of traced asymptomatic infected nodes.\\

This set of equations admits  as a stationary state the absorbing state, a configuration where all the population is susceptible. To obtain the condition for the stability of the absorbing state, i.e. the epidemic threshold, we apply a linear stability analysis around the absorbing state.

Let us now consider the case of realistic correlations between the activity and attractiveness~\cite{Pozzana2017,Ghoshal2006}: $\rho(a_S,b_S)= \rho_S(a_S) \delta(b_S-a_S)$, with generic $\rho_S(a_S)$. If we average the equations on all activity classes, we obtain the temporal evolution of the average probabilities $\overline{P}(t)$, $\overline{I}(t)$, $\overline{A}(t)$, $\overline{T}(t)$, $\overline{Q}(t)$; similarly we obtain the temporal evolution of $\overline{a_S T}(t)$, $\overline{a_S A}(t)$, $\overline{a_S P}(t)$ and $\overline{\epsilon a_S P}(t)$ multiplying the equations for $a_S \rho_S(a_S)$ or $\epsilon(a_S) a_S \rho_S(a_S)$ and integrating. Neglecting second order terms in probabilities, we  obtain a linearized set of $9$ differential equations:

\begin{align}
\partial_t \overline{I}(t)&=-\mu_I \overline{I}(t) + \gamma_P \overline{P}(t)\\
\partial_t \overline{P}(t)&= -\gamma_P \overline{P}(t) + 2 \lambda [\delta \overline{a_S P}(t)+(1-\delta)(\overline{a_S T}(t)+\overline{a_S A}(t))]\\
\partial_t \overline{Q}(t)&=-\mu \overline{Q}(t) + \gamma_A \overline{T}(t)\\
\partial_t \overline{T}(t)&=-(\mu + \gamma_A) \overline{T}(t) +2 \lambda \delta \overline{\epsilon a_S P}(t) + 2 \lambda \delta \overline{a_S A}(t) \frac{\overline{\epsilon a_S}}{\overline{a_S}}\\
\partial_t \overline{A}(t)&=-\mu \overline{A}(t) + 2 \lambda [\delta (\overline{a_S P}(t)-\overline{\epsilon a_S P}(t))+(1-\delta)(\overline{a_S T}(t)+\overline{a_S A}(t))]- 2 \lambda  \delta \overline{a_S A}(t) \frac{\overline{\epsilon a_S}}{\overline{a_S}}\\
\partial_t \overline{a_S P}(t)&= -\gamma_P \overline{a_S P}(t) + 2 \lambda \frac{\overline{a_S^2}}{\overline{a_S}} [\delta \overline{a_S P}(t)+(1-\delta)(\overline{a_S T}(t)+\overline{a_S A}(t))]\\
\partial_t \overline{\epsilon a_S P}(t)&= -\gamma_P \overline{\epsilon a_S P}(t) + 2 \lambda \frac{\overline{\epsilon a_S^2}}{\overline{a_S}} [\delta \overline{a_S P}(t)+(1-\delta)(\overline{a_S T}(t)+\overline{a_S A}(t))]\\
\partial_t \overline{a_S T}(t)&=-(\mu + \gamma_A) \overline{a_S T}(t) +2 \lambda \delta \frac{\overline{a_S^2}}{\overline{a_S}} \overline{\epsilon a_S P}(t)
+ 2 \lambda \delta \overline{a_S^2 A}(t) \frac{\overline{\epsilon a_S}}{\overline{a_S}}\\
\partial_t \overline{a_S A}(t)&=-\mu \overline{a_S A}(t) + 2 \lambda \frac{\overline{a_S^2}}{\overline{a_S}} [\delta (\overline{a_S P}(t)-\overline{\epsilon a_S P}(t))+(1-\delta)(\overline{a_S T}(t)+\overline{a_S A}(t))]
- 2 \lambda  \delta \overline{a_S^2 A}(t) \frac{\overline{\epsilon a_S}}{\overline{a_S}}
\end{align}

The linearized equations for the dynamic evolution of $\overline{a_S^n A}(t)$ and $\overline{a_S^n T}(t)$ always involve terms like $\overline{a_S^{n+1} A}(t)$, due to the contact tracing terms. This would produce an infinite set of coupled linear differential equations: to close the equations and obtain a complete set of linearized equations, we express $\overline{a_S^2 A}(t)$ in terms of the other average probabilities. By definition $\overline{a_S^2 A}(t)=\int da_S \rho_S(a_S) a_S^2 A_{a_S}(t)$: since we are interested in studying the absorbing steady state, we consider Eq.~\eqref{eq:CT_m_Ia} for $\rho(a_S,b_S)= \rho_S(a_S) \delta(b_S-a_S)$, linearized around the absorbing state and near the stationary condition $\partial_t A_{a_S}(t) \sim 0$:

\begin{equation}
  A_{a_S}(t) \simeq \frac{2 \lambda a_S [\delta(\overline{a_S P}(t)-\overline{\epsilon a_S P}(t))+(1-\delta)(\overline{a_S T}(t)+\overline{a_S A}(t))]}{\mu \overline{a_S}+2 \lambda \delta a_S \overline{\epsilon a_S}}
  \label{As}
\end{equation}

Thus, setting $r=\lambda/\mu$ and replacing Eq.~\eqref{As} into the definition of $\overline{a_S^2 A}(t)$ we obtain:

\begin{equation}
\overline{a_S^2 A}(t) \simeq \frac{2r}{\overline{a_S}} [\delta(\overline{a_S P}(t)-\overline{\epsilon a_S P}(t))+(1-\delta)(\overline{a_S T}(t)+\overline{a_S A}(t))]K
\end{equation}
where $K= \overline{\frac{a_S^3}{1+2r\delta a_S \frac{\overline{\epsilon a_S}}{\overline{a_S}}}} = \int da_S \rho_S(a_S) \frac{a_S^3}{1+2r\delta a_S \frac{\overline{\epsilon a_S}}{\overline{a_S}}}$.

In this way we obtain the following linearized equations for $\overline{a_S T}(t)$ and $\overline{a_S A}(t)$, near the absorbing stationary state:

\begin{align} 
\partial_t \overline{a_S T}(t)&=-(\mu + \gamma_A) \overline{a_S T}(t) +2 \lambda \delta \frac{\overline{a_S^2}}{\overline{a_S}} \overline{\epsilon a_S P}(t)
+ 4 \lambda r \frac{\overline{\epsilon a_S}}{\overline{a_S}^2} \delta K [\delta(\overline{a_S P}(t)-\overline{\epsilon a_S P}(t))+(1-\delta)(\overline{a_S T}(t)+\overline{a_S A}(t))]\\
\partial_t \overline{a_S A}(t) &= -\mu \overline{a_S A}(t) \begin{aligned}[t]&+ 2 \lambda \frac{\overline{a_S^2}}{\overline{a_S}} [\delta (\overline{a_S P}(t)-\overline{\epsilon a_S P}(t))+(1-\delta)(\overline{a_S T}(t)+\overline{a_S A}(t))]\\
&- 4 \lambda r \frac{\overline{\epsilon a_S}}{\overline{a_S}^2} \delta K [\delta(\overline{a_S P}(t)-\overline{\epsilon a_S P}(t))+(1-\delta)(\overline{a_S T}(t)+\overline{a_S A}(t))] \end{aligned}
\end{align}

We focus on the Jacobian matrix of this set of $9$ linearized equations:

\begin{equation}
J=\begin{bmatrix}
-\mu_I & \gamma_P & 0 & 0 & 0 & 0 & 0 & 0 & 0\\
0 & -\gamma_P & 0 & 0 & 0 & 2 \lambda \delta & 0 & 2 \lambda(1-\delta) & 2 \lambda (1-\delta)\\
0 & 0 & -\mu & \gamma_A & 0 & 0 & 0 & 0 & 0\\
0 & 0 & 0 & -\mu - \gamma_A & 0 & 0 & 2 \lambda \delta & 0 & 2 \lambda \delta \frac{\overline{\epsilon a_S}}{\overline{a_S}}\\
0 & 0 & 0 & 0 & -\mu & 2 \lambda \delta & -2 \lambda \delta & 2 \lambda (1-\delta) & 2 \lambda (1-\delta) - 2 \lambda \delta \frac{\overline{\epsilon a_S}}{\overline{a_S}}\\
0 & 0 & 0 & 0 & 0 & -\gamma_P + \Delta & 0 & \Gamma & \Gamma\\
0 & 0 & 0 & 0 & 0 & \phi & -\gamma_P & \Phi & \Phi\\
0 & 0 & 0 & 0 & 0 & \theta & \Delta-\theta  & -\mu-\gamma_A+\Psi & \Psi\\
0 & 0 & 0 & 0 & 0 & \Delta- \theta & \theta-\Delta  &  \Gamma-\Psi & -\mu+\Gamma-\Psi\\
\end{bmatrix}
=
\begin{bmatrix}
\mathbb{A}(5\text{x}5) & \mathbb{C}(5\text{x}4) \\
\mathbb{O}(4\text{x}5) & \mathbb{B}(4\text{x}4)
\end{bmatrix}
\end{equation}

where $\Delta=2 \lambda \delta \frac{\overline{a_S^2}}{\overline{a_S}}$, $\Gamma=2 \lambda (1-\delta) \frac{\overline{a_S^2}}{\overline{a_S}}$, $\phi=2 \lambda \delta \frac{\overline{\epsilon a_S^2}}{\overline{a_S}}$, $\Phi=2 \lambda (1-\delta) \frac{\overline{\epsilon a_S^2}}{\overline{a_S}}$, $\theta=4 \lambda r \delta^2 \frac{\overline{\epsilon a_S}}{\overline{a_S}^2} K$ and $\Psi= 4 \lambda r \delta (1-\delta) \frac{\overline{\epsilon a_S}}{\overline{a_S}^2} K$.\\

The Jacobian matrix is a block matrix and the condition for the stability of the absorbing state is obtained imposing all eigenvalues to be negative. We can consider separately the two blocks on the diagonal: for the first block $\mathbb{A}$ it is evident that the eigenvalues are $\xi_{1,2}=-\mu$, $\xi_3=-\mu_I$, $\xi_4=-\gamma_P$, $\xi_5=-\mu-\gamma_A$, all negative. Therefore, it is sufficient to study block $\mathbb{B}$, which is a matrix $4\text{x}4$. The characteristic polynomial of $\mathbb{B}$ is a polynomial of degree 4, thus we apply the Descartes' rule of signs to impose all roots to be negative and we obtain the condition for the stability of the absorbing state:

\begin{equation}
\begin{aligned}
8r^3 \delta^2 (1-\delta) \frac{\overline{\epsilon a_S^2} \, \overline{\epsilon a_S}}{\overline{a_S}} \overline{\frac{a_S^3}{1+2 r \delta a_S \frac{\overline{\epsilon a_S}}{\overline{a_S}}}} \frac{\gamma_A}{\mu} &-4 r^2 \delta (1-\delta) \left[ \overline{\epsilon a_S^2} \, \overline{a_S^2}+\frac{\gamma_P}{\mu} \overline{\epsilon a_S} \overline{\frac{a_S^3}{1+2 r \delta a_S  \frac{\overline{\epsilon a_S}}{\overline{a_S}}}} \right] \frac{\gamma_A}{\mu} \\
&+ 2 r \overline{a_S^2} \overline{a_S} \left(\frac{\gamma_A}{\mu}+1\right) \left(\delta+\frac{\gamma_P}{\mu}(1-\delta)\right)-\overline{a_S}^2\frac{\gamma_P}{\mu}\left(\frac{\gamma_A}{\mu}+1\right)<0
\end{aligned}
\label{eq:rc_man}
\end{equation}

By setting the equality, the equation allows to obtain a closed relation for estimating the epidemic threshold $r_C$. The epidemic threshold obtained with the mean-field approach is exact and it holds for manual contact tracing, with arbitrary delay $\tau_C$ (encapsulated in $\gamma_A$), for arbitrary $\rho_S(a_S)$ and $\epsilon(a_S)$.

\subsection{Digital CT}
We now focus on the digital CT.
  Each individual is either endowed or not with the app in the initial condition
 (with a probability $f(a_S)$ depending on her activity).
This implies that there are different
compartments for individuals without the app ($S$, $I$, $P$, etc.) and with
the app (denoted with the superscript $^{\alpha}$: $S^{\alpha}$, $I^{\alpha}$, $P^{\alpha}$, etc.).
Notice that, for pure digital CT, necessarily $T=0$ and $Q=0$.
We nevertheless write here the
transitions involving them, that may play a role for hybrid protocols.

The spontaneous transitions do not depend on whether the individual has the app or not,
but the difference with respect to the manual protocol is that $\tau_C=0$ so that $\gamma_A=\gamma_P$.
The infection events are instead as follows.
Those involving both individuals without the app (in such a case no new traced individual is generated):
\begin{align}
&P+S \xrightarrow{\lambda \delta}  P+P& \qquad \qquad \qquad A+S& \xrightarrow{\lambda \delta}  A+P& \qquad \qquad \qquad T+S& \xrightarrow{\lambda \delta} T+P&\\
&P+S \xrightarrow{\lambda (1-\delta)}  P+A& \qquad \qquad \qquad A+S& \xrightarrow{\lambda (1-\delta)} A+A& \qquad \qquad \qquad T+S& \xrightarrow{\lambda (1-\delta)} T+A&
\end{align}
Those involving only one individual with the app (in such a case no new traced individual is generated):
\begin{align}
&P^{\alpha}+S \xrightarrow{\lambda \delta} P^{\alpha}+P& \qquad \qquad A^{\alpha}+S& \xrightarrow{\lambda \delta} A^{\alpha}+P& \qquad \qquad
T^{\alpha}+S& \xrightarrow{\lambda \delta} T^{\alpha}+P& \\
&P^{\alpha}+S \xrightarrow{\lambda (1-\delta)} P^{\alpha}+A& \qquad \qquad
A^{\alpha}+S& \xrightarrow{\lambda (1-\delta)} A^{\alpha}+A& \qquad \qquad
T^{\alpha}+S& \xrightarrow{\lambda (1-\delta)} T^{\alpha}+A&\\
&P+S^{\alpha} \xrightarrow{\lambda \delta} P+P^{\alpha}& \qquad \qquad
A+S^{\alpha}& \xrightarrow{\lambda \delta} A+P^{\alpha}& \qquad \qquad
T+S^{\alpha}& \xrightarrow{\lambda \delta} T+P^{\alpha}&\\
&P+S^{\alpha} \xrightarrow{\lambda (1-\delta)} P+A^{\alpha}& \qquad \qquad
A+S^{\alpha}& \xrightarrow{\lambda (1-\delta)} A+A^{\alpha}& \qquad \qquad
T+S^{\alpha}& \xrightarrow{\lambda (1-\delta)} T+A^{\alpha}&
\end{align}
Those involving both individuals with the app (in such a case new traced individuals can be generated):
\begin{align}
&P^{\alpha}+S^{\alpha} \xrightarrow{\lambda \delta} P^{\alpha}+P^{\alpha}&  \qquad A^{\alpha}+S^{\alpha}& \xrightarrow{\lambda \delta} T^{\alpha}+P^{\alpha}&  \qquad T^{\alpha}+S^{\alpha}& \xrightarrow{\lambda \delta} T^{\alpha}+P^{\alpha}& \\
&P^{\alpha}+S^{\alpha} \xrightarrow{\lambda (1-\delta)} P^{\alpha}+T^{\alpha}&  \qquad A^{\alpha}+S^{\alpha}& \xrightarrow{\lambda (1-\delta)} A^{\alpha}+A^{\alpha}& \qquad T^{\alpha}+S^{\alpha}& \xrightarrow{\lambda (1-\delta)} T^{\alpha}+A^{\alpha}&
\end{align}
Analogously to the manual CT we apply the \textit{activity-attractiveness based mean-field} approach to the app-based CT. At the mean-field level, the epidemic dynamics is described by the probabilities:
\begin{itemize}
\item $P_{a_S,b_S}(t)$ for a symptomatic node without app to be infected pre-symptomatic at time $t$;
\item $I_{a_S,b_S}(t)$ for a symptomatic node without app to be infected symptomatic at time $t$;
\item $1-P_{a_S,b_S}(t)-I_{a_S,b_S}(t)$ to be susceptible at time $t$, for a node without app and which will develop symptoms;
\item $A_{a_S,b_S}(t)$ for an asymptomatic node without app to be infected asymptomatic at time $t$;
\item $1-A_{a_S,b_S}(t)$ to be susceptible at time $t$, for a node without app and which will not develop symptoms;
\item $P_{a_S,b_S}^{\alpha}(t)$ for a symptomatic node with app to be infected pre-symptomatic at time $t$;
\item $I^{\alpha}_{a_S,b_S}(t)$ for a symptomatic node with app to be infected symptomatic at time $t$;
\item $1-P_{a_S,b_S}^{\alpha}(t)-I_{a_S,b_S}^{\alpha}(t)$ to be susceptible at time $t$, for a node with app and which will develop symptoms;
\item $A^{\alpha}_{a_S,b_S}(t)$ for an asymptomatic node with app to be infected asymptomatic at time $t$;
\item $T_{a_S,b_S}^{\alpha}(t)$ for an asymptomatic node with app to be infected traced at time $t$;
\item $Q_{a_S,b_S}^{\alpha}(t)$ for an asymptomatic node with app to be infected isolated at time $t$;
\item $1-A_{a_S,b_S}^{\alpha}(t)-Q_{a_S,b_S}^{\alpha}(t)-T_{a_S,b_S}^{\alpha}(t)$ to be susceptible at time $t$, for a node with app and which will not develop symptoms.
\end{itemize}  
In this case the average attractiveness at time $t$ is $\langle b(t) \rangle = \overline{b_S}- (1-\delta) \overline{f b_S Q^{\alpha}}(t)- \delta (\overline{f b_S I^{\alpha}}(t)+\overline{(1-f) b_S I}(t))$.
We consider the system in the thermodynamic limit: the probabilities previously introduced evolve according to the following equations, obtained analogously to those for manual CT:

\begin{small}
\begin{align}
\partial_t I_{a_S,b_S}(t)&=-\mu_I I_{a_S,b_S}(t) + \gamma_P P_{a_S,b_S}(t)\\
\partial_t P_{a_S,b_S}(t)&=\begin{aligned}[t]& -\gamma_P P_{a_S,b_S}(t) \\
&+ \lambda a_S (1-I_{a_S,b_S}(t)-P_{a_S,b_S}(t)) \frac{\delta (\overline{f b_S P^{\alpha}}(t)+\overline{(1-f) b_S P}(t))+(1-\delta)(\overline{f b_S T^{\alpha}}(t)+\overline{f b_S A^{\alpha}}(t)+ \overline{(1-f) b_S A}(t))}{\overline{b_S}- (1-\delta) \overline{f b_S Q^{\alpha}}(t)- \delta (\overline{f b_S I^{\alpha}}(t)+\overline{(1-f) b_S I}(t))}\\
&+ \lambda b_S (1-I_{a_S,b_S}(t)-P_{a_S,b_S}(t)) \frac{\delta (\overline{f a_S P^{\alpha}}(t)+\overline{(1-f) a_S P}(t))+(1-\delta)(\overline{f a_S T^{\alpha}}(t)+\overline{f a_S A^{\alpha}}(t)+ \overline{(1-f) a_S A}(t))}{\overline{b_S}- (1-\delta) \overline{f b_S Q^{\alpha}}(t)- \delta (\overline{f b_S I^{\alpha}}(t)+\overline{(1-f) b_S I}(t))} \end{aligned}\\
\partial_t I_{a_S,b_S}^{\alpha}(t)&=-\mu_I I_{a_S,b_S}^{\alpha}(t) + \gamma_P P_{a_S,b_S}^{\alpha}(t)\\
\partial_t P_{a_S,b_S}^{\alpha}(t)&=\begin{aligned}[t]& -\gamma_P P_{a_S,b_S}^{\alpha}(t) \\
&+ \lambda a_S (1-I^{\alpha}_{a_S,b_S}(t)-P_{a_S,b_S}^{\alpha}(t)) \frac{\delta (\overline{f b_S P^{\alpha}}(t)+\overline{(1-f) b_S P}(t))+(1-\delta)(\overline{f b_S T^{\alpha}}(t)+\overline{f b_S A^{\alpha}}(t)+ \overline{(1-f) b_S A}(t))}{\overline{b_S}- (1-\delta) \overline{f b_S Q^{\alpha}}(t)- \delta (\overline{f b_S I^{\alpha}}(t)+\overline{(1-f) b_S I}(t))}\\
&+ \lambda b_S (1-I^{\alpha}_{a_S,b_S}(t)-P_{a_S,b_S}^{\alpha}(t)) \frac{\delta (\overline{f a_S P^{\alpha}}(t)+\overline{(1-f) a_S P}(t))+(1-\delta)(\overline{f a_S T^{\alpha}}(t)+\overline{f a_S A^{\alpha}}(t)+ \overline{(1-f) a_S A}(t))}{\overline{b_S}- (1-\delta) \overline{f b_S Q^{\alpha}}(t)- \delta (\overline{f b_S I^{\alpha}}(t)+\overline{(1-f) b_S I}(t))}
\end{aligned}\\
\partial_t A_{a_S,b_S}(t)&=\begin{aligned}[t]&-\mu A_{a_S,b_S}(t) \\
&+ \lambda a_S (1-A_{a_S,b_S}(t)) \frac{\delta (\overline{f b_S P^{\alpha}}(t)+\overline{(1-f) b_S P}(t))+(1-\delta)(\overline{f b_S T^{\alpha}}(t)+\overline{f b_S A^{\alpha}}(t)+ \overline{(1-f) b_S A}(t))}{\overline{b_S}- (1-\delta) \overline{f b_S Q^{\alpha}}(t)- \delta (\overline{f b_S I^{\alpha}}(t)+\overline{(1-f) b_S I}(t))}\\
&+ \lambda b_S (1-A_{a_S,b_S}(t)) \frac{\delta (\overline{f a_S P^{\alpha}}(t)+\overline{(1-f) a_S P}(t))+(1-\delta)(\overline{f a_S T^{\alpha}}(t)+\overline{f a_S A^{\alpha}}(t)+ \overline{(1-f) a_S A}(t))}{\overline{b_S}- (1-\delta) \overline{f b_S Q^{\alpha}}(t)- \delta (\overline{f b_S I^{\alpha}}(t)+\overline{(1-f) b_S I}(t))}
\end{aligned} \\ 
\partial_t A^{\alpha}_{a_S,b_S}(t)&=\begin{aligned}[t]&-\mu A^{\alpha}_{a_S,b_S}(t) \\
&+ \lambda a_S (1-A^{\alpha}_{a_S,b_S}(t)-T^{\alpha}_{a_S,b_S}(t)-Q^{\alpha}_{a_S,b_S}(t)) \frac{\delta \overline{(1-f) b_S P}(t)+(1-\delta)(\overline{f b_S T^{\alpha}}(t)+\overline{f b_S A^{\alpha}}(t)+ \overline{(1-f) b_S A}(t))}{\overline{b_S}- (1-\delta) \overline{f b_S Q^{\alpha}}(t)- \delta (\overline{f b_S I^{\alpha}}(t)+\overline{(1-f) b_S I}(t))}\\
&+ \lambda b_S (1-A^{\alpha}_{a_S,b_S}(t)-T^{\alpha}_{a_S,b_S}(t)-Q^{\alpha}_{a_S,b_S}(t)) \frac{\delta \overline{(1-f) a_S P}(t)+(1-\delta)(\overline{f a_S T^{\alpha}}(t)+\overline{f a_S A^{\alpha}}(t)+ \overline{(1-f) a_S A}(t))}{\overline{b_S}- (1-\delta) \overline{f b_S Q^{\alpha}}(t)- \delta (\overline{f b_S I^{\alpha}}(t)+\overline{(1-f) b_S I}(t))}\\
&- \lambda \delta a_S A_{a_S,b_S}^{\alpha}(t) \frac{\overline{f b_S}-\overline{f b_S P^{\alpha}}(t)-\overline{f b_S I^{\alpha}}(t)}{\overline{b_S}- (1-\delta) \overline{f b_S Q^{\alpha}}(t)- \delta (\overline{f b_S I^{\alpha}}(t)+\overline{(1-f) b_S I}(t))}\\
&- \lambda \delta b_S A_{a_S,b_S}^{\alpha}(t) \frac{\overline{f a_S}-\overline{f a_S P^{\alpha}}(t)-\overline{f a_S I^{\alpha}}(t)}{\overline{b_S}- (1-\delta) \overline{f b_S Q^{\alpha}}(t)- \delta (\overline{f b_S I^{\alpha}}(t)+\overline{(1-f) b_S I}(t))}
\end{aligned} \label{eq:CT_d_Ia}
\end{align}
\begin{align}
\partial_t Q_{a_S,b_S}^{\alpha}(t)&=-\mu Q_{a_S,b_S}^{\alpha}(t) + \gamma_P T_{a_S,b_S}^{\alpha}(t)\\
\partial_t T^{\alpha}_{a_S,b_S}(t)&=\begin{aligned}[t]&-(\mu+\gamma_P) T^{\alpha}_{a_S,b_S}(t) \\
&+ \lambda a_S (1-A^{\alpha}_{a_S,b_S}(t)-T^{\alpha}_{a_S,b_S}(t)-Q^{\alpha}_{a_S,b_S}(t)) \frac{\delta \overline{f b_S P^{\alpha}}(t)}{\overline{b_S}- (1-\delta) \overline{f b_S Q^{\alpha}}(t)- \delta (\overline{f b_S I^{\alpha}}(t)+\overline{(1-f) b_S I}(t))}\\
&+ \lambda b_S (1-A^{\alpha}_{a_S,b_S}(t)-T^{\alpha}_{a_S,b_S}(t)-Q^{\alpha}_{a_S,b_S}(t)) \frac{\delta \overline{f a_S P^{\alpha}}(t)}{\overline{b_S}- (1-\delta) \overline{f b_S Q^{\alpha}}(t)- \delta (\overline{f b_S I^{\alpha}}(t)+\overline{(1-f) b_S I}(t))}\\
&+ \lambda \delta a_S A_{a_S,b_S}^{\alpha}(t) \frac{\overline{f b_S}-\overline{f b_S P^{\alpha}}(t)-\overline{f b_S I^{\alpha}}(t)}{\overline{b_S}- (1-\delta) \overline{f b_S Q^{\alpha}}(t)- \delta (\overline{f b_S I^{\alpha}}(t)+\overline{(1-f) b_S I}(t))}\\
&+ \lambda \delta b_S A_{a_S,b_S}^{\alpha}(t) \frac{\overline{f a_S}-\overline{f a_S P^{\alpha}}(t)-\overline{f a_S I^{\alpha}}(t)}{\overline{b_S}- (1-\delta) \overline{f b_S Q^{\alpha}}(t)- \delta (\overline{f b_S I^{\alpha}}(t)+\overline{(1-f) b_S I}(t))}\end{aligned}
\end{align}
\end{small}

This set of equations admits  as a stationary state the absorbing state, a configuration where all the population is susceptible. To obtain the condition for the stability of the absorbing state, i.e. the epidemic threshold, we apply a linear stability analysis around the absorbing state.

Let us now consider the case of realistic correlations between the activity and attractiveness: $\rho(a_S,b_S)= \rho_S(a_S) \delta(b_S-a_S)$, with general $\rho_S(a_S)$. If we average on all activity classes, we obtain the temporal evolution of the average probabilities $\overline{P}(t)$, $\overline{I}(t)$, $\overline{A}(t)$, $\overline{P^{\alpha}}(t)$, $\overline{I^{\alpha}}(t)$, $\overline{A^{\alpha}}(t)$, $\overline{Q^{\alpha}}(t)$, $\overline{T^{\alpha}}(t)$; similarly we obtain the temporal evolution of $\overline{f a_S P^{\alpha}}(t)$, $\overline{f a_S T^{\alpha}}(t)$, $\overline{a_S A^{\alpha}}(t)$, $\overline{f a_S A^{\alpha}}(t)$, $\overline{(1-f) a_S P}(t)$, $\overline{(1-f) a_S A}(t)$, multiplying the equations for $f(a_S) a_S \rho_S(a_S)$ or $a_S \rho_S(a_S)$ or $(1-f(a_S)) a_S \rho_S(a_S)$ and integrating over all activity classes.  We neglect the terms of second order in probabilities obtaining a linearized set of $14$ differential equations:

\begin{small}
\begin{align}
\partial_t \overline{I}(t)&=-\mu_I \overline{I}(t)+ \gamma_P \overline{P}(t)\\
\partial_t \overline{P}(t)&=-\gamma_P \overline{P}(t)+2 \lambda [\delta (\overline{f a_S P^{\alpha}}(t)+ \overline{(1-f) a_S P}(t))+(1-\delta)(\overline{f a_S T^{\alpha}}(t)+\overline{f a_S A^{\alpha}}(t)+\overline{(1-f) a_S A}(t))]\\
\partial_t \overline{I^{\alpha}}(t)&=-\mu_I \overline{I^{\alpha}}(t)+ \gamma_P \overline{P^{\alpha}}(t)\\
\partial_t \overline{P^{\alpha}}(t)&=-\gamma_P \overline{P^{\alpha}}(t)+2 \lambda [\delta (\overline{f a_S P^{\alpha}}(t)+ \overline{(1-f) a_S P}(t))+(1-\delta)(\overline{f a_S T^{\alpha}}(t)+\overline{f a_S A^{\alpha}}(t)+\overline{(1-f) a_S A}(t))]\\
\partial_t \overline{A}(t)&=-\mu \overline{A}(t) +2 \lambda [\delta (\overline{f a_S P^{\alpha}}(t)+ \overline{(1-f) a_S P}(t))+(1-\delta)(\overline{f a_S T^{\alpha}}(t)+\overline{f a_S A^{\alpha}}(t)+\overline{(1-f) a_S A}(t))]\\
\partial_t \overline{Q^{\alpha}}(t)&=-\mu \overline{Q^{\alpha}}(t) + \gamma_P \overline{T^{\alpha}}(t)\\
\partial_t \overline{T^{\alpha}}(t)&=-(\mu + \gamma_P) \overline{T^{\alpha}}(t) + 2 \lambda \delta \overline{f a_S P^{\alpha}}(t)
+ 2 \lambda \delta  \overline{a_S A^{\alpha}}(t) \frac{\overline{f a_S}}{\overline{a_S}}\\
\partial_t \overline{A^{\alpha}}(t)&=\begin{aligned}[t]&-\mu \overline{A^{\alpha}}(t) + 2 \lambda[\delta \overline{(1-f) a_S P}(t) + (1-\delta)(\overline{f a_S T^{\alpha}}(t)+\overline{f a_S A^{\alpha}}(t)+\overline{(1-f) a_S A}(t))]\\
 &- 2 \lambda \delta  \overline{a_S A^{\alpha}}(t) \frac{\overline{f a_S}}{\overline{a_S}}\end{aligned}\\
\partial_t \overline{a_S A^{\alpha}}(t)&=\begin{aligned}[t]&-\mu \overline{a_S A^{\alpha}}(t) + 2 \lambda \frac{\overline{a_S^2}}{\overline{a_S}} [\delta \overline{(1-f) a_S P}(t)+(1-\delta)(\overline{f a_S T^{\alpha}}(t)+\overline{f a_S A^{\alpha}}(t)+\overline{(1-f) a_S A}(t))]\\
& - 2 \lambda \delta  \overline{a_S^2 A^{\alpha}}(t) \frac{\overline{f a_S}}{\overline{a_S}}\end{aligned}\\
\partial_t \overline{f a_S P^{\alpha}}(t)&=\begin{aligned}[t]&-\gamma_P \overline{f a_S P^{\alpha}}(t) \\
&+ 2 \lambda \frac{\overline{f a_S^2}}{\overline{a_S}} [\delta(\overline{f a_S P^{\alpha}}(t)+ \overline{(1-f) a_S P}(t))+(1-\delta)(\overline{f a_S T^{\alpha}}(t)+\overline{f a_S A^{\alpha}}(t)+\overline{(1-f) a_S A}(t))]\end{aligned}\\
\partial_t \overline{(1-f) a_S P}(t)&=\begin{aligned}[t]&-\gamma_P \overline{(1-f) a_S P}(t) \\
&+2 \lambda \frac{\overline{(1-f) a_S^2}}{\overline{a_S}} [\delta(\overline{f a_S P^{\alpha}}(t)+ \overline{(1-f) a_S P}(t))+(1-\delta)(\overline{f a_S T^{\alpha}}(t)+\overline{f a_S A^{\alpha}}(t)+\overline{(1-f) a_S A}(t))] \end{aligned}\\
\partial_t \overline{(1-f) a_S A}(t)&=\begin{aligned}[t]&-\mu \overline{(1-f) a_S A}(t)\\
&+ 2 \lambda \frac{\overline{(1-f) a_S^2}}{\overline{a_S}} [\delta(\overline{f a_S P^{\alpha}}(t)+ \overline{(1-f) a_S P}(t))+(1-\delta)(\overline{f a_S T^{\alpha}}(t)+\overline{f a_S A^{\alpha}}(t)+\overline{(1-f) a_S A}(t))] \end{aligned}\\
\partial_t \overline{f a_S T^{\alpha}}(t)&=-(\mu + \gamma_P) \overline{f a_S T^{\alpha}}(t) + 2 \lambda \delta \frac{\overline{f a_S^2}}{\overline{a_S}} \overline{f a_S P^{\alpha}}(t)
+ 2 \lambda \delta  \overline{f a_S^2 A^{\alpha}}(t) \frac{\overline{f a_S}}{\overline{a_S}}\\
\partial_t \overline{f a_S A^{\alpha}}(t)&=\begin{aligned}[t]&-\mu \overline{f a_S A^{\alpha}}(t) + 2 \lambda \frac{\overline{f a_S^2}}{\overline{a_S}} [\delta \overline{(1-f) a_S P}(t)+(1-\delta)(\overline{f a_S T^{\alpha}}(t)+\overline{f a_S A^{\alpha}}(t)+\overline{(1-f) a_S A}(t))]\\
&- 2 \lambda \delta  \overline{f a_S^2 A^{\alpha}}(t) \frac{\overline{f a_S}}{\overline{a_S}}\end{aligned}
\end{align}

\end{small}
Similarly to the manual CT, in order to close the equations and obtain a complete set of linearized equations, we express $\overline{f a_S^2 A^{\alpha}}(t)$ and $\overline{a_S^2 A^{\alpha}}(t)$ in terms of the other average probabilities using their own definition. By definition $\overline{f a_S^2 A^{\alpha}}(t)=\int da_S \rho_S(a_S)f(a_S) a_S^2 A_{a_S}^{\alpha}(t)$ and $\overline{a_S^2 A^{\alpha}}(t)=\int da_S \rho_S(a_S) a_S^2 A_{a_S}^{\alpha}(t)$: since we are interested in studying the steady state and its stability, we consider the Eq. \eqref{eq:CT_d_Ia} for $\rho(a_S,b_S)= \rho_S(a_S) \delta(b_S-a_S)$, linearized around the absorbing state and near to the stationary condition $\partial_t A^{\alpha}_{a_S}(t) \sim 0$:

\begin{equation}
A^{\alpha}_{a_S}(t) \simeq \frac{2 \lambda a_S [\delta \overline{(1-f) a_S P}(t)+(1-\delta)(\overline{f a_S T^{\alpha}}(t)+\overline{f a_S A^{\alpha}}(t)+\overline{(1-f) a_S A}(t))]}{\mu \overline{a_S}+2 \lambda \delta a_S \overline{f a_S}}
\end{equation}

Thus,

\begin{align}
\overline{f a_S^2 A^{\alpha}}(t)&\simeq \frac{2}{\overline{a_S}} r[\delta \overline{(1-f) a_S P}(t)+(1-\delta)(\overline{f a_S T^{\alpha}}(t)+\overline{f a_S A^{\alpha}}(t)+\overline{(1-f) a_S A}(t))] H\\
\overline{a_S^2 A^{\alpha}}(t)&\simeq \frac{2}{\overline{a_S}} r [\delta \overline{(1-f) a_S P}(t)+(1-\delta)(\overline{f a_S T^{\alpha}}(t)+\overline{f a_S A^{\alpha}}(t)+\overline{(1-f) a_S A}(t))] Z
\end{align}

where $H= \overline{\frac{f a_S^3}{1+2r\delta a_S \frac{\overline{f a_S}}{\overline{a_S}}}} = \int da_S \rho_S(a_S) \frac{f(a_S) a_S^3}{1+2r\delta a_S \frac{\overline{f a_S}}{\overline{a_S}}}$, $Z= \overline{\frac{a_S^3}{1+2r\delta a_S \frac{\overline{f a_S}}{\overline{a_S}}}}$.

In this way we obtain the following linearized equations for $\overline{f a_S T^{\alpha}}$, $\overline{f a_S A^{\alpha}}$ and $\overline{a_S A^{\alpha}}$ near the absorbing stationary state:

\begin{align}
\partial_t \overline{a_S A^{\alpha}}(t)&=\begin{aligned}[t]&-\mu \overline{a_S A^{\alpha}}(t) + 2 \lambda \frac{\overline{a_S^2}}{\overline{a_S}} [ \delta \overline{(1-f) a_S P}(t)+(1-\delta)(\overline{f a_S T^{\alpha}}(t)+\overline{f a_S A^{\alpha}}(t)+\overline{(1-f) a_S A}(t))] \\
&- 4 \lambda r \delta \frac{\overline{f a_S}}{\overline{a_S}^2} [\delta \overline{(1-f) a_S P}(t)+(1-\delta)(\overline{f a_S T^{\alpha}}(t)+\overline{f a_S A^{\alpha}}(t)+\overline{(1-f) a_S A}(t))] Z \end{aligned}
\end{align}
\begin{align}
\partial_t \overline{f a_S T^{\alpha}}(t)&=\begin{aligned}[t]&-(\mu + \gamma_P) \overline{f a_S T^{\alpha}}(t) + 2 \lambda \delta \frac{\overline{f a_S^2}}{\overline{a_S}} \overline{f a_S P^{\alpha}}(t)\\
&+ 4 \lambda r \delta \frac{\overline{f a_S}}{\overline{a_S}^2} [\delta \overline{(1-f) a_S P}(t) +(1-\delta)(\overline{f a_S T^{\alpha}}(t)+\overline{f a_S A^{\alpha}}(t)+\overline{(1-f) a_S A}(t))] H \end{aligned}\\
\partial_t \overline{f a_S A^{\alpha}}(t)&=\begin{aligned}[t]&-\mu \overline{f a_S A^{\alpha}}(t) + 2 \lambda \frac{\overline{f a_S^2}}{\overline{a_S}} [\delta \overline{(1-f) a_S P}(t)+(1-\delta)(\overline{f a_S T^{\alpha}}(t)+\overline{f a_S A^{\alpha}}(t)+\overline{(1-f) a_S A}(t))] \\
&- 4 \lambda r \delta \frac{\overline{f a_S}}{\overline{a_S}^2} [\delta \overline{(1-f) a_S P}(t) +(1-\delta)(\overline{f a_S T^{\alpha}}(t)+\overline{f a_S A^{\alpha}}(t)+\overline{(1-f) a_S A}(t))] H \end{aligned}
\end{align}

Focusing on the Jacobian matrix of this set of $14$ linearized equations:

\begin{small}
\begin{equation}
\begin{aligned}
J=&\begin{bmatrix}
-\mu_I & \gamma_P & 0 & 0 & 0 & 0 & 0 & 0 & 0 & 0 & 0 & 0 & 0 & 0 \\
0 & -\gamma_P & 0 & 0 & 0 & 0 & 0 & 0 & 0  & 2 \lambda \delta & 2 \lambda \delta & 2 \lambda(1-\delta) & 2 \lambda (1-\delta) & 2 \lambda(1-\delta)\\
0 & 0 & -\mu_I & \gamma_P & 0 & 0 & 0 & 0 & 0 & 0 & 0 & 0 & 0 & 0 \\
0 & 0 & 0 & -\gamma_P & 0 & 0 & 0 & 0 & 0 & 2 \lambda \delta & 2 \lambda \delta & 2 \lambda(1-\delta) & 2 \lambda (1-\delta) & 2 \lambda(1-\delta)\\
0 & 0 & 0 & 0 & -\mu & 0 & 0 & 0 & 0 & 2 \lambda \delta & 2 \lambda \delta & 2 \lambda(1-\delta) & 2 \lambda (1-\delta) & 2 \lambda(1-\delta)\\
0 & 0 & 0 & 0 & 0 & -\mu & +\gamma_P & 0 & 0 & 0 & 0 & 0 & 0 & 0 \\
0 & 0 & 0 & 0 & 0 & 0 & -\mu-\gamma_P & 0 & 2 \lambda \delta \frac{\overline{f a_S}}{\overline{a_S}} & 2 \lambda \delta & 0 & 0 & 0 & 0 \\
0 & 0 & 0 & 0 & 0 & 0 & 0 & -\mu & -2 \lambda \delta \frac{\overline{f a_S}}{\overline{a_S}} & 0 & 2 \lambda \delta  & 2 \lambda (1-\delta) & 2 \lambda (1-\delta) & 2 \lambda (1-\delta) \\
0 & 0 & 0 & 0 & 0 & 0 & 0 & 0 & -\mu & 0 & \Delta+\phi-\Phi \frac{Z}{H} & \theta+\Gamma-\Psi \frac{Z}{H} & \theta+\Gamma-\Psi \frac{Z}{H} & \theta+\Gamma-\Psi \frac{Z}{H} \\
0 & 0 & 0 & 0 & 0 & 0 & 0 & 0 & 0 & -\gamma_P + \Delta & \Delta & \Gamma & \Gamma & \Gamma\\
0 & 0 & 0 & 0 & 0 & 0 & 0 & 0 & 0 & \phi & -\gamma_P+\phi & \theta & \theta & \theta \\
0 & 0 & 0 & 0 & 0 & 0 & 0 & 0 & 0 & \phi & \phi  & -\mu+\theta & \theta & \theta\\
0 & 0 & 0 & 0 & 0 & 0 & 0 & 0 & 0 & \Delta & \Phi  &  \Psi & -\mu-\gamma_P+\Psi & \Psi\\
0 & 0 & 0 & 0 & 0 & 0 & 0 & 0 & 0 & 0 & \Delta-\Phi & \Gamma-\Psi & \Gamma - \Psi  & -\mu+\Gamma-\Psi 
\end{bmatrix}
\\
=&\begin{bmatrix}
\mathbb{A}(9\text{x}9) & \mathbb{C}(9\text{x}5) \\
\mathbb{O}(5\text{x}9) & \mathbb{B}(5\text{x}5)
\end{bmatrix}
\end{aligned}
\end{equation}
\end{small}
where $\Delta=2 \lambda \delta \frac{\overline{f a_S^2}}{\overline{a_S}}$, $\Gamma=2 \lambda (1-\delta) \frac{\overline{f a_S^2}}{\overline{a_S}}$, $\phi=2 \lambda \delta \frac{\overline{(1-f) a_S^2}}{\overline{a_S}}$, $\theta=2 \lambda (1-\delta) \frac{\overline{(1-f) a_S^2}}{\overline{a_S}}$, $\Phi=4 \lambda r \delta^2 \frac{\overline{f a_S}}{\overline{a_S}^2} H$ and $\Psi=  4 \lambda r \delta (1-\delta) \frac{\overline{f a_S}}{\overline{a_S}^2} H$.

The Jacobian matrix is a block matrix and we can consider separately the two blocks on the diagonal: for the first block $\mathbb{A}$ it is evident that the eigenvalues are $\xi_{1,2}=-\mu_I$, $\xi_{3,4,5,6}=-\mu$, $\xi_{7,8}=-\gamma_P$, $\xi_9=-\mu-\gamma_P$, all negative. Therefore, it is sufficient to study block $\mathbb{B}$, which is a matrix $5\text{x}5$. The characteristic polynomial of $\mathbb{B}$ is a polynomial of degree 5: we apply the Descartes' rule of signs, to impose all roots to be negative and we obtain the condition for the stability of the absorbing state:

\begin{equation}
\begin{aligned}
8r^3 \delta^2 (1-\delta) \frac{\overline{f a_S^2} \, \overline{f a_S}}{\overline{a_S}} \overline{\frac{f a_S^3}{1+2 r \delta a_S \frac{\overline{f a_S}}{\overline{a_S}}}} \frac{\gamma_P}{\mu} &-4 r^2 \delta (1-\delta) \left[ \overline{f a_S^2}^2+\frac{\gamma_P}{\mu} \overline{f a_S} \overline{\frac{f a_S^3}{1+2 r \delta a_S \frac{\overline{f a_S}}{\overline{a_S}}}} \right] \frac{\gamma_P}{\mu}\\
&+ 2 r \overline{a_S^2} \overline{a_S} \left(\frac{\gamma_P}{\mu}+1\right) \left(\delta+\frac{\gamma_P}{\mu}(1-\delta)\right)-\overline{a_S}^2\frac{\gamma_P}{\mu}\left(\frac{\gamma_P}{\mu}+1\right)<0
\end{aligned}
\label{eq:rc_app}
\end{equation}

By setting the equality, the equation allows to obtain a closed relation for an estimate of the epidemic threshold. The epidemic threshold obtained with the mean-field approach is exact and it holds for digital contact tracing, with arbitrary $\rho_S(a_S)$ and $f(a_S)$.

\subsection{Hybrid CT}
We now focus on the hybrid CT, with simultaneous implementation of both digital and manual CT. The transitions occurring in this case can be deduced starting from
  the transitions described above for manual and digital CT. A relevant difference
  is the need to consider an additional compartment, $T_M^{\alpha}$, indicating individual
  endowed with the app that are nevertheless traced manually, so that their transition
  to the $Q^{\alpha}$ state occurs with a rate $\gamma_A$.
  In a manner analogous to the previous cases, we apply the \textit{activity-attractiveness based mean-field} approach to the hybrid CT. At the mean-field level, the epidemic dynamics is described by the probabilities defined for the digital CT, with these additional/redefined probabilities:
\begin{itemize}
\item $T_{a_S,b_S|M}^{\alpha}(t)$ for an asymptomatic node with app to be infected and traced manually; 
\item $T_{a_S,b_S}(t)$ for an asymptomatic node without the app to be infected and traced manually; 
\item $Q_{a_S,b_S}(t)$ for an asymptomatic node without the app to be isolated;
\item $1-A_{a_S,b_S}(t)-Q_{a_S,b_S}(t)-T_{a_S,b_S}(t)$ to be susceptible at time $t$, for a node without app and which will not develop symptoms;
\item $1-A_{a_S,b_S}^{\alpha}(t)-Q_{a_S,b_S}^{\alpha}(t)-T_{a_S,b_S}^{\alpha}(t)-T_{a_S,b_S|M}^{\alpha}(t)$ to be susceptible at time $t$, for a node with app and which will not develop symptoms;
\end{itemize}

We consider the system in the thermodynamic limit and for $\rho(a_S,b_S)=\rho_S(a_S) \delta(b_S-a_S)$: in this case the average attractiveness at time $t$ is $\langle b(t) \rangle = \overline{a_S}- (1-\delta) (\overline{f a_S Q^{\alpha}}(t)+\overline{(1-f) a_S Q}(t))- \delta (\overline{f a_S I^{\alpha}}(t)+\overline{(1-f) a_S I}(t))$.
The probabilities introduced evolve according to the following equations, obtained analogously to those for manual and digital CT:

\begin{align}
\partial_t P_{a_S}(t)&=-\gamma_P P_{a_S}(t) \begin{aligned}[t]&+ 2 \lambda (1-P_{a_S}(t)-I_{a_S}(t)) \frac{a_S}{\langle b(t) \rangle}[\delta( \overline{a_S f P^{\alpha}}(t)+\overline{a_S (1-f) P}(t))\\&+(1-\delta)(\overline{a_S f A^{\alpha}}(t)+\overline{a_S f T^{\alpha}}(t)+ \overline{a_S f T_M^{\alpha}}(t)+ \overline{a_S (1-f) T}(t)+\overline{a_S (1-f) A}(t))]\end{aligned}\\
\partial_t I_{a_S}(t) &= -\mu_I I_{a_S}(t) + \gamma_P P_{a_S}(t)\\
\partial_t P_{a_S}^{\alpha}(t)&=-\gamma_P P_{a_S}^{\alpha}(t) \begin{aligned}[t]&+ 2 \lambda (1-P_{a_S}^{\alpha}(t)-I_{a_S}^{\alpha}(t)) \frac{a_S}{\langle b(t) \rangle}[\delta( \overline{a_S f P^{\alpha}}(t)+\overline{a_S (1-f) P}(t))\\&+(1-\delta)(\overline{a_S f A^{\alpha}}(t)+\overline{a_S f T^{\alpha}}(t)+ \overline{a_S f T_M^{\alpha}}(t)+ \overline{a_S (1-f) T}(t)+\overline{a_S (1-f) A}(t))]\end{aligned}\\
\partial_t I_{a_S}^{\alpha}(t) &= -\mu_I I_{a_S}^{\alpha}(t) + \gamma_P P_{a_S}^{\alpha}(t)\\
\partial_t A_{a_S}(t) &= \begin{aligned}[t]&-\mu A_{a_S}(t) \begin{aligned}[t]&+ 2 \lambda (1-A_{a_S}(t)-T_{a_S}(t)-Q_{a_S}(t)) \frac{a_S}{\langle b(t) \rangle} [\delta (\overline{a_S (1-\epsilon) f P^{\alpha}}(t)+\overline{a_S (1-\epsilon) (1-f) P}(t)) \\ &+ (1-\delta)(\overline{a_S f A^{\alpha}}(t)+\overline{a_S f T^{\alpha}}(t)+ \overline{a_S f T_M^{\alpha}}(t)+ \overline{a_S (1-f) T}(t)+\overline{a_S (1-f) A}(t))]\end{aligned}\\&-2 \lambda \delta A_{a_S}(t) \frac{a_S}{\langle b(t) \rangle}[(\overline{a_S \epsilon f}-\overline{a_S \epsilon f P^{\alpha}}(t)-\overline{a_S \epsilon f I^{\alpha}}(t))+(\overline{a_S \epsilon (1-f)} - \overline{a_S \epsilon (1-f) P}(t) - \overline{a_S \epsilon (1-f) I}(t))] \end{aligned}\\
\partial_t T_{a_S}(t)&=\begin{aligned}[t]&-(\mu+\gamma_A) T_{a_S}(t) + 2 \lambda \delta \frac{a_S}{\langle b(t) \rangle} (1-A_{a_S}(t)-T_{a_S}(t)-Q_{a_S}(t)) [\overline{a_S \epsilon f P^{\alpha}}(t) + \overline{a_S \epsilon (1-f) P}(t)]\\ &+ 2 \lambda \delta A_{a_S}(t) \frac{a_S}{\langle b(t) \rangle}[(\overline{a_S \epsilon f}-\overline{a_S \epsilon f P^{\alpha}}(t)-\overline{a_S \epsilon f I^{\alpha}}(t))+(\overline{a_S \epsilon (1-f)} - \overline{a_S \epsilon (1-f) P}(t) - \overline{a_S \epsilon (1-f) I}(t))] \end{aligned}\\
\partial_t Q_{a_S}(t)&= -\mu Q_{a_S}(t) + \gamma_A T_{a_S}(t)
\end{align}
\begin{align}
\partial_t A_{a_S}^{\alpha}(t) &= \begin{aligned}[t]&-\mu A_{a_S}^{\alpha}(t) \begin{aligned}[t]&+ 2 \lambda (1-A_{a_S}^{\alpha}(t)-T_{a_S}^{\alpha}(t)-T_{a_S|M}^{\alpha}(t)-Q_{a_S}^{\alpha}(t)) \frac{a_S}{\langle b(t) \rangle} [\delta \overline{a_S (1-\epsilon) (1-f) P}(t) \\ &+ (1-\delta)(\overline{a_S f A^{\alpha}}(t)+\overline{a_S f T^{\alpha}}(t)+ \overline{a_S f T_M^{\alpha}}(t)+ \overline{a_S (1-f) T}(t)+\overline{a_S (1-f) A}(t))]\end{aligned}\\&-2 \lambda \delta A_{a_S}^{\alpha}(t) \frac{a_S}{\langle b(t) \rangle}[(\overline{a_S f}-\overline{a_S f P^{\alpha}}(t)-\overline{a_S f I^{\alpha}}(t))+(\overline{a_S \epsilon (1-f)} - \overline{a_S \epsilon (1-f) P}(t) - \overline{a_S \epsilon (1-f) I}(t))] \end{aligned}\\
\partial_t T_{a_S}^{\alpha}(t)&=-(\mu+\gamma_P) T_{a_S}^{\alpha}(t) \begin{aligned}[t]&+ 2 \lambda \delta \frac{a_S}{\langle b(t) \rangle} (1-A_{a_S}^{\alpha}(t)-T_{a_S}^{\alpha}(t)-T_{a_S|M}^{\alpha}(t)-Q_{a_S}^{\alpha}(t)) \overline{a_S f P^{\alpha}}(t)\\ &+ 2 \lambda \delta A_{a_S}^{\alpha}(t) \frac{a_S}{\langle b(t) \rangle}[\overline{a_S f}-\overline{a_S f P^{\alpha}}(t)-\overline{a_S f I^{\alpha}}(t)] \end{aligned}\\
\partial_t T_{a_S|M}^{\alpha}(t)&=-(\mu+\gamma_A) T_{a_S|M}^{\alpha}(t) \begin{aligned}[t]&+ 2 \lambda \delta \frac{a_S}{\langle b(t) \rangle} (1-A_{a_S}^{\alpha}(t)-T_{a_S}^{\alpha}(t)-T_{a_S|M}^{\alpha}(t)-Q_{a_S}^{\alpha}(t)) \overline{a_S \epsilon (1-f) P}(t)\\ &+ 2 \lambda \delta A_{a_S}^{\alpha}(t) \frac{a_S}{\langle b(t) \rangle}[\overline{a_S \epsilon (1-f)} - \overline{a_S \epsilon (1-f) P}(t) - \overline{a_S \epsilon (1-f) I}(t)] \end{aligned}\\
\partial_t Q_{a_S}^{\alpha}(t)&= -\mu Q_{a_S}^{\alpha}(t) + \gamma_P T_{a_S}^{\alpha}(t) + \gamma_A T_{a_S|M}^{\alpha}(t)
\end{align}

This set of equations admits  as a stationary state the absorbing state: to obtain the epidemic threshold, we apply a linear stability analysis around the absorbing state. We consider the temporal evolution of the probabilities averaged on the activity (eventually multiplied by combinations of $a_S$, $\epsilon(a_S)$ and $f(a_S)$), obtaining a closed and complete set of $22$ non-linear differential equations. The set of equations can be linearized obtaining the following Jacobian matrix:

\begin{equation}
J=
\begin{bmatrix}
\mathbb{A}(13\text{x}13) & \mathbb{C}(13\text{x}9) \\
\mathbb{O}(9\text{x}13) & \mathbb{B}(9\text{x}9)
\end{bmatrix}
\end{equation}
where $\mathbb{O}$ is the zero matrix.
\begin{equation}
\mathbb{A}=\begin{bmatrix}
-\mu_I & \gamma_P & 0 &  0 & 0 & 0 & 0 & 0 & 0 & 0 & 0 & 0 & 0 \\
0 &-\gamma_P & 0 & 0 & 0 & 0 & 0 & 0 & 0 & 0 & 0 & 0 & 0 \\
0 & 0 &-\mu_I & \gamma_P & 0 & 0 & 0 & 0 & 0 & 0 & 0 & 0 & 0 \\
0 & 0 & 0 &-\gamma_P & 0 & 0 & 0 & 0 & 0 & 0 & 0 & 0 & 0 \\
0 & 0 & 0 & 0 &-\mu & \gamma_A &  0 & 0 & 0 & 0 & 0 & 0 & 0 \\
0 & 0 & 0 & 0 & 0 & -\mu-\gamma_A & 0 & 0 & 0 & 0 & 0 & 2 \lambda \delta \frac{\overline{a_S \epsilon}}{\overline{a_S}} & 0 \\
0 & 0 & 0 & 0 & 0 & 0 & -\mu & 0 & 0 & 0 & 0 & -2 \lambda \delta \frac{\overline{a_S \epsilon}}{\overline{a_S}} & 0 \\
0 & 0 & 0 & 0 & 0 & 0 & 0 & -\mu & \gamma_P & \gamma_A & 0 & 0 & 0 \\
0 & 0 & 0 & 0 & 0 & 0 & 0 & 0 & -\mu-\gamma_P & 0 & 0 & 0 & 2\lambda \delta \frac{\overline{a_S f}}{\overline{a_S}} \\
0 & 0 & 0 & 0 & 0 & 0 & 0 & 0 & 0 & -\mu-\gamma_A & 0 & 0 & 2\lambda \delta \frac{\overline{a_S \epsilon (1-f)}}{\overline{a_S}} \\
0 & 0 & 0 & 0 & 0 & 0 & 0 & 0 & 0 & 0 & -\mu & 0 & -2\lambda \delta \frac{\overline{a_S f}+\overline{a_S \epsilon (1-f)}}{\overline{a_S}} \\
0 & 0 & 0 & 0 & 0 & 0 & 0 & 0 & 0 & 0 & 0 & -\mu & 0 \\
0 & 0 & 0 & 0 & 0 & 0 & 0 & 0 & 0 & 0 & 0 & 0 & - \mu
\end{bmatrix}
\end{equation}
\begin{equation}
\mathbb{C}=\begin{bmatrix}
0 & 0 & 0 &  0 & 0 & 0 & 0 & 0 & 0 \\
0 & 0 & 2 \lambda \delta & 2 \lambda \delta &  2 \lambda (1-\delta) & 2 \lambda (1-\delta) & 2 \lambda (1-\delta) & 2 \lambda (1-\delta) & 2 \lambda (1-\delta) \\
0 & 0 & 0 &  0 & 0 & 0 & 0 & 0 & 0 \\
0 & 0 & 2 \lambda \delta & 2 \lambda \delta &  2 \lambda (1-\delta) & 2 \lambda (1-\delta) & 2 \lambda (1-\delta) & 2 \lambda (1-\delta) & 2 \lambda (1-\delta) \\
0 & 0 & 0 &  0 & 0 & 0 & 0 & 0 & 0 \\
2 \lambda \delta & 2 \lambda \delta & 0 & 0 &  0 & 0 & 0 & 0 & 0 \\
-2 \lambda \delta & -2 \lambda \delta& 2 \lambda \delta & 2 \lambda \delta &  2 \lambda (1-\delta) & 2 \lambda (1-\delta) & 2 \lambda (1-\delta) & 2 \lambda (1-\delta) & 2 \lambda (1-\delta) \\
0 & 0 & 0 &  0 & 0 & 0 & 0 & 0 & 0 \\
0 & 0 & 0 &  2 \lambda \delta & 0 & 0 & 0 & 0 & 0 \\
2 \lambda \delta & 0 & 0 &  0 & 0 & 0 & 0 & 0 & 0 \\
-2 \lambda \delta & 0 & 2 \lambda \delta & 0 &  2 \lambda (1-\delta) & 2 \lambda (1-\delta) & 2 \lambda (1-\delta) & 2 \lambda (1-\delta) & 2 \lambda (1-\delta) \\
-\Psi \delta & -\Psi \delta& \Psi \delta & \Psi \delta &  \Psi (1-\delta) & \Psi (1-\delta) & \Psi (1-\delta) & \Psi (1-\delta) & \Psi (1-\delta) \\
-\theta \delta & 0 & \theta \delta & 0 &  \theta (1-\delta) & \theta (1-\delta) & \theta (1-\delta) & \theta (1-\delta) & \theta (1-\delta) \\
\end{bmatrix}
\end{equation}
with $\Psi= 2 \lambda \frac{\overline{a_S^2}}{\overline{a_S}}-4\lambda r \delta\frac{\overline{a_S \epsilon}}{\overline{a_S}^2}K$, $\theta=2 \lambda \frac{\overline{a_S^2}}{\overline{a_S}}-4\lambda r \delta\frac{\overline{a_S \epsilon (1-f)} + \overline{a_S f}}{\overline{a_S}^2}J$, $K=\overline{\frac{a_S^3}{1+2 r a_S \delta \frac{\overline{a_S \epsilon}}{\overline{a_S}}}}$ and $J=\overline{\frac{a_S^3}{1+2 r a_S \delta \frac{\overline{a_S f}+\overline{a_S \epsilon (1-f)}}{\overline{a_S}}}}$.
\begin{equation}
\mathbb{B}=\begin{bmatrix}
-\gamma_P & 0 & \Xi \delta &  \Xi \delta & \Xi \sigma & \Xi \sigma & \Xi \sigma & \Xi \sigma & \Xi \sigma \\
0 &-\gamma_P & \Pi \delta &  \Pi \delta & \Pi \sigma & \Pi \sigma & \Pi \sigma & \Pi \sigma & \Pi \sigma \\
0 & 0 & -\gamma_P+ \Delta \delta & \Delta \delta & \Delta \sigma & \Delta \sigma & \Delta \sigma & \Delta \sigma & \Delta \sigma \\
0 & 0 & \beta \delta & -\gamma_P+\beta \delta & \beta \sigma & \beta \sigma & \beta \sigma & \beta \sigma & \beta \sigma \\
\Omega\delta & \Omega\delta & \phi \delta & \phi \delta & -\mu-\gamma_A +\phi \sigma& \phi  \sigma & \phi  \sigma & \phi  \sigma & \phi  \sigma \\
-\Gamma \delta & 0 & \Gamma \delta & \beta \delta &  \Gamma \sigma&  -\mu-\gamma_P+\Gamma  \sigma & \Gamma  \sigma & \Gamma  \sigma & \Gamma  \sigma \\
\Sigma\delta & 0 & \Phi \delta & 0 &  \Phi \sigma& \Phi  \sigma & -\mu-\gamma_A+\Phi  \sigma & \Phi  \sigma & \Phi  \sigma \\
-\Omega\delta & -\Omega\delta & \Omega \delta & \Omega \delta & \Omega \sigma& \Omega  \sigma & \Omega  \sigma & -\mu+\Omega  \sigma & \Omega  \sigma \\
-\Lambda\delta & 0 & \Lambda \delta & 0 & \Lambda \sigma& \Lambda  \sigma & \Lambda  \sigma & \Lambda  \sigma & -\mu+\Lambda  \sigma \\
\end{bmatrix}
\end{equation}
with $\sigma=(1-\delta)$, $\Xi=2 \lambda \frac{\overline{a_S^2 \epsilon(1-f)}}{\overline{a_S}}$, $\Pi=2 \lambda \frac{\overline{a_S^2\epsilon f}}{\overline{a_S}}$, $\Delta=2 \lambda \frac{ \overline{a_S^2(1-f)}}{\overline{a_S}}$, $\beta=2 \lambda \frac{\overline{a_S^2 f}}{\overline{a_S}}$, $\phi=4 \lambda r \delta \frac{\overline{a_S \epsilon}}{\overline{a_S}^2}H$, $\Gamma=4 \lambda r \delta \frac{\overline{a_S f}}{\overline{a_S}^2}Y$, $\Phi=4 \lambda r \delta \frac{\overline{a_S \epsilon (1-f)}}{\overline{a_S}^2}Y$, $H= \overline{\frac{(1-f)a_S^3}{1+2r\delta a_S \frac{\overline{ a_S \epsilon}}{\overline{a_S}}}}$, $Y= \overline{\frac{f a_S^3}{1+2r\delta a_S \frac{\overline{ a_S \epsilon(1-f)}+\overline{ a_S f}}{\overline{a_S}}}}$, $\Omega=\Delta-\phi$, $\Sigma=\beta-\Phi$ and $\Lambda=-\Gamma-\Phi+\beta$.\\

The Jacobian matrix is a block matrix and we can consider separately the two blocks on the diagonal: the first block $\mathbb{A}$ is triangular and the eigenvalues are $\xi_{1,2}=-\mu_I$, $\xi_{3,4,5,6,7,8}=-\mu$, $\xi_{9,10}=-\gamma_P$, $\xi_{11}=-\mu-\gamma_P$, $\xi_{12,13}=-\mu-\gamma_A$ all negative. Therefore, it is sufficient to study block $\mathbb{B}$, which is a matrix $9\text{x}9$. The epidemic threshold in the hybrid case is therefore obtained by numerically diagonalizing the matrix $\mathbb{B}$ and imposing all its eigenvalues to be negative: this allows to obtain analytically the epidemic threshold for the hybrid CT for arbitrary $\rho_S(a_S)$, $\epsilon(a_S)$ and $f(a_S)$.

\subsection{Limit cases}\label{sez:limitcases}
The obtained closed relations for the stability of the absorbing state in the manual and digital CT hold for arbitrary $\rho_S(a_S)$, for completely general $f(a_S)$ and $\epsilon(a_S)$ and for general delays: this allows to introduce complicated effects, such as delay in isolation, activity heterogeneities and limited scalability of the system. 

Due to the complicated structure of the equations for the stability (Eq. \eqref{eq:rc_man} and Eq. \eqref{eq:rc_app}), it is possible to derive the epidemic threshold $r_C$ only by solving the equation numerically. However, there are some simple limit cases in which the equations are considerably simplified, allowing to obtain the epidemic threshold in an explicit analytic form.

\subsubsection{Non-adaptive case (NA)}
Here we consider the non-adaptive case, in which no adaptive behaviour are implemented, i.e. infected nodes behave as if they were susceptible, with $(a_I,b_I)=(a_S,b_S)$. Thus, in this case $f(a_S)=\epsilon(a_S)=0$, $\forall a_S$ and $\gamma_P/\mu=1$.
Replacing these values either in Eq. \eqref{eq:rc_man} or in Eq. \eqref{eq:rc_app} we obtain, as expected:

$$2 r \overline{a_S^2} \overline{a_S}-\overline{a_S}^2=0$$

So we obtain an explicit form for the epidemic threshold $r_C$ in the non-adaptive case:

\begin{equation}
r_C^{NA}=\frac{\overline{a_S}}{2 \overline{a_S^2}}
\label{eq:Na}
\end{equation}
which is the Eq. (3) in the main paper. It reproduces the results previously obtained in Refs.~\cite{Pozzana2017,Mancastroppa2020}.

\subsubsection{Isolation of only symptomatic nodes}
Here we consider the case in which only symptomatic nodes are isolated as soon as they develops symptoms, i.e. no CT is implemented. Thus, in this case $f(a_S)=\epsilon(a_S)=0$, $\forall a_S$, while $(a_I,b_I)=(0,0)$ . Replacing these values either in Eq. \eqref{eq:rc_man} or in Eq. \eqref{eq:rc_app} we obtain, as expected:

$$2 r \overline{a_S^2} \overline{a_S}\left(\delta+(1-\delta)\frac{\gamma_P}{\mu}\right)-\overline{a_S}^2 \frac{\gamma_P}{\mu}=0$$

So we obtain an explicit form for the epidemic threshold $r_C$:

\begin{equation}
r_C^{SYMPTO}=r_C^{NA}\frac{\frac{\gamma_P}{\mu}}{\delta+(1-\delta)\frac{\gamma_P}{\mu}}
\end{equation}
which is the Eq. (4) in the main paper. For instantaneous symptoms development $\gamma_P/\mu \to \infty$ it reproduces the results previously obtained in Ref.~\cite{Mancastroppa2020}, and for $\gamma_P/\mu=1$ it reproduces the NA case (Eq. \eqref{eq:Na}).

\subsubsection{Homogeneous case without delays and without limited scalability}
Here we consider the case in which the population is homogeneous, i.e. $\rho(a_S,b_S)=\delta(a_S-a) \delta(b_S-b)$, with constant probability of downloading the app in digital CT, i.e. $f(a_S)=f$, and constant probability for a contact to be traced in manual CT, i.e. $\epsilon(a_S)=\epsilon$. Moreover we assume $\tau_C=0$, that is $\gamma_A=\gamma_P$. Replacing these values in the Eq. \eqref{eq:rc_man}, for the manual CT we obtain a quadratic equation in $r$:

$$4 a^2 \delta^2 \epsilon r^2 + 2 a \left(\delta+\frac{\gamma_P}{\mu}(1-\delta-\delta \epsilon)\right) r-\frac{\gamma_P}{\mu}=0$$

The equation can be solved and we obtain:

\begin{equation}
r_C^{MANUAL}=r_C^{NA} \frac{2 \frac{\gamma_P}{\mu}}{\delta+(1-\delta-\epsilon \delta)\frac{\gamma_P}{\mu}+\sqrt{(\delta+(1-\delta-\epsilon \delta)\frac{\gamma_P}{\mu})^2+4 \delta^2 \epsilon \frac{\gamma_P}{\mu}}}
\end{equation}
which is the Eq. (5) in the main paper.

Analogously, replacing the values in the Eq. \eqref{eq:rc_app} for the digital CT we obtain an equation of second degree in $r$:

$$4 a^2 \delta f \left(\delta+\frac{\gamma_P}{\mu}(1-\delta)(1-f)\right) r^2 + 2 a \left(\delta+\frac{\gamma_P}{\mu}(1-\delta-\delta f)\right) r-\frac{\gamma_P}{\mu}=0$$

The equation can be solved and we obtain:

\begin{equation}
r_C^{APP}=r_C^{NA} \frac{2 \frac{\gamma_P}{\mu}}{\delta+(1-\delta-f \delta)\frac{\gamma_P}{\mu}+\sqrt{(\delta+(1-\delta-f \delta)\frac{\gamma_P}{\mu})^2+4 \delta f \frac{\gamma_P}{\mu} (\delta + \frac{\gamma_P}{\mu} (1-f)(1-\delta))}}
\end{equation}
which is the Eq. (6) in the main paper.

\subsubsection{Heterogeneous case}

\begin{figure}
  \includegraphics[width=0.5\textwidth]{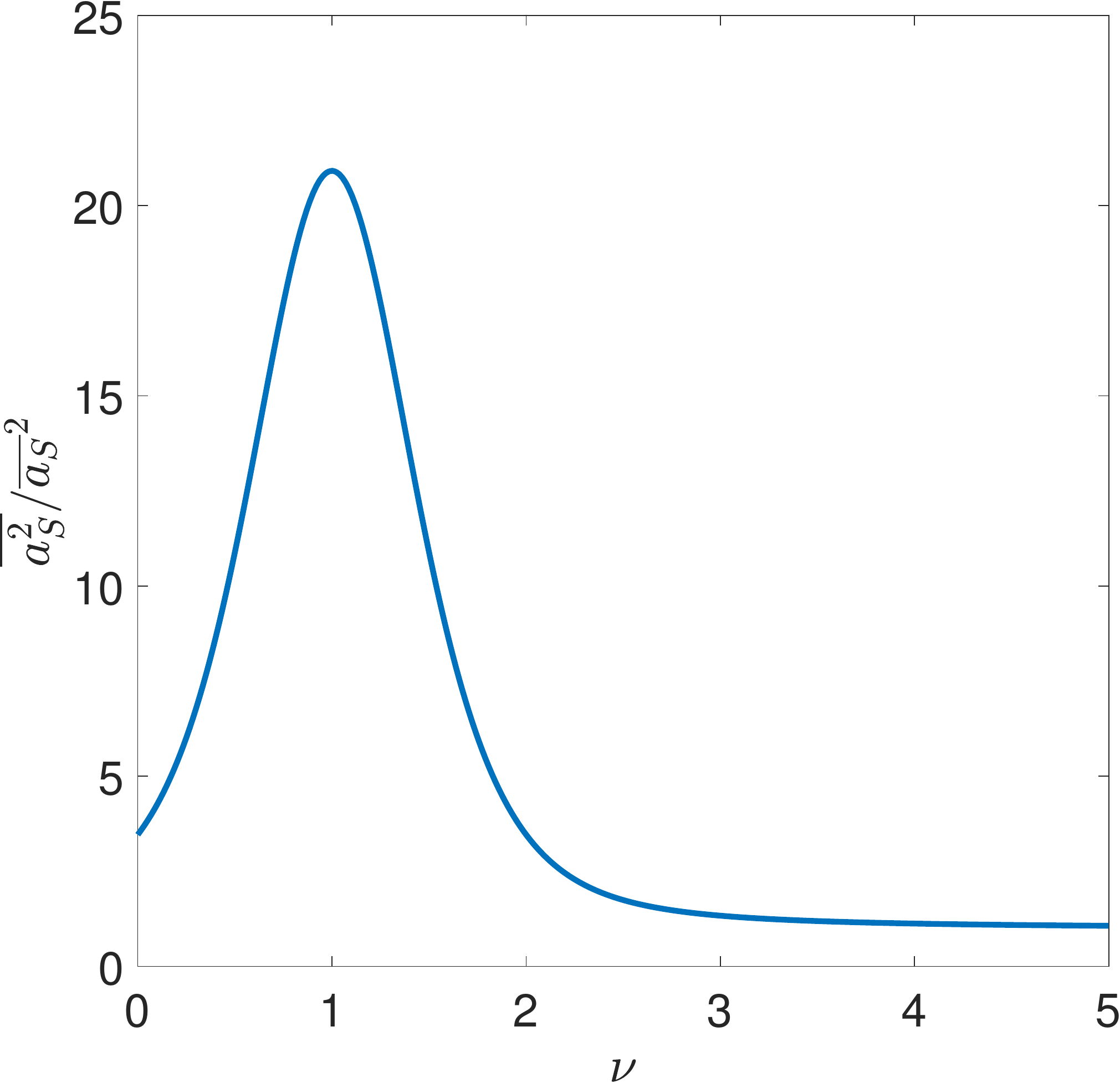}
  \caption{\textbf{Heterogeneity of the activity distribution.} We plot the ratio $\overline{a_S^2}/\overline{a_S}^2$ as a function of exponent $\nu$, considering $\rho(a_S,b_S)$ given by Eq.~\eqref{eq:heterog_act_rho} with $a_S \in [a_m,a_M]$, $\eta=a_M/a_m$ and fixing $\eta=10^3$.}
  \label{fig:heterog_act}
\end{figure}

Hereafter and in the main text we consider a realistic heterogeneous population, i.e. a power-law activity distribution:
\begin{equation}
\rho(a_S,b_S) \sim a_S^{-(\nu+1)} \delta(b_S-a_S)
\label{eq:heterog_act_rho}
\end{equation}
with $a_S \in [a_m,a_M]$ and $\eta=a_M/a_m$, where $a_m$ and $a_M$ are respectively activity lower and upper cut-off.

Supplementary Fig.~\ref{fig:heterog_act} shows that the considered distribution is maximally broad for $\nu = 1$: indeed, the ratio $\overline{a_S^2}/\overline{a_S}^2$, which we plot, provides a estimate of the heterogeneity and fluctuations of the distribution and it is maximized for $\nu=1$. On the contrary for $\nu \to 0$ the distribution is counterintuitively more homogeneous, due to cut-off effects and constraints imposed on the distribution ($\overline{a_S}$ and $\eta$ fixed), which thus set the maximum heterogeneous condition at $\nu=1$.

In this heterogeneous case it is possible to derive the epidemic threshold $r_C$ only numerically, due to the complex structure of the equations for the stability (Eq. \eqref{eq:rc_man} and Eq. \eqref{eq:rc_app}). 
Fig. 3(a,b) and Fig. 4(a,b) of the main text show that both manual and digital CT protocols are more effective when the population is heterogeneous, indeed the maximum gain in the epidemic threshold is for $\nu \sim 1-1.5$. However, the maximum does not occur exactly at $\nu = 1$, i.e. when the distribution is maximally broad. Indeed, the epidemic thresholds do not depend solely on activity fluctuations, i.e. $\overline{a_S^2}$, but also on higher order moments (as $\overline{a_S^3}$), as shown by Eq. \eqref{eq:rc_man} and Eq. \eqref{eq:rc_app}. Thus, the CT protocols are more effective in heterogeneous populations and the maximum gain in $r_C$ also depends on higher order moments of $\rho_S$ and on other model parameters, as shown in Supplementary Fig.~\ref{fig:rob_rc}(b-c).

\section{Supplementary Method 2: Continuous time Gillespie-like algorithm for network dynamics and epidemic evolution}
The network dynamics is performed by a continuous-time Gillespie-like algorithm~\cite{gillespie1976general}: we assign to each node the activity $a_S$ and attractiveness $b_S$ drawn from the joint distribution $\rho(a_S,b_S)$.
Initially, the network evolves in the absorbing state, i.e. all nodes are susceptible and infection does not propagate, up to a relaxation time $t_0$, to reach the equilibrium of activation dynamics:
\begin{enumerate} 
\item The first activation time, $t_i$, of each node $i$ is drawn from $\Psi_{a_S^i}(t_i)= a_S^i e^{-a_S^i t_i}$ at time $t=0$. 
\item \label{item:3a} The node $i$ with the lowest $t_i$ activates and connects $m$ randomly-selected nodes, with probability proportional to their attractiveness $b_S$.
\item The next activation time $t_i$ for node $i$ is set to $t_i+\tau$, with $\tau$ inter-event time drawn from $\Psi_{a_S^i}(\tau)$.
\item All links are destroyed and the process is iterated from point~\ref{item:3a}.
\end{enumerate}

Then we start the epidemic and contact tracing dynamics as follows: 
\begin{enumerate} 
\item At time $t = t_0$ the population is divided into a configuration of susceptible ($S$) and infected ($P$ or $A$) nodes, moreover each node has an activation time $t_i>t_0=t$ obtained from the initial relaxation dynamics. 
\item \label{item:2} Node $i$ with the lowest $t_i$ activates. Asymptomatic ($A$, $T$ and $Q$) and symptomatic nodes ($I$) at time $t$ recover at $t_i$ with probability respectively $1-e^{-\mu (t_i-t)}$ and $1-e^{-\mu_I (t_i-t)}$: recovered nodes change their activity and attractiveness into $(a_R,b_R)=(a_S,b_S)$. Traced nodes at time $t$ are isolated at $t_i$ with probability $1-e^{-(t_i-t)/\tau_C}$ (with $\tau_C>0$ for manual CT and $\tau_C=0$ for digital CT) and set their activity and attractiveness to zero $(a_Q,b_Q)=(0,0)$.
\item Pre-symptomatic nodes at time $t$ develop symptoms at $t_i$ with probability $1-e^{-\gamma_P (t_i-t)}$: they set to zero their activity and attractiveness $(a_I,b_I)=(0,0)$ and the contact tracing is activated.\\ 

\textbf{Manual CT:} the protocol is enabled for every symptomatic node. Every contact made in the last $T_{CT}$ period has $\epsilon(a_S)$ probability of being identified and tested, with $a_S$ activity of the symptomatic node. Every node tested and found infected asymptomatic $A$ becomes traced $T$. \\

\textbf{Digital CT:} the protocol is enabled only if the symptomatic node has downloaded the app. Each contact made in the last $T_{CT}$ period with an individual who downloaded the app is identified and tested. Every node tested and found infected asymptomatic $A$ becomes traced $T$.

\item We set the actual time $t=t_i$ and the active agent $i$ generates exactly $m$ links with $m$ nodes randomly-chosen with probability proportional to their attractiveness $b$ at time $t$ (depending on their status and isolation). The contacts are registered in the contact list of both nodes engaged in the connection. If the link involves a susceptible and an infected node ($P$, $A$ or $T$), a contagion can occur with probability $\lambda$ and the susceptible node becomes pre-symptomatic with probability $\delta$ or asymptomatic with probability $1-\delta$.
\item The new activation time of node $i$ is $t_i=t+\tau$ and it is obtained drawing the inter-event time $\tau$ from the inter-event time distribution $\Psi_{a^i}(\tau) = a^i e^{-a^i \tau}$, where $a^i$ is the activity of node $i$ at time $t$.  All the links are deleted and the process is iterated from point~\ref{item:2}.

\end{enumerate}

\section{Supplementary Notes: Robustness of the results}

The advantage of the manual CT is robust  relaxing many assumptions and changing most  parameters in the modelling scheme. 

\subsection{Activity-attractiveness distribution, limited scalability parameters and delays}

\begin{figure}
\centering
\includegraphics[width=0.445\textwidth]{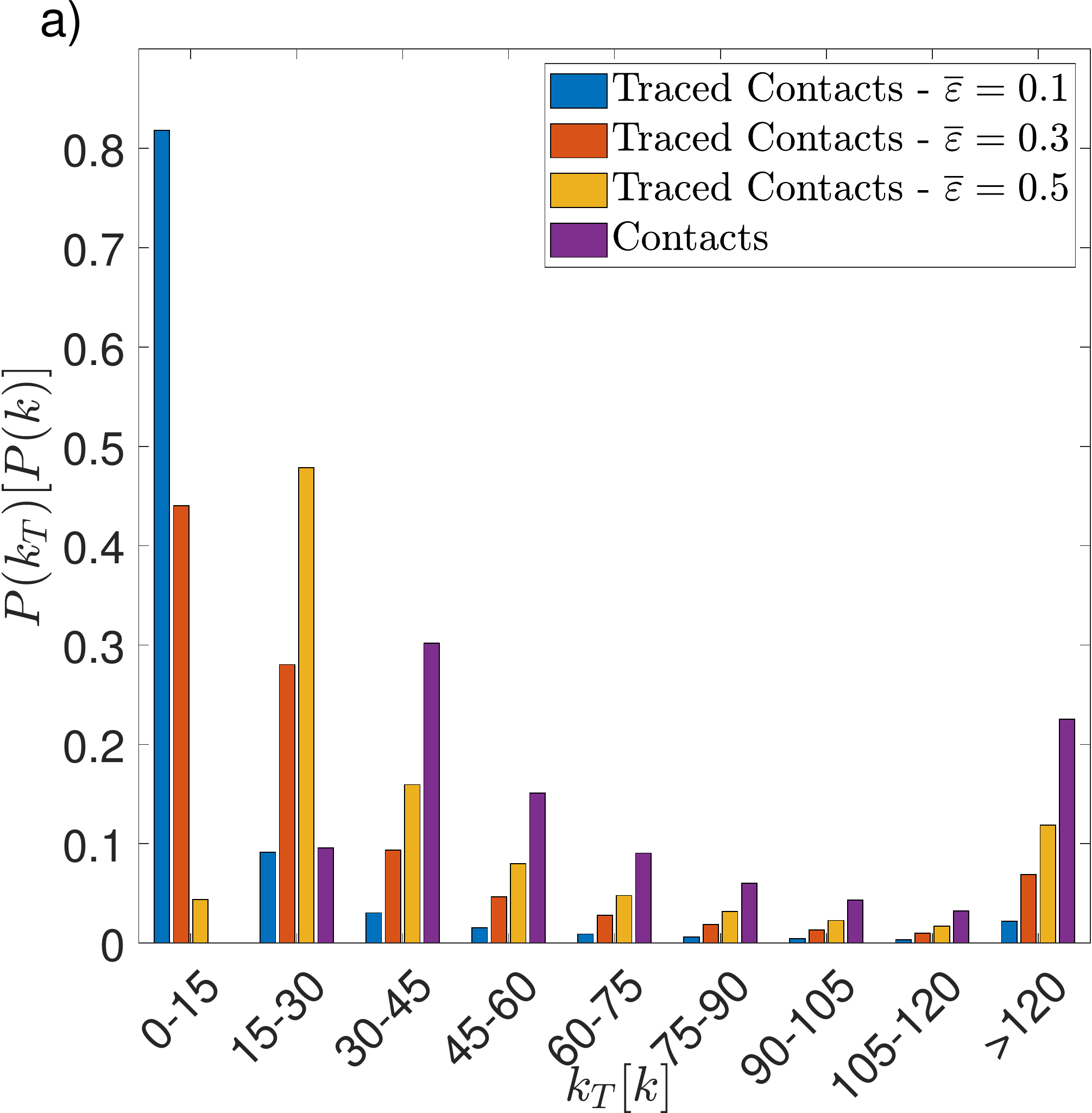}
\includegraphics[width=0.45\textwidth]{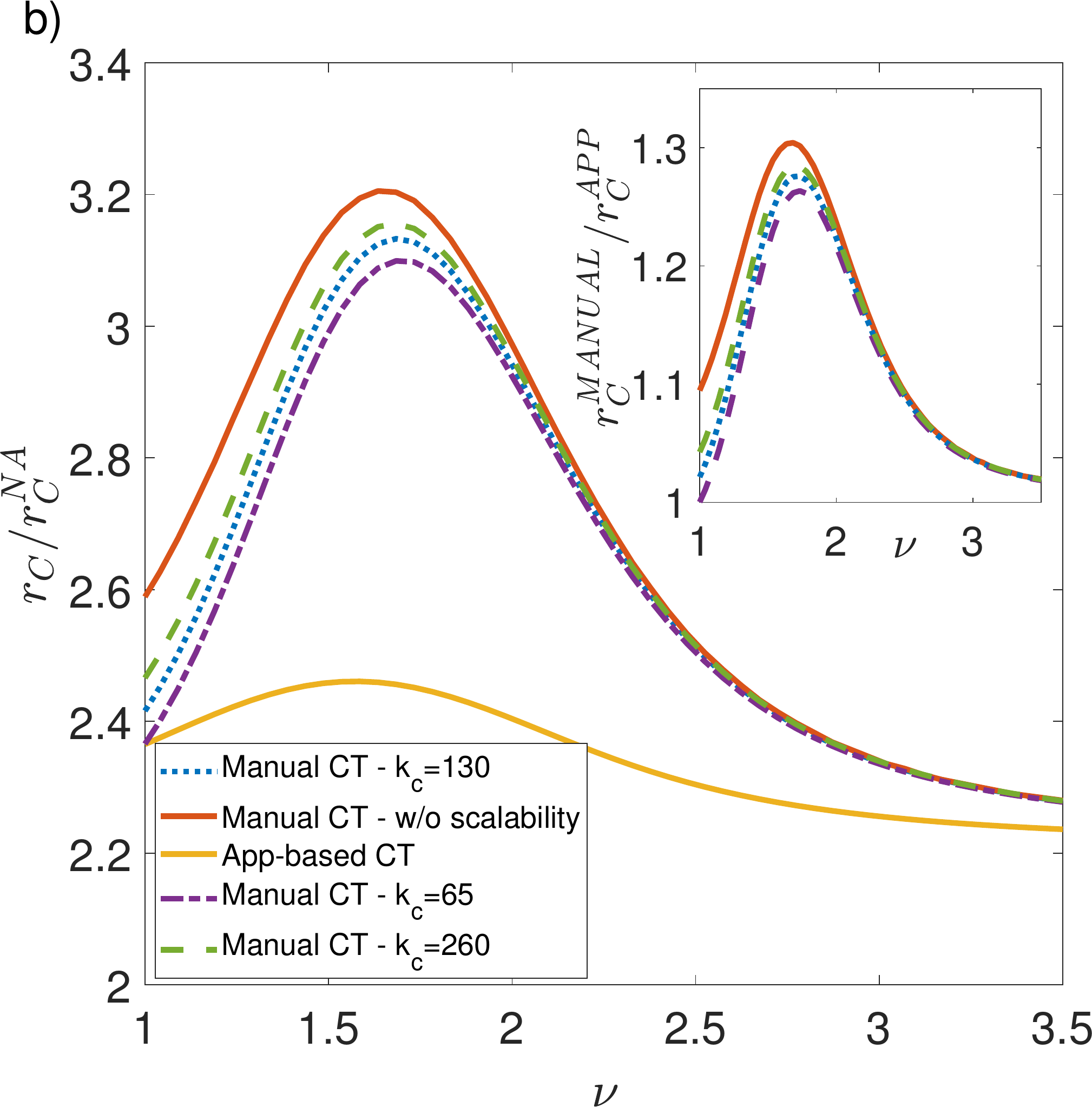}\\
\includegraphics[width=0.45\textwidth]{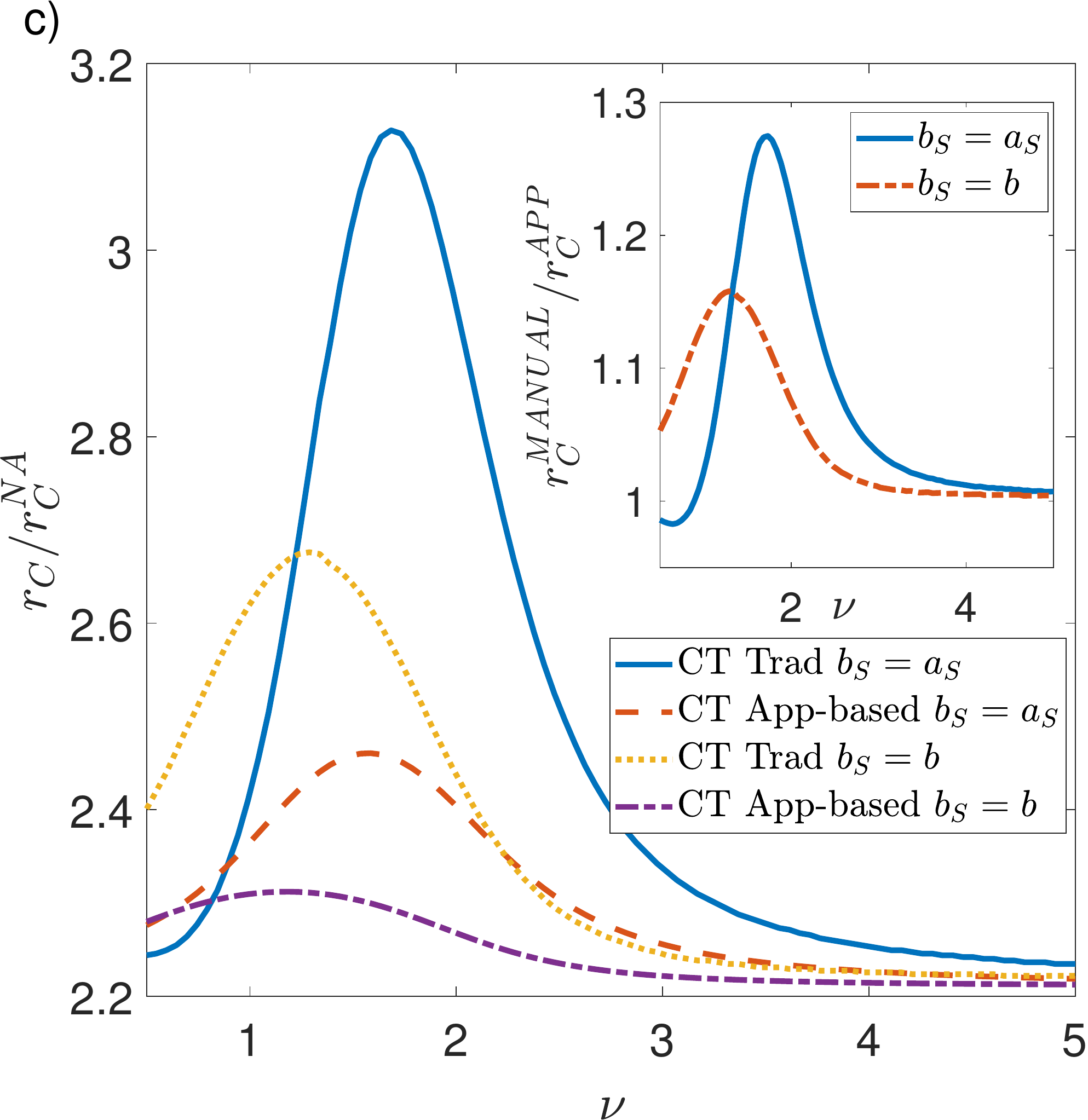}
\includegraphics[width=0.45\textwidth]{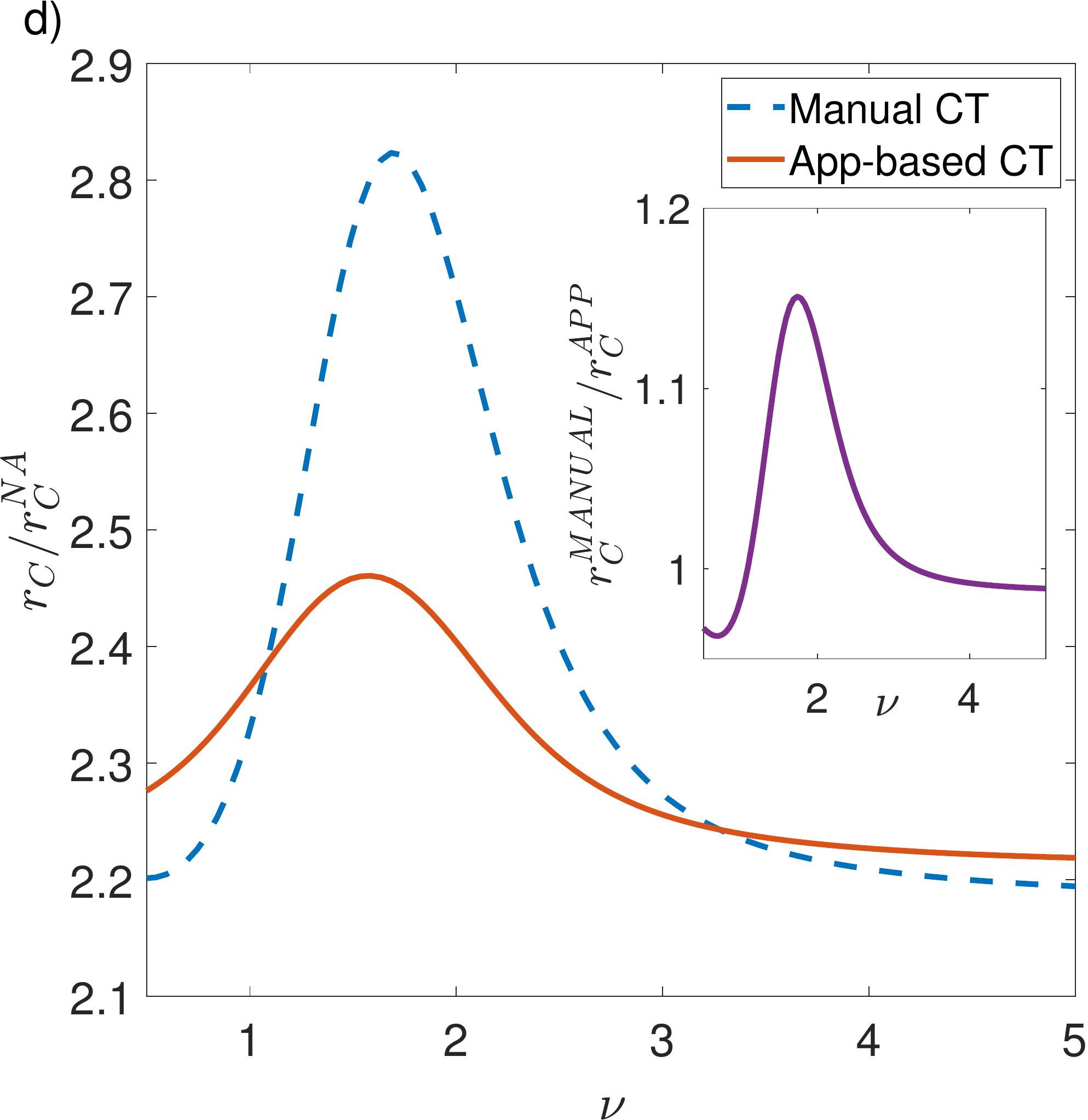}
  \caption{{\bf Robustness of the results for the epidemic threshold.} In panel (a) we plot the distribution $P(k)$ of contacts made by an individual over a $T_{CT}$ period and the distribution $P(k_T)$ of contacts traced by an index case with manual CT in the presence of limited scalability ($k_C=130$) for fixed $\overline{\epsilon}$ (see legend) and for $\rho(a_S,b_S)$ given by Eq.~\eqref{eq:heterog_act_rho} with $\nu=1$. In panels (b)-(d) we plot the ratio between the epidemic threshold $r_C$ in the presence of contact tracing protocols and the epidemic threshold of the non-adaptive case $r_C^{NA}$. The ratio is plotted for both manual and digital contact tracing, as a function of $\nu$. In all the insets we plot the ratio between the epidemic threshold of the manual contact tracing $r_C^{MANUAL}$ and that of the app-based contact tracing $r_C^{APP}$, as a function of $\nu$. In panel (b) the ratio for the manual CT is plotted for several values of $k_c$ and also when scalability is not considered (legend), setting $\tau_C=3$ days and $\rho(a_S,b_S)$ given by Eq.~\eqref{eq:heterog_act_rho}. In panel (c) the ratio is plotted both for $\rho(a_S,b_S) \sim a_S^{-(\nu+1)} \delta(b_S-a_S)$ and for $\rho(a_S,b_S) \sim a_S^{-(\nu+1)} \delta(b_S-b)$ (legend), setting $\tau_C=3$ days and $k_c=130$. In panel (d) the ratio is plotted for $\rho(a_S,b_S)$ given by Eq.~\eqref{eq:heterog_act_rho} setting $\tau_C=7$ days and $k_c=130$. In all panels $a_S \in [a_m,a_M]$ with $a_M/a_m=10^3$, $\overline{a_S}=6.7 \, days^{-1}$, $T_{CT}=14$ days, and in panels (b)-(d) $\overline{\epsilon}=f^2=0.1$, $\delta=0.57$, $\tau_P=1.5$ days, $\tau=14$ days.}
 \label{fig:rob_rc}
\end{figure}

Supplementary Fig.~\ref{fig:rob_rc}(a) shows the distribution $P(k)$ of the number of contacts made by an individual over a $T_{CT}$ period and the distribution $P(k_T)$ of the number of contacts $k_T$ traced by an index case with manual CT in the presence of limited scalability (recall probability as Eq.(2) of the main paper with $k_c=130$~\cite{Keeling2020}), for a realistically heterogeneous population. The average number of contacts traced by an index case is approximately $10-60$ (depending on $\overline{\epsilon}\sim 0.1-0.5$), consistent with the reported data~\cite{Keeling2020,Guardian2020}. Analogous contact patterns are generated by changing $k_C$ or other parameters of the activity distribution.

The results on the comparison between manual and digital CT are robust to changes in the maximum number of traceable contacts $k_c$ in the manual CT: an increase in $k_c$ reduces the effects of limited scalability; similarly, its reduction makes the effects of limited scalability stronger. However, modifying $k_c$ with realistic values changes only slightly the epidemic threshold of the manual contact tracing protocol, without changing qualitatively the results, since the manual protocol remains more effective than the digital one, for small $\overline{\epsilon}=f^2$ values (Supplementary Fig.~\ref{fig:rob_rc}(b)).

The results are also robust when considering changes in the social properties of the population, i.e. assuming a different functional form of the $\rho(a_S,b_S)$ distribution. For example, we can assume that all nodes feature equal attractiveness $b$ and different activity: $\rho(a_S,b_S)=\rho_S(a_S) \delta(b_S-b)$. In this case the correlations between activity and attractiveness are removed: however, again the manual CT is more effective than the digital one in heterogeneous populations and for small $\overline{\epsilon}=f^2$ (Supplementary Fig.~\ref{fig:rob_rc}(c)). The differences between the two methods are reduced, compared to the case with correlations (see the inset of Supplementary Fig.~\ref{fig:rob_rc}(c)), due to the reduction in heterogeneities, since homogeneous terms are introduced assuming all the nodes having the same attractiveness. However, differences between the two protocols remain evident, even in the presence of delays $\tau_C>0$ and limited scalability in the manual CT.

Similarly, the advantage of manual contact tracing holds also considering limited scalability and stronger delays in manual contact tracing $\tau_C$: even for a delay of $7$ days the manual CT remains more effective than the digital one, for small $\overline{\epsilon}=f^2$ and in heterogeneous populations (Supplementary Fig.~\ref{fig:rob_rc}(d)). 

\begin{figure}
\centering
\includegraphics[width=0.45\textwidth]{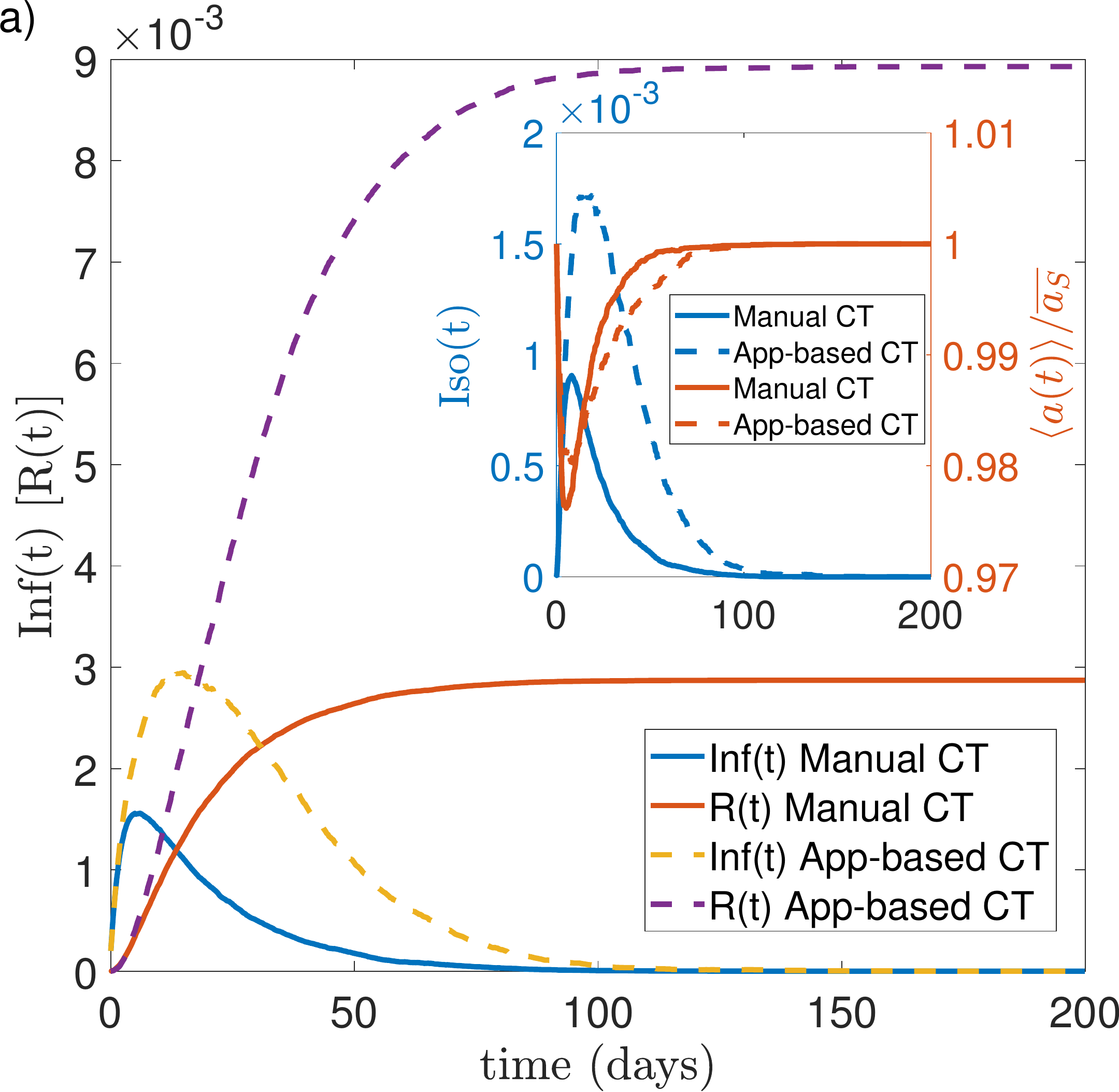}
\includegraphics[width=0.45\textwidth]{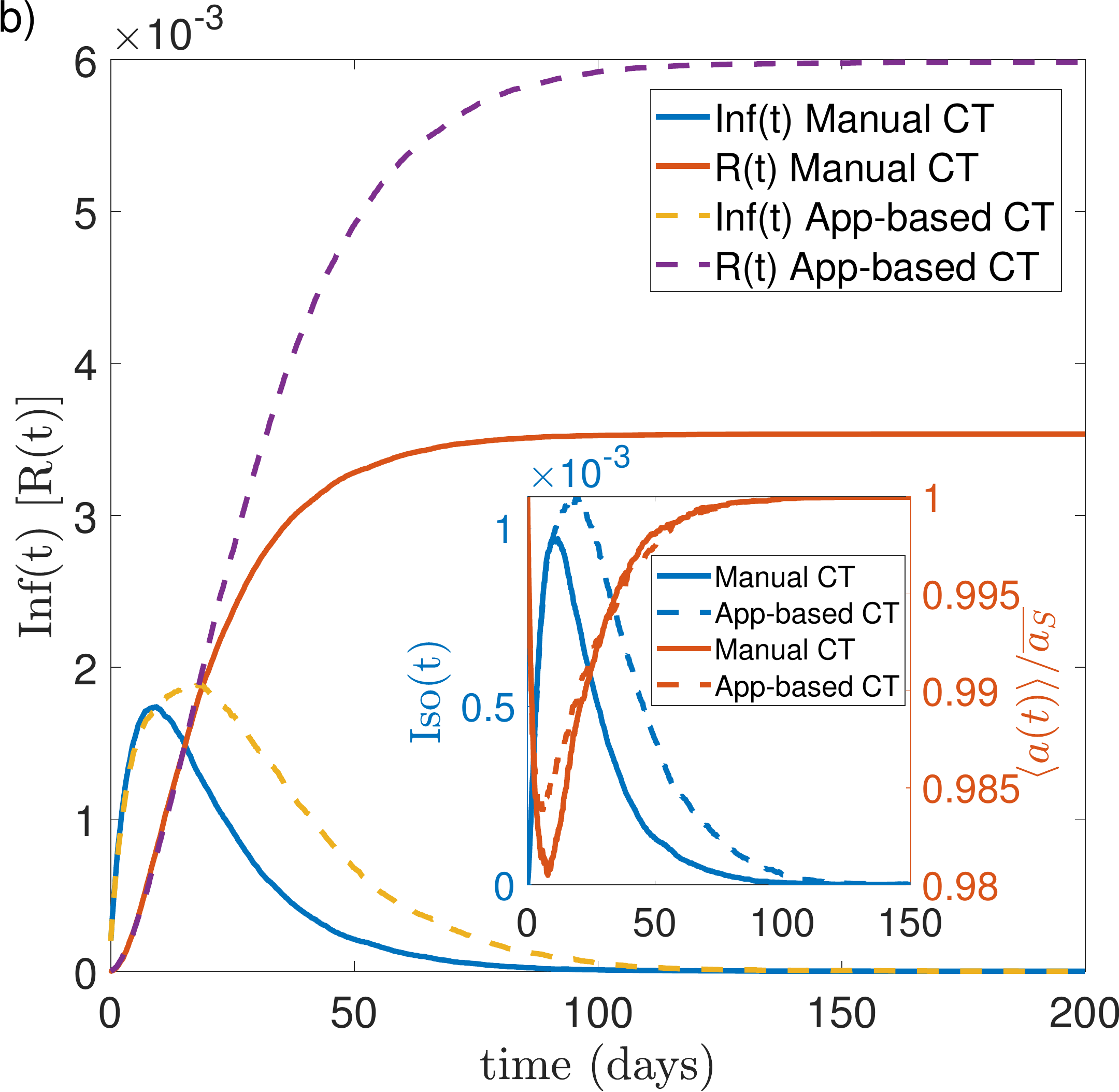}\\
\includegraphics[width=0.45\textwidth]{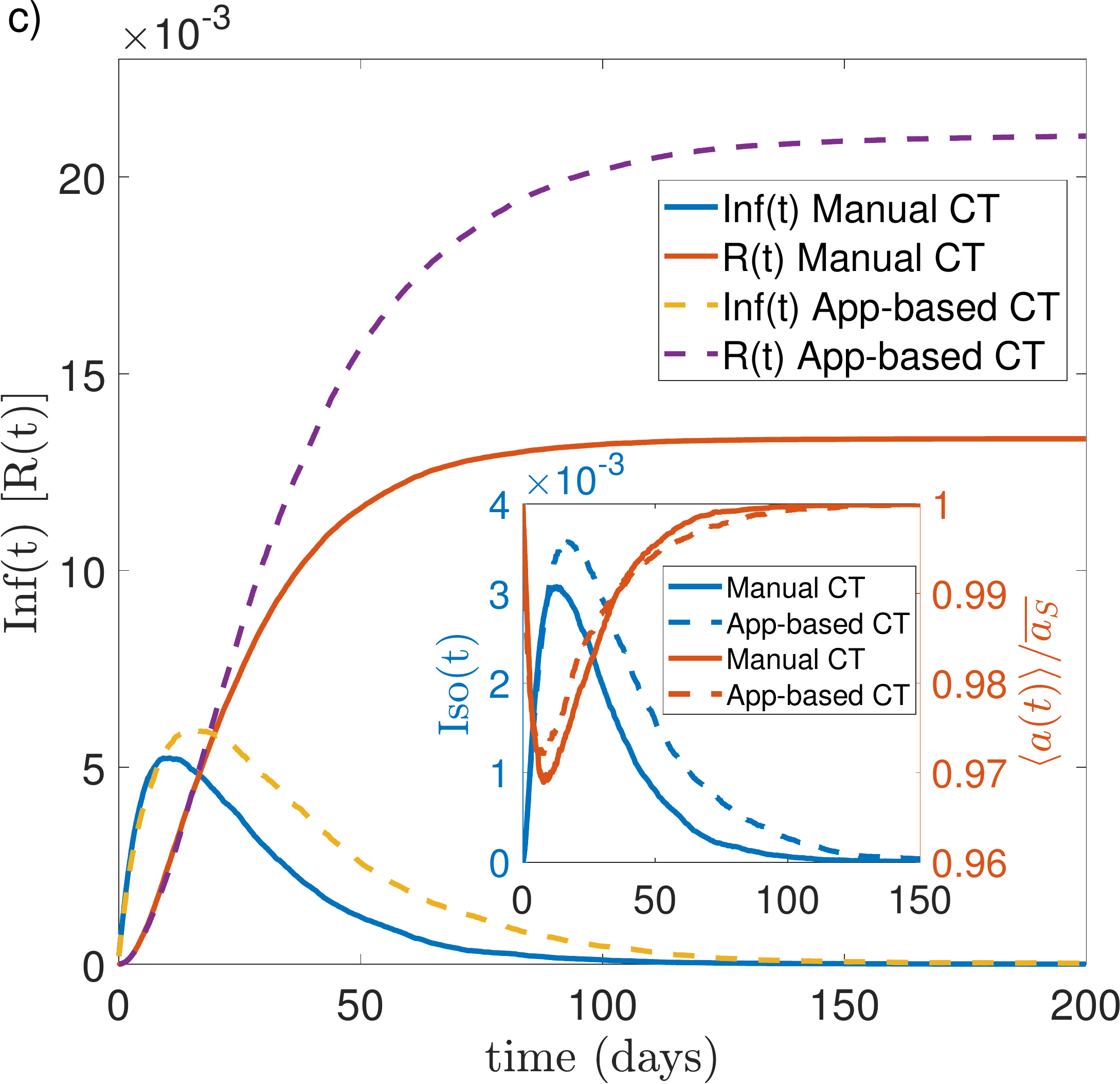}
\includegraphics[width=0.45\textwidth]{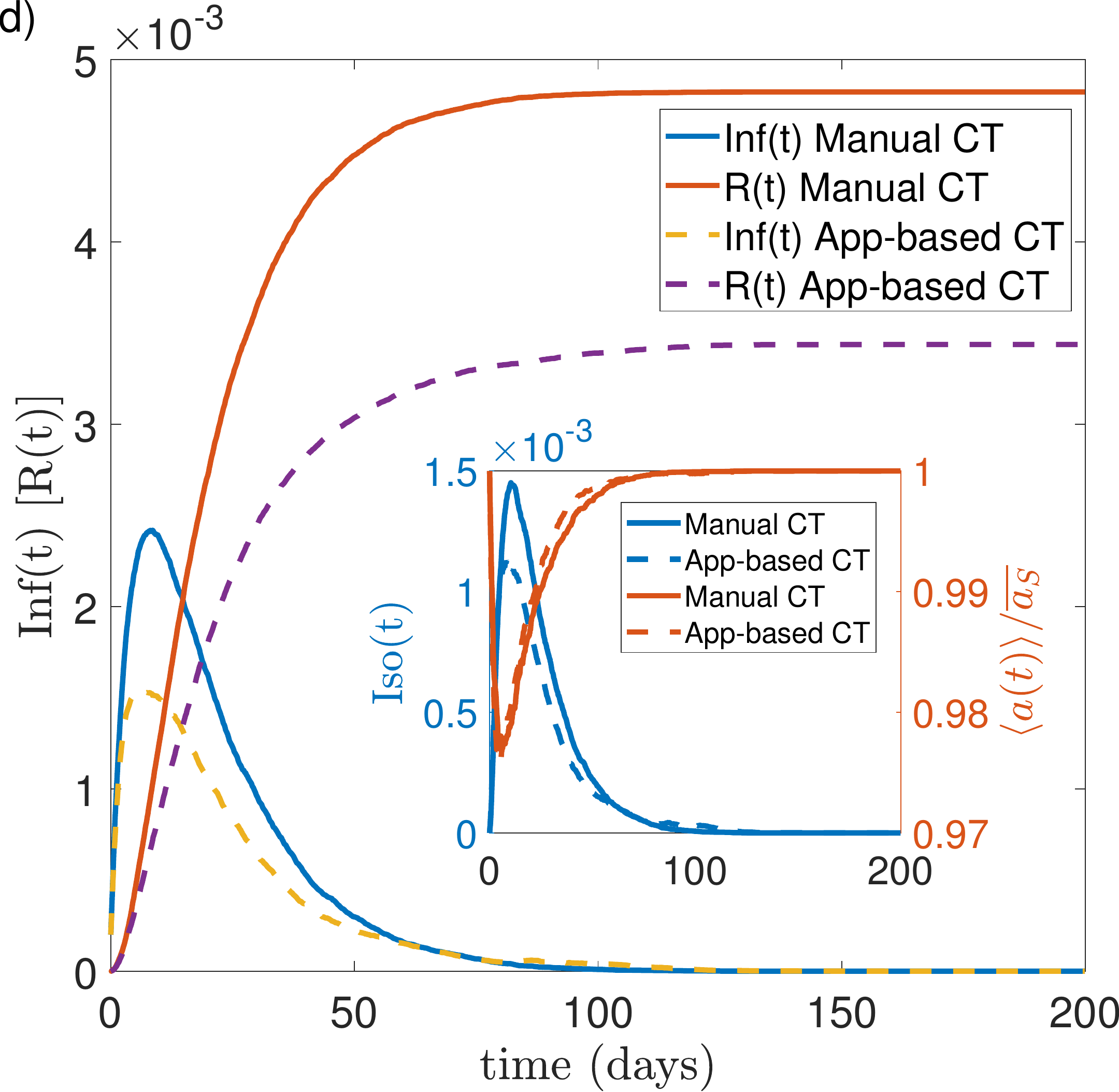}
  \caption{{\bf Robustness of the results for the epidemic active phase.} In all panels we plot the temporal evolution of the fraction of infected nodes $Inf(t)$, i.e. infected asymptomatic and infected symptomatic, and of the fraction of removed nodes $R(t)$, for both manual and digital CT. In the insets we plot the temporal evolution of the fraction of isolated nodes $Iso(t)$ (right y-axis) and of the average activity of the population $\langle a(t) \rangle$ (left y-axis), normalized with $\overline{a_S}$. All curves are averaged on several realization of disorder and temporal evolution. In panel (a) we set $\tau_C=0$, $\overline{\epsilon}=f^2=0.1$, $r/r_C^{NA}=4.0$ and the curves for digital and manual CT are averaged respectively over $379$ and $445$ realizations; in panel (b) we set $\tau_C=5$ days, $\overline{\epsilon}=f^2=0.1$, $r/r_C^{NA}=3.1$ and the curves for digital and manual CT are averaged respectively over $554$ and $604$ realizations; in panel (c) we set $\tau_C=5$ days, $\overline{\epsilon}=f^2=0.1$, $r/r_C^{NA}=7.0$ and the curves for digital and manual CT are averaged respectively over $311$ and $348$ realizations; in panel (d) we set $\tau_C=5$ days, $\overline{\epsilon}=f^2=0.6$, $r/r_C^{NA}=4.5$ and the curves for digital and manual CT are averaged respectively over $552$ and $469$ realizations. In all panels $\rho(a_S,b_S)$ is given by Eq.~\eqref{eq:heterog_act_rho} with $a_S \in [a_m,a_M]$, $a_M/a_m=10^3$, $\nu=1.5$, $\overline{a_S}=6.7 \, days^{-1}$, $N=5 \, 10^3$, $\delta=0.57$, $\tau_P=1.5$ days, $\tau=14$ days, $k_c=130$, $T_{CT}=14$ days and the errors, evaluated through the standard deviation, are smaller or comparable with the curves thickness.}
 \label{fig:rob_din_t}
\end{figure}

\subsection{Epidemic active phase}
The results are also robust when considering the effects of contact tracing on the active phase of the epidemic: as expected from the analysis on the epidemic threshold, the differences between the two methods are reduced increasing $\tau_C$, however even with considerable delays $\tau_C$ the manual CT for small values of $\overline{\epsilon}=f^2$ is more effective in flattening the infection peak and in lowering the epidemic final-size (Supplementary Fig.~\ref{fig:rob_din_t}). Indeed, its effectiveness, compared to the digital CT, is maximized for $\tau_C=0$ (Supplementary Fig.~\ref{fig:rob_din_t}(a)); the differences are reduced but still present considering strong delays in manual CT, such as $\tau_C=5$ days (Supplementary Fig.~\ref{fig:rob_din_t}(b)). The differences remain even if we consider the system deeply in the active phase, that is for $r \gg r_C$: in this case the differences are sightly reduced due to the high infectivity of the system, however again the manual method for small $\overline{\epsilon}$ is more effective than digital CT (Supplementary Fig.~\ref{fig:rob_din_t}(c)). Finally, as observed for the epidemic threshold, for very large $\overline{\epsilon}=f^2=0.6$ and strong delays $\tau_C=5$ days, the digital protocol becomes more effective in reducing the impact of the epidemic, further flattening the infection peak and reducing the epidemic final-size (Supplementary Fig.~\ref{fig:rob_din_t}(d)). This again confirms that the effects of the protocols on the active phase are similar to those observed on the epidemic threshold, including the differences in the two approaches.

\begin{figure}
\centering
\includegraphics[width=0.45\textwidth]{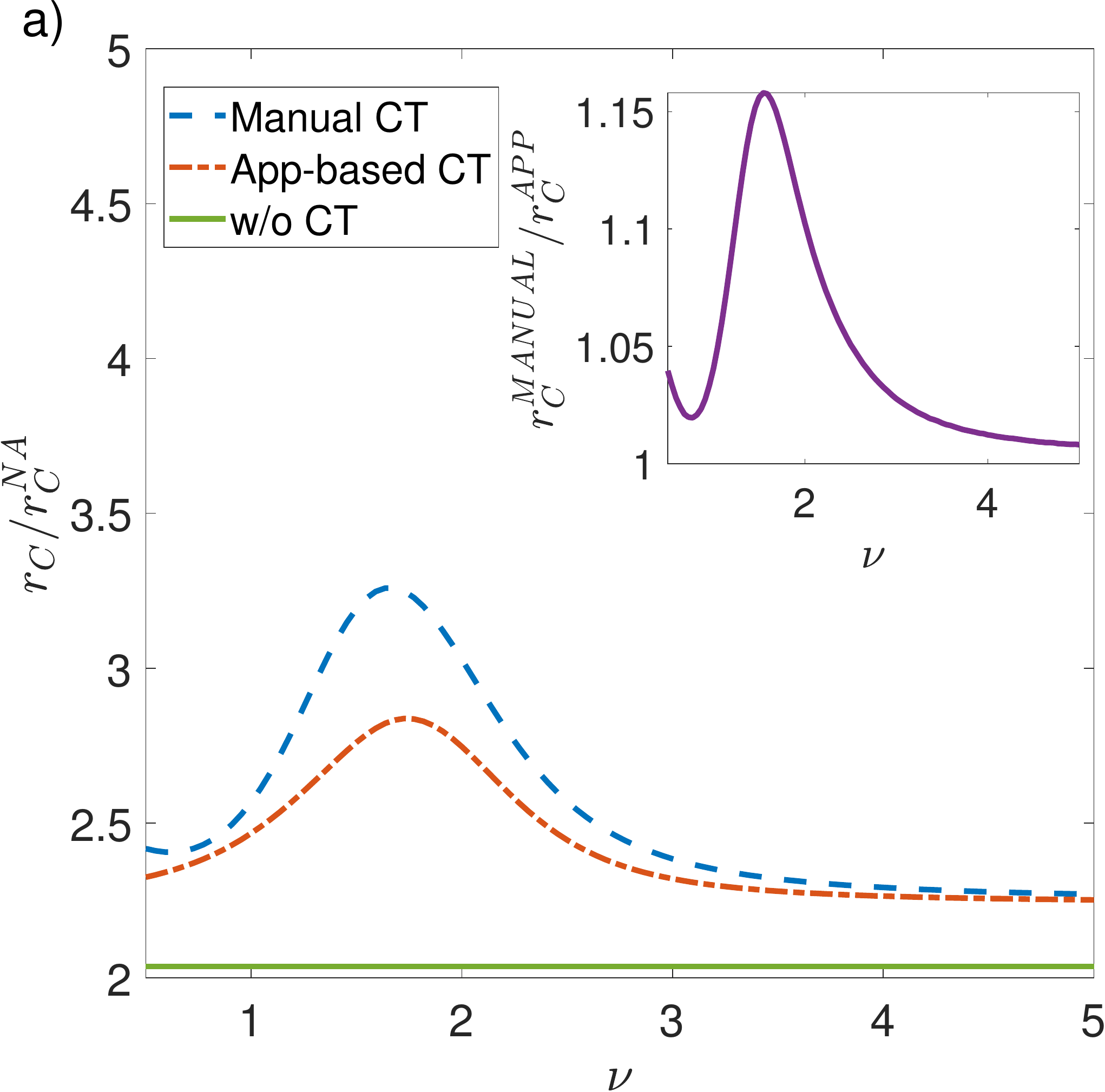}
\includegraphics[width=0.45\textwidth]{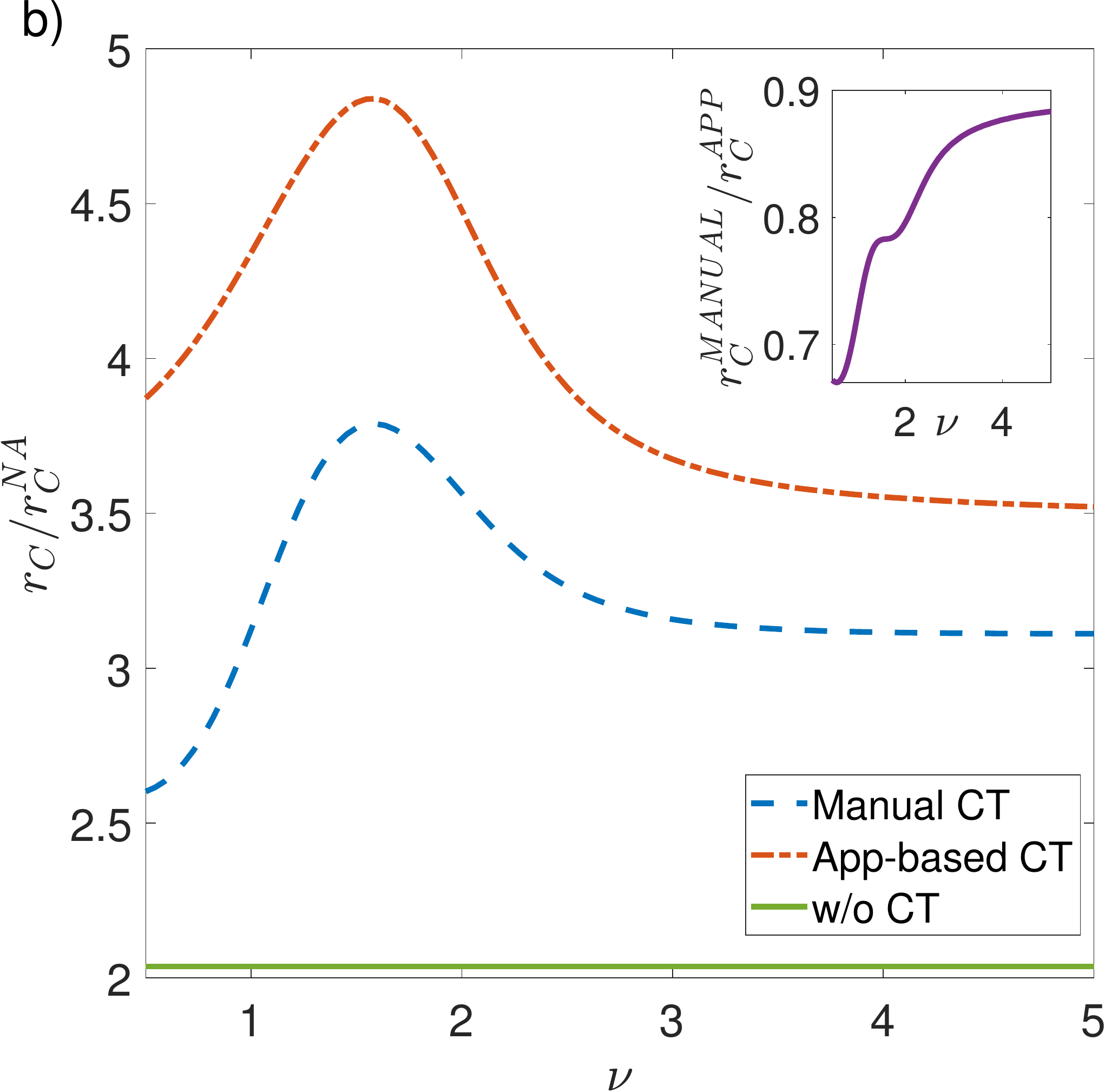}\\
\includegraphics[width=0.45\textwidth]{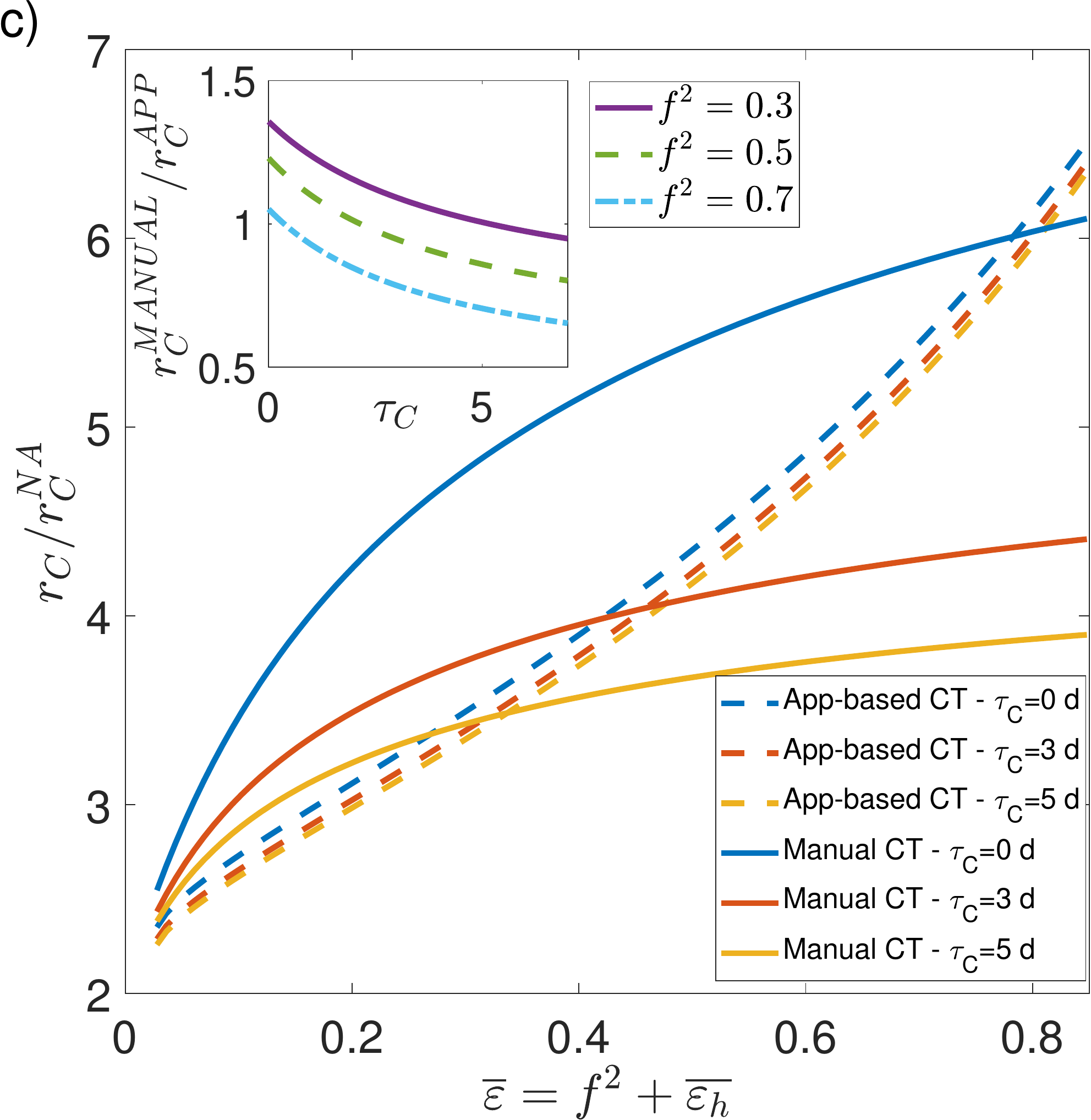}\hspace{0.1cm}\includegraphics[width=0.45\textwidth]{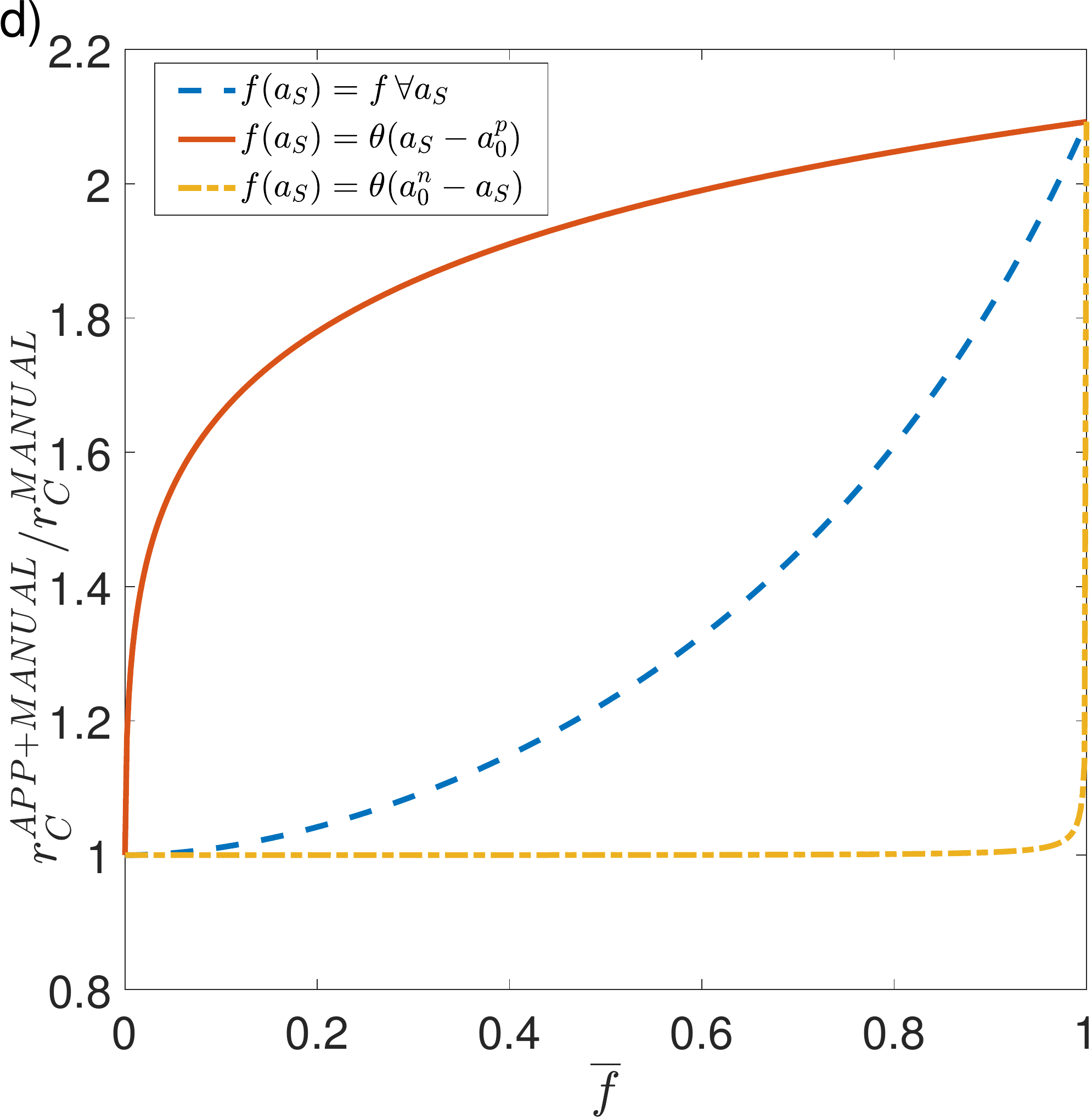}
  \caption{{\bf Effects of deterministic household contact tracing
        and effects of correlations between app adoption and
        activity.} In panels (a) and (b) we plot, as a function of the
      exponent $\nu$, the ratio between the epidemic threshold $r_C$
      for both CT protocols and the epidemic threshold of the
      non-adaptive case $r_C^{NA}$, with limited scalability, delay in
      manual CT and the additional deterministic household CT.  In the
      insets we plot the ratio between the epidemic threshold of the
      augmented manual CT, $r_C^{MANUAL}$, and that of the augmented
      app-based CT, $r_C^{APP}$, as a function of $\nu$. In panel (a)
      $f^2=0.1$ and $\tau_C=3$ days, while in panel (b) $f^2=0.6$ and
      $\tau_C=5$ days. In panel (c) we plot the ratio $r_C/r_C^{NA}$
      as a function of $\overline{\epsilon}=f^2+\overline{\epsilon_h}$
      for both augmented CT protocols, setting $\nu=1.5$ and for several values
      of $\tau_C$: in the inset we plot the ratio
      $r_C^{MANUAL}/r_C^{APP}$ as a function of $\tau_C$ for several
      $f^2$ values.  In all panels (a)-(c) we fix $s=3$ (size of
      household) and $\epsilon_h^*=1$, corresponding to
      $\overline{\epsilon_h}=0.028$. In panel (d) we plot, as a function
      of $\overline{f}$, the ratio
      between $r_C^{MANUAL+APP}$, the epidemic threshold when the
      hybrid CT protocol is implemented, and $r_C^{MANUAL}$, the
      epidemic threshold when only the manual CT is implemented. The
      curves correspond to the uncorrelated, positively correlated and
      negatively correlated $f(a_S)$ (see legend), and are obtained fixing $\overline{\epsilon}=0.4$, $\tau_C=3$ days, $\nu=1.5$. In all
      panels the distribution $\rho(a_S,b_S)$ is given by Eq.~\eqref{eq:heterog_act_rho}, $a_M/a_m=10^3$, $\overline{a_S}=6.7 \,
      days^{-1}$, $\delta=0.57$, $\tau_P=1.5$ days, $\tau=14$ days,
      $k_c=130$, $T_{CT}=14$ days.}
 \label{fig:social_circ}
\end{figure}

\subsection{Deterministic household CT}
  Typically when an individual becomes an index case,
  developing symptoms, her household is always traced and isolated,
  regardless of the CT protocol implemented. To take into account this
  effect,
  we augment both CT protocols with a deterministic contribution:
  a number of contacts, corresponding to the household, is always
  traced, both in digital and in manual CT.
  The augmented digital tracing is implemented by means of the
  hybrid CT formalism,
  assuming that the index case manually traces at least $s$ contacts
  (household size) or, if she has had less, traces them all.
  The recall probability of this manual part is
  $\epsilon_h(a_S)$ defined as in Eq.~(2) of the main paper, with
  $k_C=s$ and $\epsilon_h^*=1$.
  The probability of tracing a contact within this augmented
  digital protocol is $f^2+\overline{\epsilon_h}$
  (where $\overline{\epsilon_h} = \int d a_S \epsilon_h(a_S) \rho_S(a_S)$):
  thus, to compare its efficacy with the manual CT, we fix the same
  probability to trace a contact, that is
  $\overline{\epsilon}=f^2+\overline{\epsilon_h}$. In this way, even
  for the manual CT at least the tracing of the household is necessarily
  performed.

Supplementary Fig.~\ref{fig:social_circ}(a)-(c) shows that for low
  values of $\overline{\epsilon}=f^2+\overline{\epsilon_h}$
  (i.e. small $f^2$) manual CT is more advantageous than the
  digital one, even taking into account the deterministic tracing
  of contacts in the household.
  The digital protocol becomes more effective only for high app
  adoption rates and unrealistic long delays: therefore, our results
  are robust to the addition of this deterministic household CT contribution.

\subsection{Correlation between probability of app adoption and individual activity}
We considered the probability of downloading the app
  uniform over the population, since economic and personal factors can
  produce opposite forces which correlate and anticorrelate the
  probability of downloading the app with the individuals'
  activity. Moreover personal data of CT app users are not available,
  due to privacy issues. However, evidences are currently emerging
  that those who download the app are individuals who engage very
  cautious behaviors, i.e. $f$ and $a_S$ are
  anticorrelated~\cite{Wyl2020,Saw2020}.

We implement the hybrid CT protocol on a heterogeneous
  population, investigating the effectiveness of digital CT varying
  the app adoption level $\overline{f}=\int d a_S f(a_S) \rho_S(a_S)$
  and the correlations between $f$ and $a_S$. We consider three
  extreme cases of correlations: the uncorrelated case in which
  $f(a_S)=\overline{f}=f \, \forall a_S$;
    the completely positively correlated case
    $f(a_S)=\theta(a_S-a_0^p)$, with $\theta(x)$ the
  Heaviside step function;
    the completely negatively correlated
  case $f(a_S)=\theta(a_0^n-a_S)$. These borderline cases are hardly
  realistic, however they allow to obtain useful information on the
  role of correlations: more realistic shapes of $f(a_S)$ are
  interesting directions for future work.

To compare these three cases, we set the thresholds $a_0^n$ and $a_0^p$ by fixing $\overline{f}$: Supplementary Fig.~\ref{fig:social_circ}(d) shows that correlations have a strong impact on the effectiveness of digital CT. If the app is downloaded from all hubs (positive correlations) a low level of adoption $\overline{f} \sim 0$ is enough to obtain a significant increase in the epidemic threshold; on the contrary, if the app is downloaded only by very cautious people (negative correlations) the effect of the digital CT becomes significant only for very high adoption level $\overline{f} \sim 1$, worse than the uncorrelated case.

The effects of correlations in app adoption further strengthens our results, highlighting the dominant role of manual CT in the current situation of negative correlations~\cite{Wyl2020,Saw2020}. Furthermore these results suggest future directions to make digital CT more effective exploiting heterogeneities.
